\documentclass[12pt,a4paper]{article}
\pdfoutput=1
\usepackage{graphicx,epsfig}
\usepackage{dcolumn} 
\usepackage{slashed,xcolor,amsmath,amssymb}
\usepackage{jheppub}
\usepackage{enumitem}
\usepackage{slashbox}
\usepackage{gensymb}
\usepackage{comment}
\usepackage[normalem]{ulem}
\usepackage{caption, subcaption}
\usepackage{multirow}
\usepackage{bigcenter}
\usepackage{verbatim}
\newcommand{\be}{\begin{equation}}
\newcommand{\ee}{\end{equation}}
\newcommand{\bea}{\begin{eqnarray}}
\newcommand{\eea}{\end{eqnarray}}
\newcommand{\bit}{\begin{itemize}}
\newcommand{\eit}{\end{itemize}}

\def\gsim{\lower0.5ex\hbox{$\:\buildrel >\over\sim\:$}}
\def\lsim{\lower0.5ex\hbox{$\:\buildrel <\over\sim\:$}}

\hypersetup{
   colorlinks=true,       
   linkcolor=blue,        
   citecolor=red,         
   filecolor=magenta,     
}

\usepackage{array}
\newcolumntype{x}[1]{>{\centering\arraybackslash\hspace{0pt}}p{#1}}

\usepackage{bm}

\preprint{BONN-TH-2025-10}

\title{Long-lived Light Mediators in a Higgs Portal Model at the FCC-ee}

\author{Biplob Bhattacherjee$^1$, Camellia Bose$^1$, Herbi K. Dreiner$^2$, Nivedita Ghosh$^3$, \\ Shigeki Matsumoto$^3$, Rhitaja Sengupta$^2$}

\affiliation{\vspace*{0.1in}$^1$ Centre for High Energy Physics, Indian Institute of Science, Bengaluru 560012, India}

\affiliation{\vspace*{0.1in}$^2$ Bethe Center for Theoretical Physics and Physikalisches Institut der Universit\"at Bonn, Nu\ss allee 12, Bonn 53115, Germany}

\affiliation{\vspace*{0.1in}$^3$ Kavli IPMU (WPI), UTIAS, University of Tokyo, Kashiwa, Chiba, 277-8583, Japan}

\emailAdd{biplob@iisc.ac.in}
\emailAdd{camelliabose@iisc.ac.in}
\emailAdd{dreiner@uni-bonn.de}
\emailAdd{nivedita.ghosh@ipmu.jp}
\emailAdd{shigeki.matsumoto@ipmu.jp}
\emailAdd{rsengupt@uni-bonn.de}

\abstract{In the search for beyond the Standard Model (SM) physics, long-lived particles (LLPs) have emerged as potential candidates and are being explored in various ongoing experiments. Future lepton colliders, such as the FCC-ee, shall provide an excellent opportunity to probe LLPs, owing to their clean environment and improved particle identification. This study investigates the potential of the proposed {\bf I}nnovative {\bf D}etector for an {\bf E}lectron-Positron {\bf A}ccelerator (IDEA) detector at FCC-ee in the detection of LLPs produced from $B$-meson and Higgs boson decays. We explore benchmark scenarios for different final states resulting from LLP decays, including a detailed analysis of the SM long-lived hadronic background. Additionally, we propose dedicated LLP detectors with different configurations, dimensions, and locations with respect to the IDEA detector. DELIGHT B, originally proposed as a dedicated LLP detector for the FCC-hh, stands out as the detector with the maximum efficiency for detecting LLPs produced at FCC-ee.
We find that cylindrical detector configurations, if feasible to construct around the IDEA detector, would also enhance sensitivity for LLPs mostly decaying outside it.
}

\begin{document}

\maketitle


\section{Introduction}
\label{sec:intro}

Our quest to find any hint of physics beyond the Standard Model of particle physics (SM) is fueled by open 
questions, like what is a feasible candidate for dark 
matter, how to generate the observed small neutrino masses, 
and why is there a baryon asymmetry in the Universe?
A decade after the Higgs boson discovery at the LHC, we are 
at a juncture where the lack of significant observation of 
Beyond the Standard Model (BSM) physics 
has led to a 
paradigm shift in our search strategies. A growing interest is to expand the 
parameter space of BSM physics searches beyond the conventional regime by letting go of some of our 
fundamental assumptions. One such assumption is the prompt decay of a hypothetical newly 
produced particle in a collider experiment. Even within SM, we have a wide spectrum of particle lifetime, including some quite long lifetimes. For similar reasons, we can naturally have long-lived particles (LLP) in many well-motivated BSM models, like supersymmetry (SUSY)\,
\cite{Dimopoulos:1996vz,Farrar:1996rg,Giudice:1998bp, 
Baer:1998pg, Choudhury:1999tn, Mafi:1999dg, Kraan:2004tz, 
Mackeprang:2006gx,Arkani-Hamed:2004ymt, Giudice:2004tc, 
Hewett:2004nw, Arvanitaki:2005nq, Meade:2010ji, 
Dreiner:2009ic, Conley:2010jk, Fan:2012jf, Bhattacharyya:2012ct, Bhattacherjee:2012ed, Arvanitaki:2012ps, Banerjee:2016uyt, Nagata:2017gci, Banerjee:2018uut, Ito:2018asa, Dreiner:2020qbi, Dreiner:2023yus}, dark matter models\,\cite{Thomas:1998wy,Cirelli:2005uq,FileviezPerez:2008bj,Tucker-Smith:2001myb,Bai:2011jg,Weiner:2012cb,Izaguirre:2015zva,Griest:1990kh,Baker:2015qna,Khoze:2017ixx,Garny:2017rxs,Hochberg:2015vrg,Hall:2009bx,Co:2015pka,Hessler:2016kwm,Ghosh:2017vhe,Belanger:2018sti,Goudelis:2018xqi,Garny:2018ali}, models with heavy neutral leptons \cite{Minkowski:1977sc,Yanagida:1979as,Gell-Mann:1979vob,Mohapatra:1979ia,Mohapatra:1986bd,Keung:1983uu,Ferrari:2000sp,Helo:2013esa, Maiezza:2015lza,Izaguirre:2015pga,Batell:2016zod,Nemevsek:2016enw,Accomando:2017qcs, Helo:2018qej,Chakraborty:2018khw,  Cottin:2021lzz, Abdullahi:2022jlv, Batell:2022ogj,Pati:1974yy, Mohapatra:1974gc,PhysRevD.12.1502,Bhardwaj:2018lma,Das:2022rbl,Das:2019fee,Chiang:2019ajm,Das:2018tbd}, gauge and Higgs portal models\,\cite{Holdom:1985ag,Langacker:2008yv,Boehm:2003ha,Pospelov:2007mp,Arkani-Hamed:2008hhe,Buckley:2009in,Silveira:1985rk,Curtin:2013fra,Craig:2015pha}.
The unique nature of long-lived particle signatures, especially for neutral LLPs, may explain why they have been elusive until now.
Several phenomenological studies have explored a plethora of LLP signatures arising in a wide variety of models\,\cite{deVries:2015mfw, Banerjee:2017hmw, Dercks:2018wum, Bhattacherjee:2019fpt, Banerjee:2019ktv, DeVries:2020jbs, Bhattacherjee:2020nno, Bhattacherjee:2021rml, Bhattacherjee:2021qaa, Adhikary:2022pni, Ovchynnikov:2022its, Bandyopadhyay:2022mej, Bandyopadhyay:2023joz, Bhattacherjee:2023plj, Bhattacherjee:2023evs, Bhattacherjee:2023kxw, Bandyopadhyay:2023lvo, Gunther:2023vmz, deVries:2024mla, Wang:2024ieo}. The ATLAS, CMS and LHCb\,
\cite{ATLAS:2015xit,ATLAS:2018rjc,ATLAS:2018niw,ATLAS:2018tup,ATLAS:2019fwx,ATLAS:2019tkk,ATLAS:2019jcm,ATL-PHYS-PUB-2019-002,ATLAS:2020xyo,ATLAS-CONF-2021-032,ATLAS:2021jig,CMS:2014hka,CMS:2017kku,CMS:2018bvr,CMS-PAS-FTR-18-002,CMS:2019zxa,CMS:2020atg,CMS-PAS-EXO-19-021,CMS:2021juv,CMS:2021kdm,CMS:2021yhb,LHCb:2016buh,LHCb:2016inz,LHCb:2016awg,LHCb:2017xxn,LHCb:2019vmc,LHCb:2020akw} experimental collaborations at the LHC have searched for different LLP signatures in various sub-detectors for a range of benchmark models, and set limits in the mass and 
lifetime plane. Apart from these collider detectors, several beam dump experiments also contribute to LLP searches\,\cite{E949:2008btt,BNL-E949:2009dza,NA62:2020pwi,NA62:2020xlg,NA62:2021zjw,Gorbunov:2021ccu,CHARM:1985anb,Egana-Ugrinovic:2019wzj}. Additionally, there are several proposals for dedicated detectors for neutral LLPs, like  
FASER-2 \cite{Feng:2022inv}, FACET
\cite{Cerci:2021nlb}, MAPP-MoEDAL \cite{MoEDAL-MAPP:2022kyr}, MATHUSLA \cite{Curtin:2018mvb,MATHUSLA:2019qpy,Curtin:2023skh}, CODEX-b \cite{Aielli:2019ivi}, ANUBIS \cite{Bauer:2019vqk}, which are proposed to be placed around different collider interaction points (IP) at 
the High Luminosity LHC (HL-LHC) in the forward or transverse directions. There is also a recent proposal of a gaseous fixed target experiment, called SHIFT\,\cite{Niedziela:2024khw}, around 160\,m from the CMS IP. These detectors are placed far enough from the IP to ensure that we shield most of the SM backgrounds, providing a cleaner environment where with the 
observation of a few events (3-4), one can claim discovery. The FASER detector placed along the ATLAS beam line has already started collecting data and has constrained some regions of the displaced dark photon parameter space\,\cite{FASER:2023tle}. These dedicated detector proposals 
make use of existing empty shafts or caverns around the IP in the present collider complex. However, this might not be optimal for LLPs motivated by BSM models. For future colliders, we have the opportunity to optimize the LLP dedicated detector designs to maximize sensitivity for various LLP models. There are few studies which explore the idea of dedicated detectors for future colliders\,\cite{Blondel:2022qqo,Wang:2019xvx,Chrzaszcz:2020emg,Schafer:2022shi,Boyarsky:2022epg,MammenAbraham:2024gun,Lu:2024fxs}. We propose the DELIGHT and FOREHUNT dedicated detectors for the Future Circular Hadronic Collider (FCC-hh)\,\cite{Bhattacherjee:2021rml,Bhattacherjee:2023plj}. 

Given the rich program to look for hints of long-lived particles, we have bounds and projections on the lifetime frontier of various BSM models from present and future collider searches, beam dump experiments, and dedicated LLP detectors. 
At this stage, it is important to study the unique role that future colliders will play in this search program. This would be instrumental in aiding the physics case of these colliders. The next collider would most likely be a lepton 
collider, given the crucial role of precision measurements in particle physics, which can even be sensitive to new physics at very high energy scales. It is very common in particle physics that precision measurements precede and hint towards major discoveries, like the discovery of weak gauge bosons, 
the top quark, and the Higgs boson. Electron-positron colliders provide a cleaner environment for performing precision studies. There are a number of proposals for electron-positron colliders $-$ the International Linear Collider (ILC)\,\cite{ILC:2013jhg,Behnke:2013lya}, the Circular Electron Positron Collider (CEPC)\,\cite{CEPCStudyGroup:2018rmc,CEPCStudyGroup:2018ghi,CEPCStudyGroup:2023quu}, the Compact Linear Collider (CLIC)\,\cite{CLIC:2018fvx,CLICdp:2018cto,Roloff:2018dqu} and the Future Circular $e^+e^-$ Collider (FCC-ee)\,\cite{FCC:2018byv,FCC:2018evy}. 
FCC is a proposal for next-generation collider experiments at CERN after the completion of the Large Hadron Collider 
(LHC) and High Luminosity LHC (HL-LHC) runs. It is proposed to have two stages $-$ an electron-positron collider (FCC-ee) followed by a proton-proton collider (FCC-hh)\,\cite{FCC:2018vvp}. FCC-ee plans to operate at center of mass energies corresponding to the $Z$ pole (91.2\,GeV), $W^+W^-$ threshold (161\,GeV), $HZ$ production peak (240\,GeV), 
and the $t\bar{t}$ threshold (350/365\,GeV), while FCC-hh is 
aiming for a center of mass energy of 100\,TeV. 

Owing to a cleaner environment and better particle 
identification capabilities, the FCC-ee could play a multifaceted role in exploring long-lived particles\,\cite{Alipour-Fard:2018lsf,Cheung:2019qdr,Ripellino:2024tqm}. To illustrate the unique power of lepton colliders, we 
investigate the minimal extension of the SM 
with a light scalar particle, which couples to the SM 
particles only through the Higgs boson. This is a 
well-motivated model in the context of dark matter\,\cite{Matsumoto:2018acr}, and the light scalar, which 
mediates interactions with the dark matter, 
can be long-lived. It is also phenomenologically attractive 
since the LLP has a wide range of production and decay 
modes, depending on its mass and couplings with the 125\,GeV 
Higgs boson. There is some uncertainty in 
the computation of the decay modes 
of the LLP in a mass range where there is a transition between 
the use of chiral perturbation theory, dispersive relations, 
and the perturbative spectator model\,\cite{Winkler:2018qyg}.
Experimental collaborations have extensively studied their 
sensitivities for this model, covering different regions in the mass and mixing angle plane. However, many of the 
dedicated detector proposals have yet to be approved, and we don't know which of them will be realized in the future. In this situation, we identify some important benchmarks and study the role of electron-positron colliders in probing them. For simplicity, we perform the analyses in the context of 
the FCC-ee.
The benchmarks are motivated by one of 
the following questions:
\begin{enumerate}
    \item Can the FCC-ee be sensitive to any region in the parameter space that lies outside the coverage of any currently proposed experiment?
    \item Can the FCC-ee help distinguish this model from others and measure its parameters if we observe a signal in one or more of the proposed future detectors?
    \item Can we use the particle identification capabilities of the FCC-ee to be sensitive to the mediator masses, where the dominant decay mode is to mesons, like pions and kaons, where LHC or other hadron colliders lose sensitivity?
\end{enumerate}

We also discuss the possibility of dedicated LLP detectors around the FCC-ee interaction point and their optimal design and position for probing LLPs. 
  
The rest of the paper is organized as follows: in Sec.\,\ref{sec:model}, we discuss the Higgs portal model that we 
consider here and the various production and decay processes of the dark Higgs boson it entails. In 
Sec.\,\ref{sec:benchmark}, we discuss the current and future 
projected sensitivity of various experiments to the parameter space of the dark Higgs model and choose the benchmark points that can be interesting for the electron-positron colliders. 
We discuss the detection prospect of these benchmarks at the FCC-ee in Sec.\,\ref{sec:analysis}. 
We extensively explore various options of dedicated detectors for the FCC-ee in Sec.\,\ref{sec:dedicated}. 
Finally, we conclude in Sec.\,\ref{sec:concl}.


\section{Scalar LLPs in the Higgs portal}
\label{sec:model}

One of the minimal extensions of the SM is the addition of a scalar field, which is a singlet under the SM gauge group. 
It only interacts with the SM fields by mixing with the SM Higgs field.
The generic renormalizable Lagrangian for this minimal model is as follows\,\cite{Feng:2017vli}:
\begin{equation}
    \mathcal{L} = \mathcal{L}_{\rm SM} + \mu_1^3 S + \mu^2 S^2 - \mu_3 S^3 - \mu_{12} S |H|^2 - \frac{1}{4}\lambda_S S^4 - \epsilon S^2 |H|^2,
    \label{eq:Lmodel}
\end{equation}
where $S$ is the new scalar field and $H$ is the SM  
complex scalar Higgs doublet. Imposing a discrete $\mathbb 
{Z}_2$ symmetry on $S$ sets $\mu_1=\mu_3=\mu_{12}=0$. Minimizing 
the scalar potential and diagonalizing the mass matrix lead to the physical states $h$ and $\phi$, where the former 
denotes the discovered 125\,GeV SM-like Higgs boson, and the latter is the dark Higgs boson. Thus, after the spontaneous 
symmetry breaking of the SM Higgs field, we are left with the following Lagrangian for $\phi$\,\cite{Feng:2017vli}:
\begin{equation}
    \mathcal{L} = -m_{\phi}^2\phi^2 -\sin\theta\,\frac{m_f}{v} \phi\bar{f}f - \lambda v\,h\phi\phi + \dots,
    \label{eq:Lphi}
\end{equation}
where $m_\phi$ is the mass of the dark Higgs boson, $\theta$ is the mixing angle between $h$ and $\phi$, $\lambda$ is the 
trilinear coupling between the two scalars, $f$ denotes the SM fermions, and $v\approx 246$\,GeV is the {\it vev} of the Higgs doublet. 
We have $\lambda \simeq \epsilon v \cos^3{\theta} + \mathcal{O}(\sin^2{\theta}\cos{\theta})$, where the first term is not suppressed by small $\sin{\theta}$, while the rest of the terms have a $\sin^2{\theta}\cos{\theta}$ dependence.
In Eq.\,(\ref{eq:Lphi}), we have omitted further cubic and quartic interactions of $\phi$ and $h$, irrelevant to the phenomenology of the dark Higgs boson discussed in the present work. 

\begin{figure}[hbt!]
    \centering
    \includegraphics[width=0.8\textwidth]{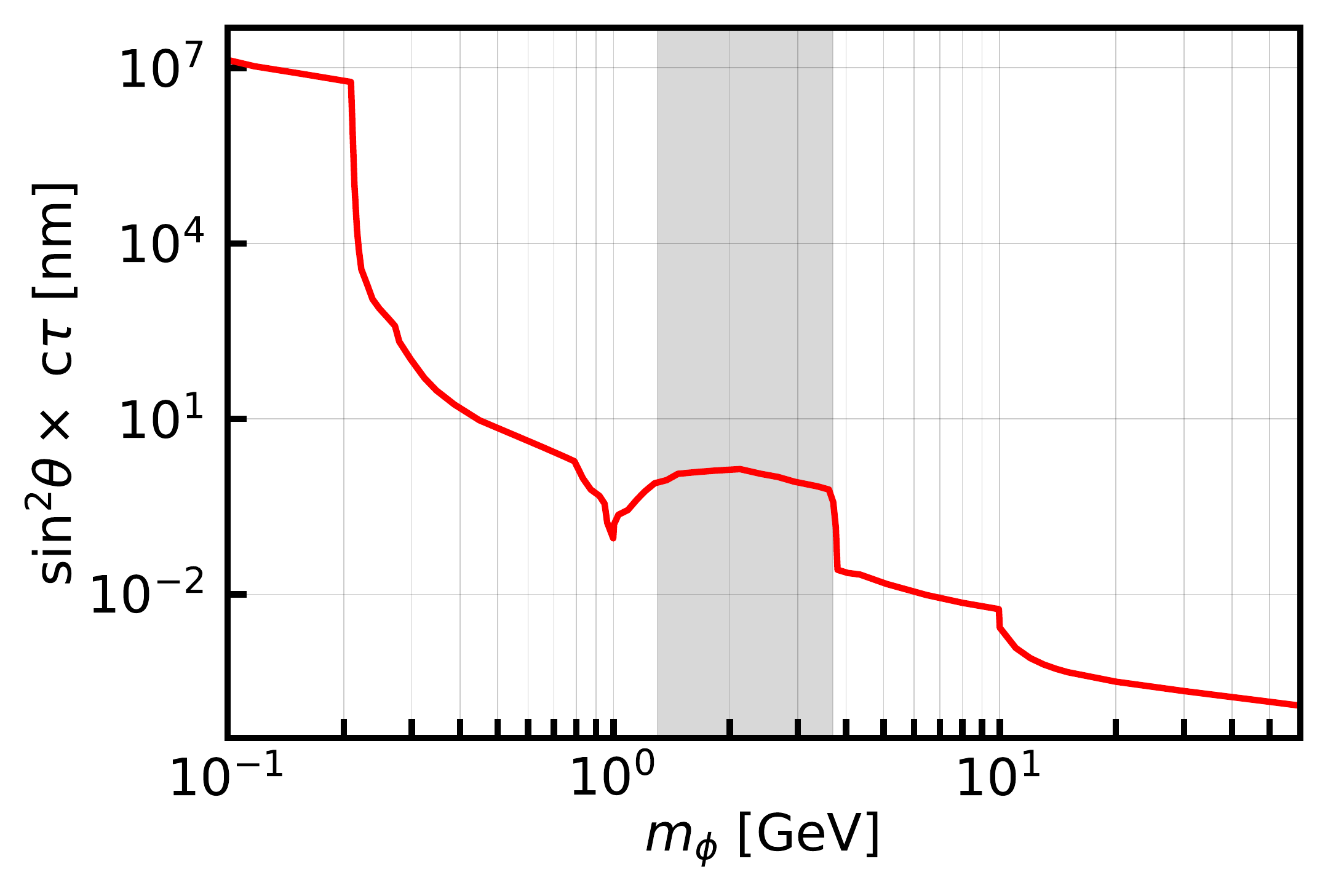}
    \caption{Variation of the $\sin^2\theta\times c\tau$ as a function of the dark Higgs boson mass in the minimal model.}
    \label{fig:model_sintheta}
\end{figure}

The second term in Eq.\,(\ref{eq:Lphi}) denotes the Yukawa interactions of the SM fermions with the dark Higgs boson, where the Yukawa couplings have the same structure as those of the SM-like Higgs boson, 
suppressed by $\sin\theta$. The 
various decay modes of $\phi$ depending on its mass and the relative branching fractions have been discussed extensively in multiple studies\,\cite{Dolan:2014ska,Winkler:2018qyg,Matsumoto:2018acr,FASER:2018eoc,Ferber:2023iso}.
The total decay width of the new dark Higgs boson, $\Gamma_{\phi, {\rm tot}}$, scales as:
\begin{equation}
    \Gamma_{\phi, {\rm tot}} (m_\phi) = \Gamma_{h, {\rm tot}}(m_h=m_\phi) \times \sin^2\theta
    \label{eq:total_gamma}
\end{equation}
where $\Gamma_{h, {\rm tot}}(m_h=m_\phi)$ denotes the total 
decay width estimated assuming SM-like Higgs boson couplings 
with its mass set equal to the dark Higgs boson mass. 
Eq.\,(\ref{eq:total_gamma}) shows that reducing the 
mixing angle $\theta$, reduces $\Gamma_{\phi, {\rm tot}}$, which increases the decay length of $\phi$. Since the mean 
proper decay length of $\phi$, \textit{i.e.}, $c\tau$\,\footnote{We refer to the mean proper decay length as simply the decay length from hereafter for simplicity, unless stated otherwise.}, is 
the inverse of $\Gamma_{\phi, {\rm tot}}$, the product $\sin^2\theta \times c\tau$ depends only on the mass 
of $\phi$, and is proportional to the inverse of the quantity $\Gamma_{h, {\rm tot}}(m_h=m_\phi)$.
Fig.\,\ref{fig:model_sintheta} shows the variation of $
\sin^2\theta \times c\tau$ with $m_\phi$. 
Depending on the mass of $\phi$, it can have various 
decay modes, such as $\mu^+\mu^-$, $\pi^+\pi^-$, $K^+K^-$, $c\bar{c}$, $\tau^+\tau^-$, and $b\bar{b}$.
The branching fraction of the dark Higgs boson to mesons, like pions when $2m_\pi<m_\phi<2.0$\,GeV and kaons when $2m_K<m_\phi<2.0$\,GeV, is estimated using dispersive analysis\,\cite{Winkler:2018qyg}. For $m_\phi>2.0$\,GeV, the branching fractions to SM particles are calculated using \texttt{HDECAY}\,\cite{Djouadi:1997yw,Djouadi:2018xqq}.
The {\it gray} shaded mass range around 2.0\,GeV in Fig.\,\ref{fig:model_sintheta} denotes the region where the total decay width is uncertain due to QCD non-perturbative effects and the transition from the non-perturbative to perturbative regime\,\cite{Ferber:2023iso}\,\footnote{Ref.\,\cite{Blackstone:2024ouf} estimated theoretical uncertainties on the di-meson decay widths and their effect on the experimental sensitivities is studied in Ref.\,\cite{DallaValleGarcia:2025aeq}.}. The decay length of $\phi$ for a particular $m_\phi$ and $\sin\theta$ in the minimal model can be extracted from Fig.\,\ref{fig:model_sintheta}.
In the present work, we consider the parameter region where $\phi$ is lighter than half the Higgs boson mass, \textit{i.e.}, from $\sim 0.1$\,GeV to 60\,GeV.

\vspace{0.2cm}

Depending on the value of the trilinear coupling, we can have different production processes for $\phi$ at the colliders. 

\subsection*{Case A: Negligible Trilinear Coupling}

When the trilinear coupling is negligible, $\lambda \approx 0$, $\phi$ can only 
be produced via its Yukawa couplings with the SM fermions. These couplings have 
the same structure as those of the SM-like Higgs boson but are suppressed by 
the sin of the mixing angle. Therefore, the dominant production is from $B$ 
meson decays, where the dark Higgs boson is radiated off from the top quark in 
a $b\to s$
transition via the $t-W$ loop. The inclusive branching fraction of the $B$ 
mesons to the scalar LLP is given 
by\,\cite{Chivukula:1988gp,Feng:2017vli}:\footnote{We consider the inclusive 
branching fraction of $B$ mesons decay to any strange meson, $X_s$, and the dark Higgs boson to reduce the theoretical uncertainties arising from the form 
factors\,\cite{Chivukula:1988gp,Dolan:2014ska,Feng:2017vli,FASER:2018eoc,Boiarska:2019jym}.}
\begin{eqnarray}
 \text{Br}(B \to X_S \phi) = 5.7~ \sin^2\theta\,\left(1-\frac{m_{\phi}^2}{m_b^2}\right)^2,
 \quad \text{for}\; m_\phi< m_B-m_K.
 \label{eq:br_BtoKphi}
\end{eqnarray}
where $m_b=4.75$\,GeV.
The branching ratio is inclusive with respect to both the initial state $B$-mesons and the final state hadrons with strange quark\,\footnote{The factor $5.7$ comes from a combination of the masses of the top quark, bottom quark and $W$ boson, the CKM elements, and the phase space factor in the semi-leptonic $B$ decay, $f_{c/b}$ (Ref.\,\cite{Feng:2017vli}).}.
Although the 
production rate of kaons and light mesons ($\eta$, $\pi$, etc.) is larger compared to the production rate of $B$ mesons, the branching fraction of kaons to $\phi$ is much smaller than $\text{Br}(B \to X_S \phi)$. It 
is further suppressed for $\eta$, $\eta'$, and $\pi$ decays to the dark Higgs boson\,\cite{Feng:2017vli}. Moreover, production from kaons provides sensitivity to a smaller mass range of $\phi$, where already strong constraints from beam 
dump experiments, like E949\,\cite{E949:2008btt,BNL-E949:2009dza}, are present. Hence, we do not consider these production modes in our study.
We consider $\phi$ masses up to 4.5\,GeV in this case, as they can be produced from $B$ meson decays.

Another production mode for $\phi$ in this case can be from the decay of the $Z$ boson via the process $Z\rightarrow Z^* 
\phi$, where the off-shell $Z$ boson decays to leptons. 
This is analogous to Higgs-Strahlung at LEP.
This decay mode of the $Z$ boson has a very small partial width 
proportional to the square of the mixing angle and decreases rapidly with increasing mass of $\phi$. However, due to the 
large luminosity of FCC-ee at the $Z$ pole, we briefly discuss the sensitivity of this production mode of $\phi$
and the mass range it covers.

\subsection*{Case B: Large Trilinear Coupling}

When the trilinear coupling, $\lambda$, is large, $\phi$ can also be pair-produced from an on-shell or off-shell Higgs boson. 
One possible production process is again via $B$ meson decays, involving an off-shell Higgs boson ($b\rightarrow s\phi\phi$), which has a branching fraction around $2.1 \times 10^{-4}\, \lambda^2$ for an $m_\phi=1\,$GeV dark 
Higgs boson\,\cite{Feng:2017vli}.
Another possibility is Higgs boson production at colliders, 
where the Higgs boson decays to a pair of $\phi$. This process has a much 
larger branching fraction to the dark Higgs bosons, 
$\simeq 4700\,\lambda^2$ for $m_\phi=1$\,GeV\,\cite{Feng:2017vli}\,\footnote{Note that the 
current bound of Br$(h\to {\rm inv}) < 0.11$ from ATLAS\,\cite{ATLAS-CONF-2020-052} 
implies that $\lambda < 4.8 \times 10^{-3}$ for $m_\phi< m_h/2$, which is the mass range of $\phi$ considered in this work.
}, which is seven orders of magnitude larger than the former.
Moreover, the $B$ meson decay process is sensitive to a much smaller range of $m_\phi$ than the second process, which is sensitive up to $m_\phi \lesssim m_h/2 \sim 60$\,GeV. Therefore, in the present work, we only consider the process $h\rightarrow\phi\phi$ for probing the large trilinear coupling scenario.


\section{Current status and benchmark choice}
\label{sec:benchmark}

In the previous section, we discussed the minimal dark Higgs model and the 
possible production and decay modes of the new particle. The dominant 
production mode depends on whether the trilinear coupling can be neglected or 
not, where for the former, we study the process $B\rightarrow X_s\phi$, and for the latter, we consider the process $h\rightarrow \phi\phi$ at colliders. 
The production mode also depends on the mass $m_\phi$. Since this model has been explored 
extensively by various experiments, we briefly discuss the current  
status in this section and select an interesting benchmark 
points for probing this model in the future, proposed electron-positron 
colliders.

\subsection*{Case A: Negligible Trilinear Coupling}

\begin{figure}[hbt!]
    \centering
    \includegraphics[width=0.8\textwidth]{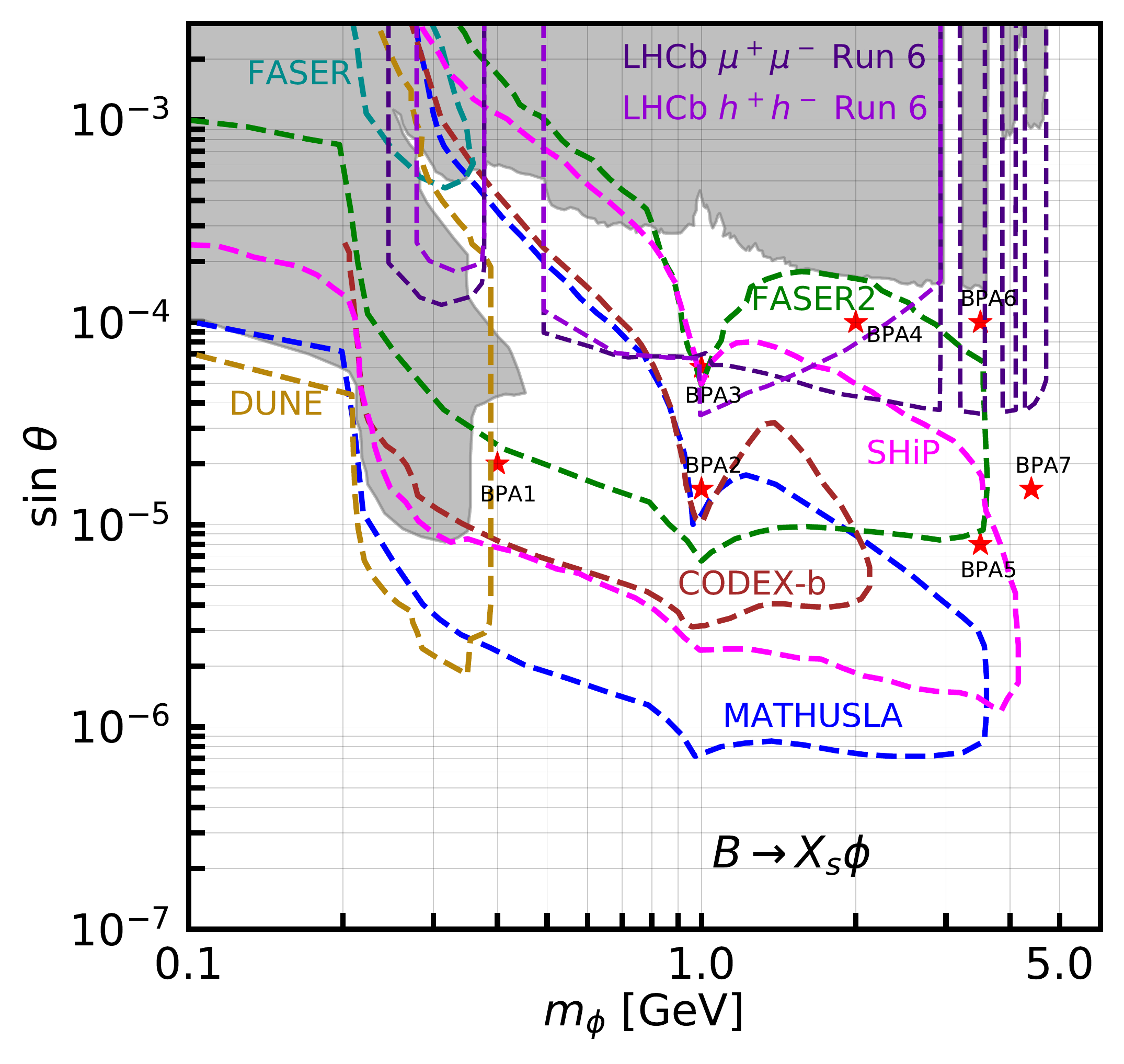}
    \caption{Current status (in {\it gray}) and future projected sensitivities of proposed experiments (in {\it dashed lines}) for the dark Higgs boson model when produced from meson decays and the trilinear coupling is negligible.
    Benchmark points for the present study are marked with {\it red stars}.}
    \label{fig:benchmarks_BtoKphi}
\end{figure}

This scenario has been extensively studied in various beam dump experiments, 
neutrino detectors, as well as collider main detectors (see 
Refs.\,\cite{Batell:2022dpx,Ferber:2023iso} for an overview). The recently 
proposed dedicated LLP detectors also have sensitivity to the dark Higgs 
parameter space. Fig.\,\ref{fig:benchmarks_BtoKphi} shows in {\it gray} the 
combined bound on the $m_\phi-\sin\theta$ parameter space from 
the existing experimental results for the mass range $m_\phi\in[0.1,5]\,$GeV. We also show the projected 
sensitivity of proposed future experiments and phenomenological projections for 
the HL-LHC to the parameter space in {\it dashed lines}. 
We briefly discuss these bounds below.

Beam dump experiments play a major role in searching for new light physics from 
various meson decays. Experiments like the E949\,\cite{E949:2008btt, 
BNL-E949:2009dza}, NA62\,\cite{NA62:2020pwi,NA62:2020xlg,NA62:2021zjw}, 
PS191\,\cite{Gorbunov:2021ccu}, CHARM\,\cite{CHARM:1985anb}, and 
KOTO\,\cite{Egana-Ugrinovic:2019wzj} have already placed bounds in the 
parameter space of the mixing angle and mass of the light scalar\,\footnote{The Big European Bubble Chamber (BEBC) CERN-WA-066
beam dump experiment also places bounds on various light new physics models\,\cite{Marocco:2020dqu,Barouki:2022bkt}, which can be translated to the dark Higgs boson model as well.}. The recently 
approved future experiment SHiP (\textbf{S}earch for \textbf{Hi}dden \textbf{P}articles)\,\cite{Alekhin:2015byh,SHiP:2015vad} is a proposed fixed target facility at the CERN SPS accelerator. It aims to explore 
hidden particles with masses less than 10\,GeV. For a scalar mass of 1\,GeV, SHiP is expected to probe mixing angles between $5\times 10^{-5}$ and $2.4\times 10^{-6}$. It is the most sensitive at a mass of 4\,GeV, probing a minimum mixing angle of $1.5\times 10^{-6}$. 
   
Neutrino experiments like LSND\,\cite{LSND:1997vqj,LSND:2001aii,Foroughi-Abari:2020gju} and 
MicroBOONE\,\cite{MicroBooNE:2021usw} also probe regions of the dark Higgs boson parameter space. Excluded regions from these experiments are included in the {\it gray} region of Fig.\,\ref{fig:benchmarks_BtoKphi}. 
The proposed DUNE\,\cite{Berryman:2019dme} experiment is expected to improve the sensitivity to lower mixing angles for masses of $\phi$ below $\sim 400$\,MeV, \textit{i.e.}, where $\phi$ can be produced from kaon decays. 
    
Searches for new physics via displaced vertices at the various collider 
experiments, like L3 at LEP\,\cite{L3:1996ome,Ferber:2023iso}, 
LHCb\,\cite{LHCb:2015nkv,LHCb:2016awg} and CMS\,\cite{Evans:2020aqs} at the 
LHC, BaBar\,\cite{BaBar:2015jvu}, and Belle-II\,\cite{Belle-II:2023ueh, 
Kachanovich:2020yhi}, can probe $\phi$ masses up to 4.5\,GeV, 
produced from the process $B\rightarrow X_s\phi$. 
For example, Ref.\,\cite{Evans:2020aqs} projects that, at the HL-LHC, one could probe mixing angles down to the $10^{-4}-10^{-5}$ range for $m_\phi$ between 0.3 and 4.5\,GeV utilizing the CMS L1 track trigger.
The bound from LHCb at the 
LHC has been reinterpreted and projected for the HL-LHC in 
Refs.\,\cite{Gligorov:2017nwh,Craik:2022riw}. While the projected limit in 
Ref.\,\cite{Gligorov:2017nwh} assumes a constant scale factor to project the present LHCb limits, Ref.\,\cite{Craik:2022riw} improves the projection by taking into account the strong lifetime dependence of the LHCb 
limits and including the hadronic decay modes to pions and kaons\,\footnote{The reason why a constant scale factor for increased luminosity cannot give a reliable projected sensitivity is based on two factors: (a) the increase in the 
center of mass energy from 8\,TeV to 14\,TeV, and (b) different mixing angle corresponds to different c$\tau$ for varying mass of the LLP, making the detection probability different for the two different energy scales.
The projected limit can become stronger or weaker for different masses than that predicted in Ref.\,\cite{Gligorov:2017nwh}.}.
Here, we consider the projected limits given for the muon and hadronic final states in the theory paper, Ref.\,\cite{Craik:2022riw}.
Being closer to the IP, the collider experiments are sensitive to relatively larger values of mixing angles. For probing smaller values of mixing angles, where $\phi$ is highly displaced and decays outside the main collider detectors, dedicated LLP detectors, like FASER2 at FPF, MATHUSLA, and CODEX-b, have been proposed. 
The FASER detector is already collecting data at the ongoing Run-3 of the LHC.

With this vast program of searches for dark Higgs boson, we choose seven benchmark points to explore the sensitivity of the future electron-positron colliders. They are described in Table\,\ref{tab:bpchoice_a} along with the dominant decay modes of the dark scalar and the existing approved or proposed experiments that are sensitive to them. The benchmark points are also marked in Fig.\,\ref{fig:benchmarks_BtoKphi} with {\it red stars}. 

\begin{table}[hbt!]
    \centering
     \resizebox{\textwidth}{!}{  
    \begin{tabular}
    {|c|x{1.2cm}|c|c|x{2.5cm}|c|c|c|c|}
    \hline
    \multirow{2}{*}{Benchmark} & $m_{\phi}$ & \multirow{2}{*}{sin\,$\theta$} & $c\tau$  & Dominant & \multicolumn{4}{c|} {Potential experiments to probe}  \\ \cline{6-9}
    & (GeV) & & (mm) & decay modes & FASER2 & LHCb (projected) & MATHUSLA & SHIP \\ \hline\hline
    \multirow{2}{*}{BPA1} & \multirow{2}{*}{0.4} & \multirow{2}{*}{$2.0\times 10^{-5}$} & \multirow{2}{*}{39666.6} & $\pi^+\pi^-:$ 76\% & \multirow{2}{*}{$\times$}  & \multirow{2}{*}{$\times$} & \multirow{2}{*}{$\checkmark$} & \multirow{2}{*}{$\checkmark$} \\ 
    & & & & $\mu^+\mu^-:$ 10\% & & & & \\
    \hline
    
    \multirow{2}{*}{BPA2} & \multirow{2}{*}{1.0} & \multirow{2}{*}{$1.5\times 10^{-5}$} & \multirow{2}{*}{554.3} & $\pi^+\pi^-:$ 50\% & \multirow{2}{*}{$\checkmark$} & \multirow{2}{*}{$\times$} & \multirow{2}{*}{$\times$} & \multirow{2}{*}{$\checkmark$} \\ 
    & & & & $K^+K^-:$ 50\% & & & & \\
    \hline
    
    \multirow{2}{*}{BPA3} & \multirow{2}{*}{1.0} & \multirow{2}{*}{$6\times 10^{-5}$} & \multirow{2}{*}{34.6} & $\pi^+\pi^-:$ 50\% & \multirow{2}{*}{$\times$} & \multirow{2}{*}{$\times$} & \multirow{2}{*}{$\times$} & \multirow{2}{*}{$\times$} \\ 
    & & & & $K^+K^-:$ 50\% & & & & \\
    \hline

      \multirow{3}{*}{BPA4} & \multirow{3}{*}{2.0} & \multirow{3}{*}{$10^{-4}$} & \multirow{3}{*}{135.2} & $\pi^+\pi^-:$ 41\% & \multirow{3}{*}{$\checkmark$} & \multirow{3}{*}{$\checkmark$} & \multirow{3}{*}{$\times$} & \multirow{3}{*}{$\times$ } \\ 
    & & & & $K^+K^-:$ 41\% & & & & \\
    & & & & $\mu^+\mu^-:$ 12\% & & & & \\
    \hline

    \multirow{3}{*}{BPA5} & \multirow{3}{*}{3.5} & \multirow{3}{*}{$8\times 10^{-6}$} & \multirow{3}{*}{10285.4} & $\pi^+\pi^-:$ 53\% & \multirow{3}{*}{$\times$} & \multirow{3}{*}{$\times$} & \multirow{3}{*}{$\times$} & \multirow{3}{*}{$\checkmark$ } \\ 
    & & & & $K^+K^-:$ 21\% & & & & \\
    & & & & $\mu^+\mu^-:$ 5\% & & & & \\
    \hline
    
    \multirow{3}{*}{BPA6} & \multirow{3}{*}{3.5} & \multirow{3}{*}{$10^{-4}$} & \multirow{3}{*}{65.8} & $\pi^+\pi^-:$ 53\% & \multirow{3}{*}{$\times$} & \multirow{3}{*}{$\checkmark$} & \multirow{3}{*}{$\times$} & \multirow{3}{*}{$\times$} \\
    & & & & $K^+K^-:$ 21\% & & & & \\
    & & & & $\mu^+\mu^-:$ 5\% & & & & \\
    \hline
    
    \multirow{2}{*}{BPA7} & \multirow{2}{*}{4.4} & \multirow{2}{*}{$1.5\times 10^{-5}$} & \multirow{2}{*}{95.0} & $c\bar{c}:$ 65\% & \multirow{2}{*}{$\times$} & \multirow{2}{*}{$\times$} & \multirow{2}{*}{$\times$} & \multirow{2}{*}{$\times$} \\ 
    & & & & $\tau^+\tau^-:$ 20\% & & & & \\
    \hline\hline
    \end{tabular}}
    \caption{Choice of benchmark points for negligible trilinear coupling.}
    \label{tab:bpchoice_a}
\end{table}

These benchmarks are chosen to represent a number of scenarios. For example, SHiP is expected to be sensitive to the benchmarks BPA1, BPA2, and BPA5. 
On the other hand, benchmarks BPA4 and BPA6 are chosen to have a relatively higher mixing angle, falling 
within the projected sensitivity of the HL-LHC LHCb analysis. In the scenario 
where the FASER2 experiment is approved, signals for benchmarks BPA2 and BPA4 
are expected to be testable. Similarly, if MATHUSLA is 
approved, BPA1 can be probed at SHiP and MATHUSLA. 

For such benchmark points, where one or more of the existing proposals are 
sensitive, we want to study whether FCC-ee can provide an additional probe that 
might be helpful in the identification of the model and an estimate of the model parameters. Some of these benchmarks are chosen 
just at the sensitivity boundary of specific experiments. For example, BPA3 is just at the boundary of the SHiP, FASER2, and LHCb projected 
sensitivities. The motivation is that while these experiments might just 
observe a few signal events, a main collider detector would still be needed to 
investigate the signals further. Since FCC-ee is possibly the next future 
collider, we selected these benchmarks for the FCC-ee analysis. Moreover, it is 
worth studying whether FCC-ee can probe any region of the parameter space 
beyond the sensitivity of the above-discussed experiments, which motivates the 
choice of benchmarks BPA3 and BPA7.

\subsection*{Case B: Large Trilinear Coupling}

\begin{figure}[hbt!]
    \centering
    \includegraphics[width=0.5\textwidth]{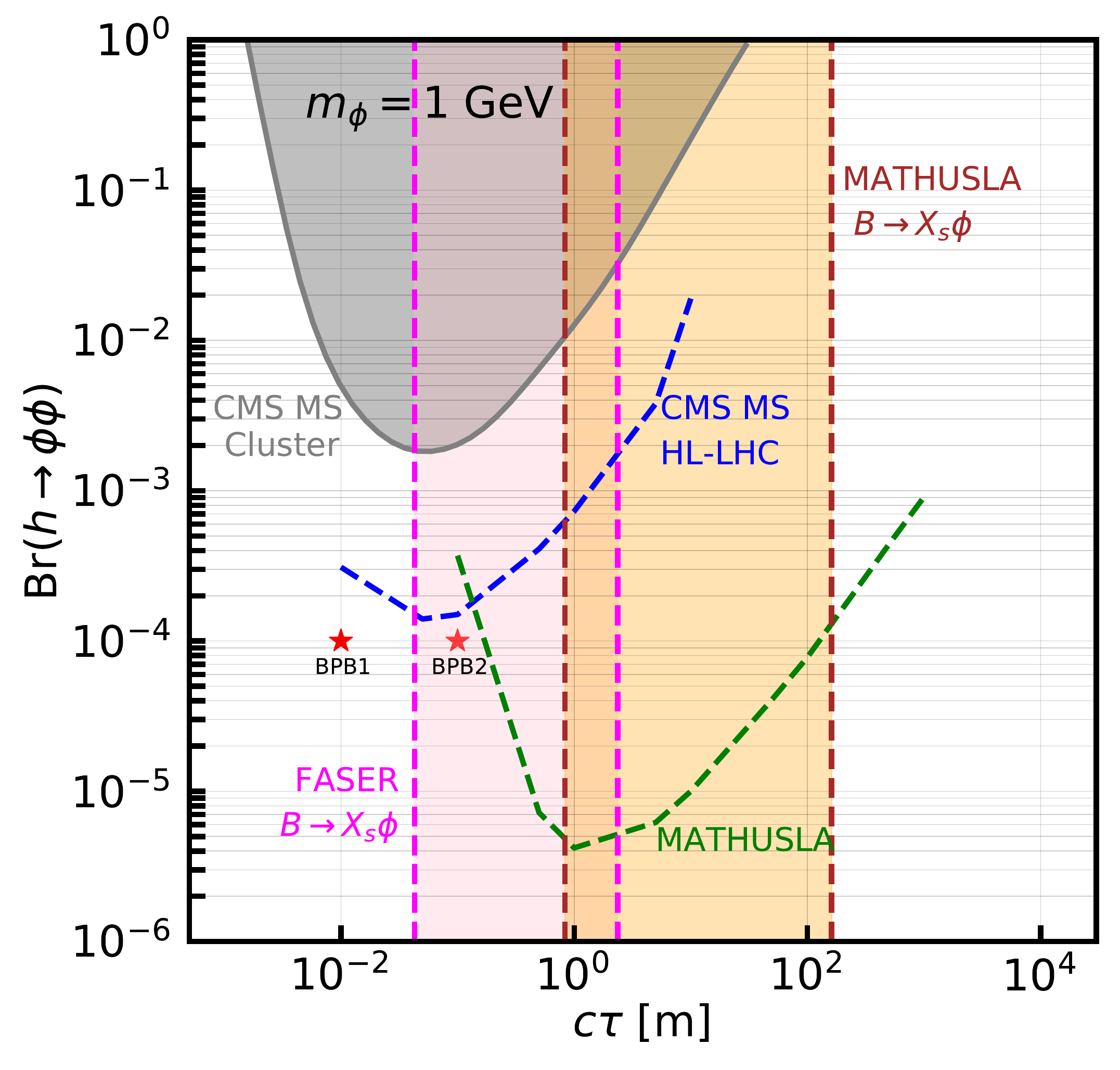}~
    \includegraphics[width=0.5\textwidth]{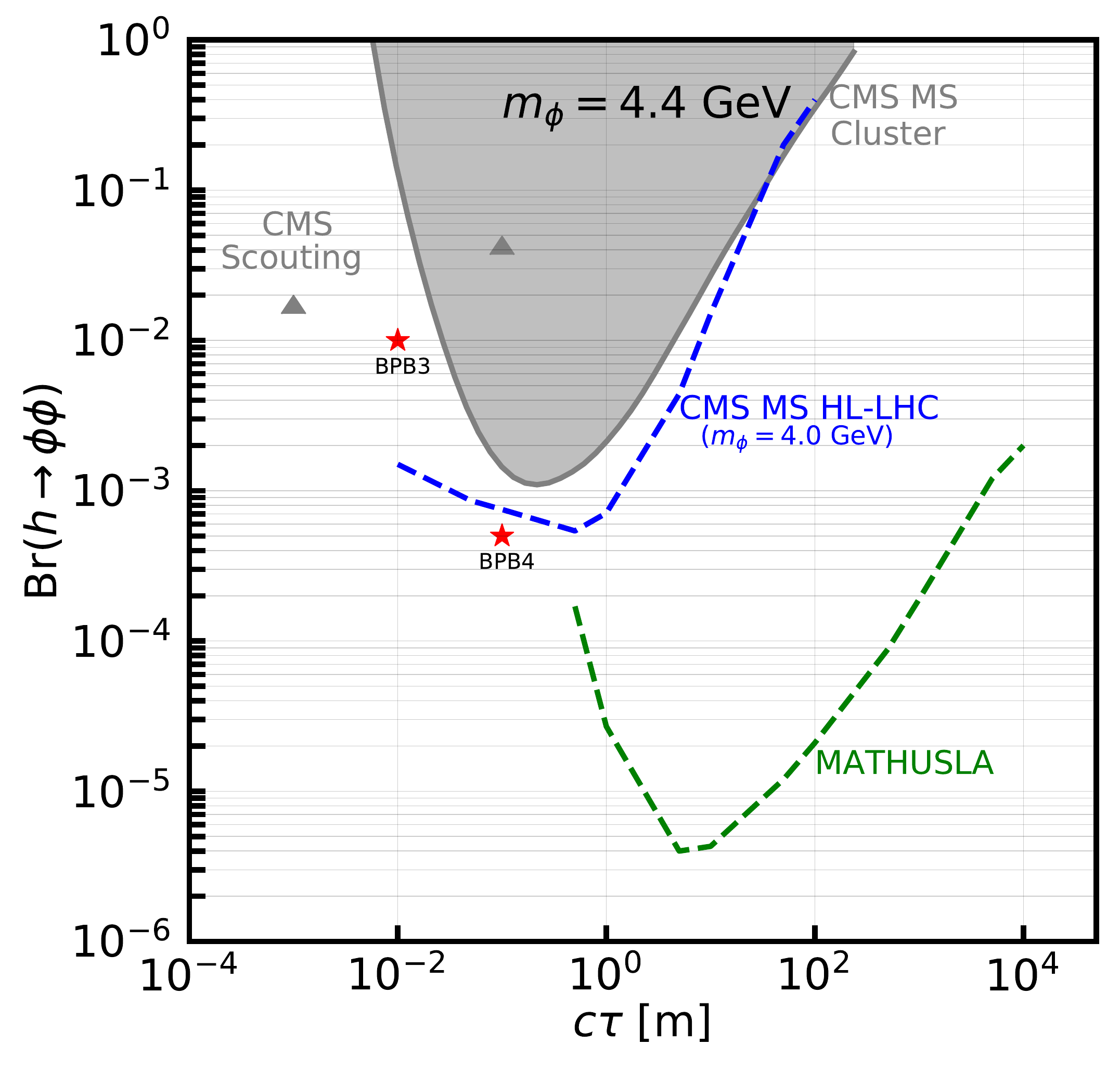}\\
    \includegraphics[width=0.5\textwidth]{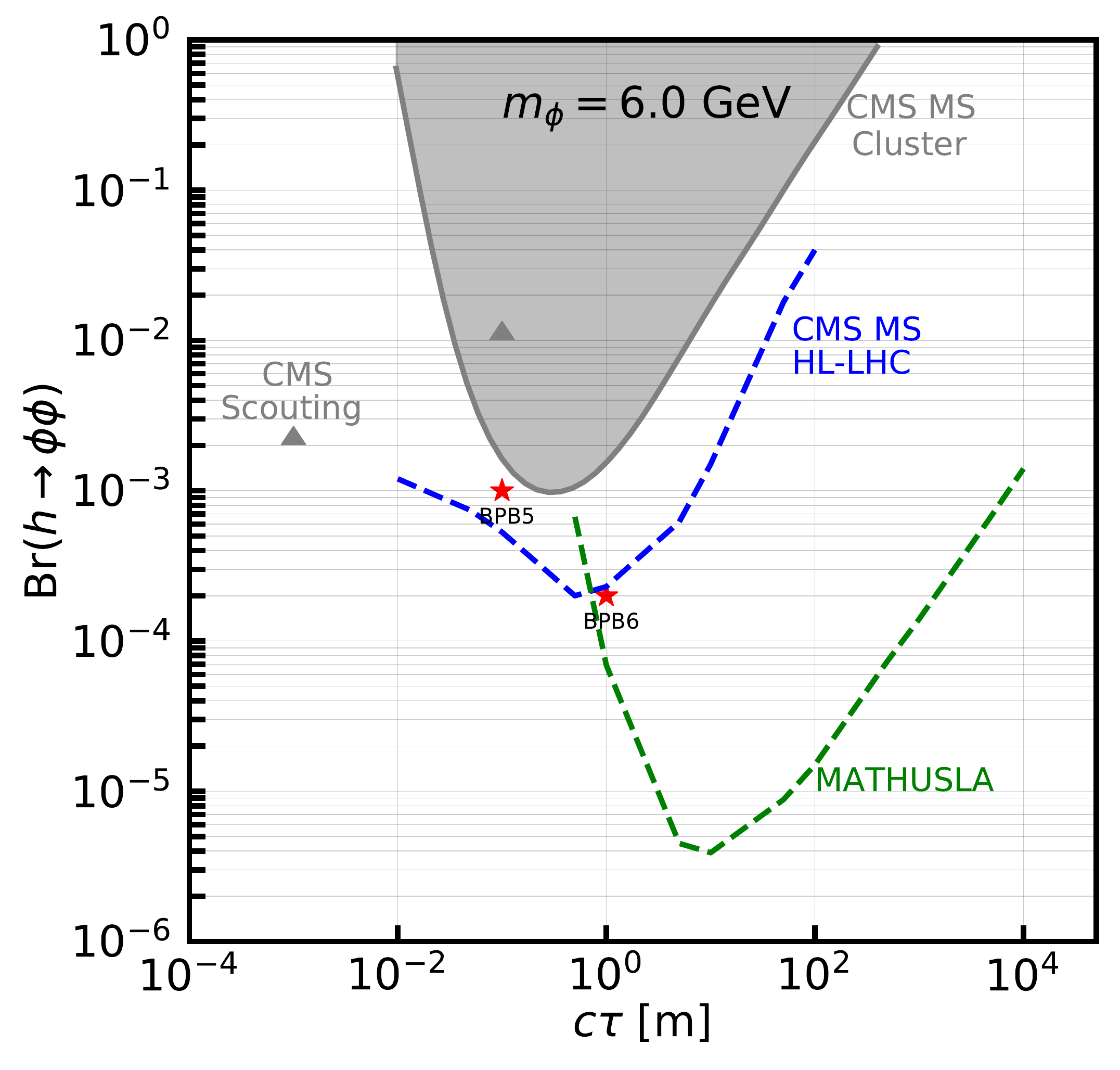}~
    \includegraphics[width=0.5\textwidth]{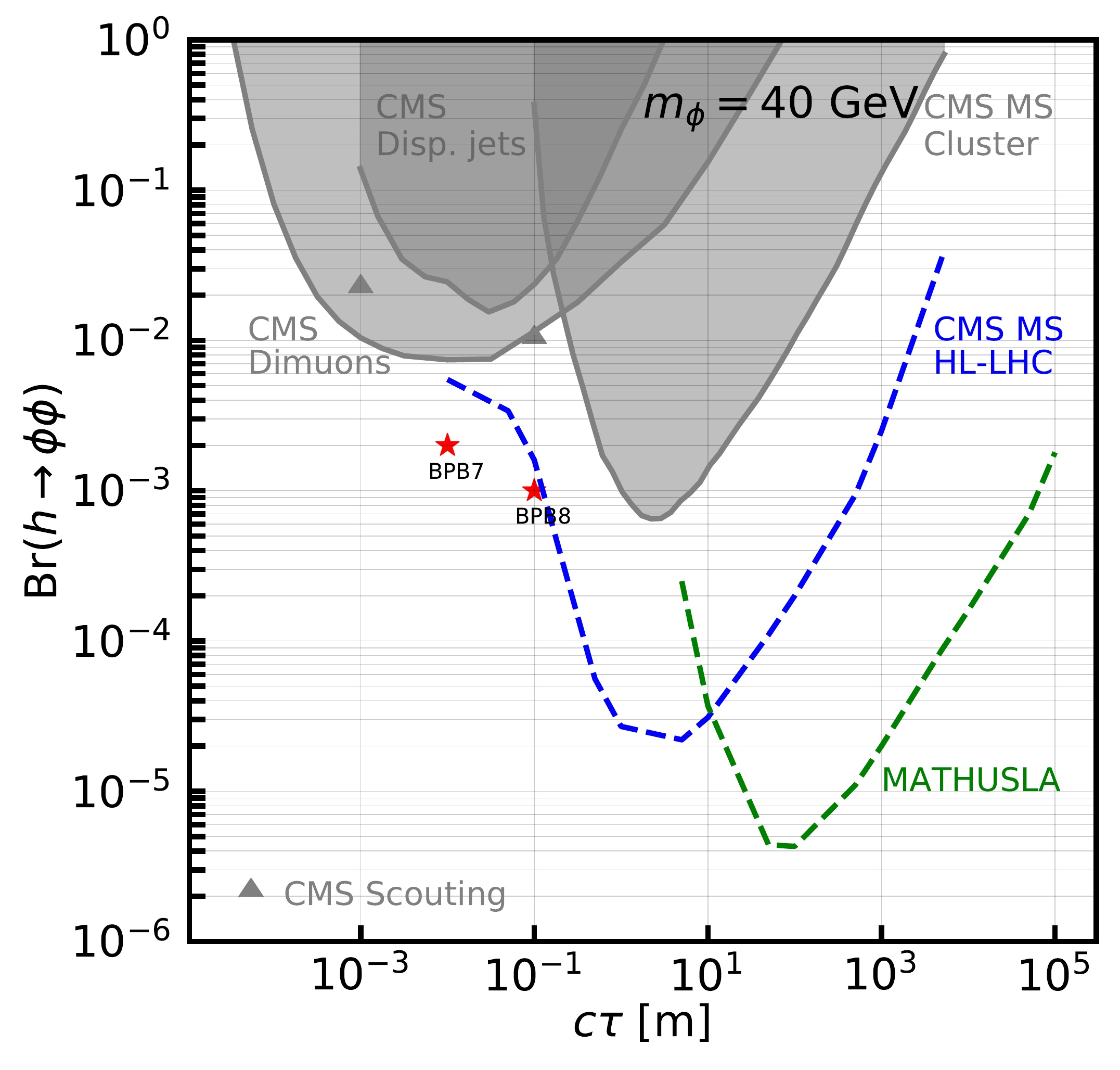}
    \caption{Current status (in {\it gray}) and future projected sensitivities of proposed experiments (in {\it dashed lines}) for the dark Higgs boson model when produced from $h$ boson decays in the process, $h\rightarrow \phi\phi$.
    The {\it gray triangles} show the upper limit on the Br($h\to\phi\phi$) for two values of $c\tau$ from the CMS scouting dataset.
    Benchmark points for the present study are marked as {\it red stars}.
    }
\label{fig:benchmarks_htophiphi}
\end{figure}

In this case, as discussed in Section\,\ref{sec:model}, the additional 
trilinear coupling $\lambda$ opens up the production of the dark Higgs boson 
$\phi$ from the decay of the 125\,GeV Higgs boson. Since this is the
dominant production mode of $\phi$, the most relevant limits come from various 
searches for LLPs at the main collider detectors, like 
ATLAS\,\cite{ATLAS:2022qex} and CMS\,\cite{CMS:summary}. Experiments put an 
upper limit on the Br($h\rightarrow\phi\phi$) for a specific mass and decay 
length of $\phi$, which is related to the parameter $\lambda$.
In this case, we can extend the mass range of $\phi$ up to 60\,GeV as they are produced from the Higgs boson.
Fig.\,\ref{fig:benchmarks_htophiphi} shows the combined excluded $c\tau-$Br($h\rightarrow\phi\phi$) parameter space for 
four different values of $m_\phi$: 1\,GeV, 4.4\,GeV, 6\,GeV and 40\,GeV.
The existing limits from the CMS scouting\,\footnote{Scouting refers to selecting events with lower $p_T$ thresholds to improve sensitivity to light physics at the cost of storing less information for offline analyses.}
\,\cite{CMS:2021sch}, searches for displaced 
leptons\,\cite{CMS:2021kdm,CMS:2024qxz}, displaced 
jets\,\cite{CMS:2020iwv,CMS:2021yhb}, delayed 
jets\,\cite{Bhattacherjee:2021qaa}, and displaced muon spectrometer (CMS MS) 
clusters\,\cite{CMS:2024bvl} are shown in {\it gray}, while the projected 
limits from HL-LHC\,\cite{Bhattacherjee:2021rml}, or proposed experiments like 
MATHUSLA and FASER, are shown in {\it dashed lines}\,\footnote{Note that some of the bounds in Fig.\,\ref{fig:benchmarks_htophiphi} end at specific $c\tau$ values since outside this range the particular search loses sensitivity drastically, either since the decay is not displaced enough or it escapes the decay volume.}.

The CMS MS HL-LHC line shows the projected sensitivity estimated in 
Ref.\,\cite{Bhattacherjee:2021rml} assuming 50 observed events with a specific set of cuts. These include demanding a prompt object from 
the Higgs boson production and displaced activity from the dark 
Higgs boson decay in the muon spectrometer. The {\it superscript} $S$ 
is to indicate a relatively softer set of cuts on both the prompt and the displaced object, defined in Ref.\,\cite{Bhattacherjee:2021rml}.
Refs.\,\cite{Bhattacherjee:2020nno,Bhattacherjee:2021qaa} discuss the reach of 
dedicated LLP triggers for probing various regions of the dark Higgs model.
For example, for a 40\,GeV $\phi$ with $c\tau = 100$\,cm, the level-1 trigger 
in Ref.\,\cite{Bhattacherjee:2021qaa} can select at least 50 signal events even for branching as low as Br$(h \to \phi\phi)=5.4\times 10^{-6}$.
However, these discussions do not consider the acceptance of the high-level trigger system and the offline analyses.

\begin{table}[hbt!]
    \centering
     \resizebox{\textwidth}{!}{  
    \begin{tabular}
    {|c|x{1.2cm}|c|c|x{2.5cm}|c|c|c|c|}
    \hline
    \multirow{2}{*}{Benchmark} & $m_{\phi}$ & \multirow{2}{*}{sin\,$\theta$} & $c\tau$  & Br($h\rightarrow\phi\phi$) & Dominant & \multicolumn{3}{c|}{Potential experiments to probe}  \\ \cline{7-9}
    & (GeV) & & (mm) &  & decay modes & HL-LHC & FASER & MATHUSLA \\ \hline\hline
    \multirow{2}{*}{BPB1} & \multirow{2}{*}{1.0} & \multirow{2}{*}{$1.1\times 10^{-4}$} & \multirow{2}{*}{10} & \multirow{2}{*}{$10^{-4}$} & $\pi^+\pi^-:$ 50\% & \multirow{2}{*}{$\times$} & \multirow{2}{*}{$\times$} & \multirow{2}{*}{$\times$} \\ 
    & & & & & $K^+K^-:$ 50\%  & & & \\ \hline
    \multirow{2}{*}{BPB2} & \multirow{2}{*}{1.0} & \multirow{2}{*}{$3.5\times 10^{-5}$} & \multirow{2}{*}{100} & \multirow{2}{*}{$10^{-4}$} & $\pi^+\pi^-:$ 50\% & \multirow{2}{*}{$\times$} & \multirow{2}{*}{$\checkmark$} & \multirow{2}{*}{$\times$} \\ 
    & & & & & $K^+K^-:$ 50\% & & & \\ \hline
    \multirow{2}{*}{BPB3} & \multirow{2}{*}{4.4} & \multirow{2}{*}{$4.6\times 10^{-5}$} & \multirow{2}{*}{10} & \multirow{2}{*}{$10^{-2}$} & $c\bar{c}:$ 65\% & \multirow{2}{*}{$\checkmark$} & \multirow{2}{*}{$\times$} & \multirow{2}{*}{$\times$} \\ 
    & & & & & $\tau^+\tau^-:$ 20\% & & & \\ \hline
    \multirow{2}{*}{BPB4} & \multirow{2}{*}{4.4} & \multirow{2}{*}{$1.5\times 10^{-5}$} & \multirow{2}{*}{100} & \multirow{2}{*}{$5\times10^{-4}$} & $c\bar{c}:$ 65\% & \multirow{2}{*}{$\times$} & \multirow{2}{*}{$\times$} & \multirow{2}{*}{$\times$} \\ 
    & & & & & $\tau^+\tau^-:$ 20\% & & & \\ \hline
    \multirow{2}{*}{BPB5} & \multirow{2}{*}{6.0} & \multirow{2}{*}{$1.1\times 10^{-5}$} & \multirow{2}{*}{100} & \multirow{2}{*}{$10^{-3}$} & $c\bar{c}:$ 49\% & \multirow{2}{*}{$\checkmark$} & \multirow{2}{*}{$\times$} & \multirow{2}{*}{$\times$} \\ 
    & & & & & $\tau^+\tau^-:$ 30\% & & &  \\ \hline
    \multirow{2}{*}{BPB6} & \multirow{2}{*}{6.0} & \multirow{2}{*}{$3.4\times 10^{-6}$} & \multirow{2}{*}{1000} & \multirow{2}{*}{$2\times10^{-4}$} & $c\bar{c}:$ 49\% & \multirow{2}{*}{$\times$} & \multirow{2}{*}{$\times$} & \multirow{2}{*}{$\checkmark$} \\ 
    & & & & & $\tau^+\tau^-:$ 30\% & & & \\ \hline
    \multirow{2}{*}{BPB7} & \multirow{2}{*}{40.0} & \multirow{2}{*}{$4.2\times 10^{-6}$} & \multirow{2}{*}{10} & \multirow{2}{*}{$2\times10^{-3}$} & \multirow{2}{*}{$b\bar{b}:$ 100\%} & \multirow{2}{*}{$\times$} & \multirow{2}{*}{$\times$} & \multirow{2}{*}{$\times$} \\ 
    & & & & & & & & \\ \hline
    \multirow{2}{*}{BPB8} & \multirow{2}{*}{40.0} & \multirow{2}{*}{$1.3\times 10^{-6}$} & \multirow{2}{*}{100} & \multirow{2}{*}{$10^{-3}$} & \multirow{2}{*}{$b\bar{b}:$ 100\%} & \multirow{2}{*}{$\times$} & \multirow{2}{*}{$\times$} & \multirow{2}{*}{$\times$} \\ 
    & & & & & & & &  \\
    \hline\hline
    \end{tabular}}
    \caption{Choice of benchmark points for large trilinear coupling.}
    \label{tab:bpchoice_b}
\end{table}

Based on the above discussion, we select eight benchmark points for the large 
trilinear coupling scenario where the dark scalar is pair-produced from the Higgs boson decay. We describe these benchmarks in 
Table\,\ref{tab:bpchoice_b} and mark them in 
Fig.\,\ref{fig:benchmarks_htophiphi} with {\it red stars}. Here, we choose benchmarks BPB1, BPB4, and BPB7 beyond any present bounds or projections of other proposed experiments, to study the unique prospect of the FCC-ee. 
Benchmark BPB2 can be probed at 
FASER2, while BPB3 and BPB5 lie within the projected sensitivity of the HL-LHC 
muon spectrometer analysis\,\cite{Bhattacherjee:2021rml}. BPB6 lies just at the 
edge of the HL-LHC MS search range and MATHUSLA sensitivities, while BPB8
is chosen to be very close to the HL-LHC MS clusters projection for their 
search reach. For these benchmarks, we intend to explore the complementary 
role the FCC-ee can play in observing and disentangling 
the signal in case we find some hints at the HL-LHC and, if approved, the 
MATHUSLA detector.


\section{Role of $e^+e^-$ Colliders in exploring the Benchmarks}
\label{sec:analysis}

In this section, we discuss our analysis strategy at the electron-positron 
collider to probe the various benchmarks that we selected in the previous 
section, for both the cases with negligible and large trilinear coupling, 
respectively. Among the various options for electron-positron colliders, 
we focus here on the FCC-ee machine; however, our results can be extended 
to the other options by using the respective detector 
parameters. As mentioned earlier, the FCC-ee collider is expected to first run 
at the $Z$ pole with $\sqrt{s}=91.2$\,GeV and later at the peak of the 
$HZ$ production cross section ($\sqrt{s}=240$\,GeV). We use these two options
in the present study $-$ the $Z$ factory to examine the sensitivity for 
the production of $\phi$ from $B$ meson decays, and the Higgs factory to 
explore the case where the LLP can be produced from the Higgs boson decay.  

For \textbf{Case A}, the number of signal events at the FCC-ee is calculated using  
\begin{eqnarray}
 N = \sigma_{b\bar{b}} \times 2 \times \text{Br}(B \to X_s \phi) \times \cal{L} \times \mathcal{A} \times \epsilon,
 \label{eq:numA}
\end{eqnarray}
where $\sigma_{b\bar{b}}$ is the production cross-section of $b$ partons at the
FCC-ee $Z$-pole. The factor of $2$ arises because either of the two $b$ partons 
can hadronize into a $B$ meson that decays to $\phi$. 
For the FCC-ee at the $Z$-pole, $\sigma_{b\bar{b}}=\sigma_{Z}\times {\rm Br}
(Z\to b\bar{b})$, where we use $\sigma_Z=59.29$\,nb and Br$(Z\to b\bar{b})=
0.156$, and the integrated luminosity, $\mathcal{L}=150$\,ab$^{-1}$\,
\cite{FCC:2018evy}. The factor $\mathcal{A}$ takes into account the detector 
acceptance, which is the probability of the LLP decay within a detector element, and $\epsilon$ is the signal efficiency after our analyses, including the detector efficiency of the LLP decay products. We calculate $\text{Br}(B \to X_s \phi)$ from Eq.\,(\ref{eq:br_BtoKphi}).

For \textbf{Case B}, the number of signal events at the FCC-ee becomes  
\begin{eqnarray}
 N = \sigma_{HZ} \times \text{Br}(h \to \phi\phi) \times \cal{L} \times \mathcal{A} \times \epsilon,
 \label{eq:numB}
\end{eqnarray}
where $\sigma_{HZ}$ is the peak of the Higgs boson production cross-
section at the FCC-ee (237\,fb) and $\mathcal{L}$ is the integrated luminosity 
(10.8\,ab$^{-1}$\,\cite{higgs_lumi}). The factor Br($h \to \phi\phi$) is a free 
parameter in this case, and we obtain an upper limit on this branching fraction 
from our analyses at the FCC-ee, which can then be translated into an upper limit 
on the trilinear coupling, $\lambda$. Here, $\mathcal{A}$ and $\epsilon$ have 
the same meaning as in the previous case. 

In this section, we first introduce the IDEA detector and its various 
components. We then show the maximum number of events that we expect to have in 
each of the detector elements for our chosen benchmarks, using 
Eqs.\,(\ref{eq:numA}) and (\ref{eq:numB}), while assuming a 100\% efficiency, 
\textit{i.e.}, $\epsilon=1$. Following this, we discuss the realistic efficiencies we consider for the various decay products in different detector components. Additionally, we provide a brief overview of the potential backgrounds and their simulation. 
Finally, we delve into the details and results of our analyses for the benchmarks in both cases.

\subsection{The IDEA Detector}

The {\bf I}nnovative {\bf D}etector for an {\bf E}lectron-Positron 
{\bf A}ccelerator (IDEA)
detector~\cite{FCC:2018byv,FCC:2018evy,Bernardi:2022hny} is a cylinder
designed to be centered at one of the IPs of the FCC-ee 
collider. It includes a vertex detector (VTX), a drift chamber (DCH), a dual-
readout calorimeter (DRC), and a muon system (MS). The VTX is the 
innermost part of the detector that surrounds the beampipe and is made of 
silicon pixels, which have excellent tracking resolution and high efficiency 
for charged particles. The DCH is designed for good tracking, high-
precision momentum measurement, and excellent particle identification. The 2\,T 
solenoid for the IDEA detector is positioned outside the 
DCH. There is also a preshower detector placed between the magnet and the 
calorimeter in the barrel region, as well as in front of the end-cap 
calorimeters in the forward regions to identify charged particles and photons. 
In addition, it can also tag a $\pi^0$ from its decay to two photons. The DRC 
measures the energy of electrons/photons and hadrons, providing good intrinsic 
discrimination between muons, electrons/photons, and hadrons for isolated 
particles. The muon system is characterized by a high-resolution 
momentum measurement for muons, and consists of layers of 
chambers~\cite{Bencivenni:2017wee} embedded in the magnet yoke.

\begin{table}[hbt!]
    \centering 
    \resizebox{0.9\textwidth}{!}{
    \centering 
    \begin{tabular}{|c|c|c|c|c|}
    \hline
    Detector component & $R_{in}$ (mm) & $R_{out}$ (mm) & $Z_{in}^{\rm half}$ (mm) & $Z_{out}^{\rm half}$ (mm) \\
    \hline\hline
    Silicon pixel detector (VTX) & 17 & 340 & 400 & 2000 \\
    Drift chamber (DCH) & 345 & 2020 & 2125 & 2125 \\
    Solenoid & 2100 & 2400 & 2500 & 2500 \\
    Preshower & 2400 & 2500 & 2500 & 2600 \\
    Dual-readout calorimeter (DRC) & 2500 & 4500 & 2600 & 4500 \\
    Muon system (MS) & 4500 & 5500 & 4500 & 6500 \\
    \hline\hline
    \end{tabular}
    }
    \caption{The detector dimensions for the IDEA detector used in this work, based on Refs.\,\cite{FCC:2018evy,IDEA_slide1,IDEA_slide2}.}
    \label{tab:IDEA_det}
\end{table}

We show the 
dimensions of these various detector components in Table\,\ref{tab:IDEA_det}, 
and employ these parameters to model the 
geometry of the IDEA detector. Using the coordinates of
the LLP decay products, we reconstruct the position of the LLP decay 
vertex in both the transverse ($d_T$) and the $z$ direction ($d_z$). These ($d_T,\,d_z$) coordinates help us identify the specific component of the IDEA detector where the LLP decays or whether the decay occurs outside the main detector.

\begin{table}[hbt!]
    \centering
    \resizebox{\textwidth}{!}{
    \begin{tabular}{|c|c|c|c|c|c|c|c|}
    \hline
    \multirow{2}{*}{Benchmark} & \multicolumn{7}{c|}{Number of LLP decays 
    for $\mathcal{L}=150\,\text{ab}^{-1}$ (assuming 100\% efficiency)} \\\cline{2-8}
    & VTX & DCH & Solenoid & Preshower & DRC & MS & Outside \\
    \hline\hline     
    BPA1 & $5.21 \pm 0.89$ & $16.85 \pm 1.61$ & $1.38 \pm 0.46$ & $1.23 \pm 0.43$ & $20.99 \pm 1.79$ & $13.18 \pm 1.42$ & $6174.71 \pm 30.76$ \\
    BPA2 & $315.00 \pm 5.02$ & $720.20 \pm 7.59$ & $31.32 \pm 1.58$ & $28.21 \pm 1.50$ & $484.87 \pm 6.22$ & $227.89 \pm 4.27$ & $1348.34 \pm 10.38$ \\ 
    BPA3 & $32750.06 \pm 204.62$ & $16977.21 \pm 147.33$ & $184.10 \pm 15.34$ & $134.24 \pm 13.10$ & $1013.85 \pm 36.00$ & $84.38 \pm 10.39$ & $49.86 \pm 7.98$ \\ 
    BPA4 & $42924.01 \pm 336.09$ & $47166.15 \pm 352.31$ & $860.53 \pm 47.59$ & $834.22 \pm 46.85$ & $8534.27 \pm 149.86$ & $1936.86 \pm 71.39$ & $1610.54 \pm 65.10$ \\
    BPA5 & $1.80 \pm 0.10$ & $6.44 \pm 0.18$ & $0.49 \pm 0.05$ & $0.31 \pm 0.04$ & $7.71 \pm 0.20$ & $5.14 \pm 0.16$ & $188.37 \pm 0.99$ \\ 
    BPA6 & $22573.27 \pm 135.40$ & $10028.50 \pm 90.25$ & $32.49 \pm 5.14$ & $24.37 \pm 4.45$ & $124.27 \pm 10.05$ & $4.06 \pm 1.82$ & $0$ \\
    BPA7 & $41.61 \pm 0.27$ & $27.87 \pm 0.22$ & $0.24 \pm 0.02$ & $0.14 \pm 0.02$ & $0.79 \pm 0.04$ & $0.05 \pm 0.01$ & $0.01 \pm 0.004$ \\

    \hline\hline
    \end{tabular}}
    \caption{Number of LLP decays (with statistical uncertainties) in various components of the IDEA detector for the chosen benchmark points from \textbf{Case A}, assuming 100\% efficiency.}
    \label{tab:decay_events_a}
\end{table}

\begin{table}[hbt!]
    \centering
    \resizebox{\textwidth}{!}{
    \begin{tabular}{|c|c|c|c|c|c|c|c|}
    \hline
    \multirow{2}{*}{Benchmark} & \multicolumn{7}{c|}{Number of LLP decays 
    for $\mathcal{L}=10.8\,\text{ab}^{-1}$ (assuming 100\% efficiency)} \\\cline{2-8}
    & VTX & DCH & Solenoid & Preshower & DRC & MS & Outside \\
    \hline\hline     
    BPB1 & $193.84 \pm 1.58$ & $178.62 \pm 1.52$ & $2.24 \pm 0.17$ & $1.50 \pm 0.14$ & $8.81 \pm 0.34$ & $0.59 \pm 0.09$ & $0.19 \pm 0.05$ \\
    BPB2 & $38.41 \pm 0.71$ & $105.89 \pm 1.18$ & $6.29 \pm 0.29$ & $5.42 \pm 0.27$ & $86.69 \pm 1.07$ & $43.24 \pm 0.76$ & $154.58 \pm 1.43$ \\
    BPB3 & $25487.22 \pm 180.61$ & $3355.64 \pm 65.53$ & 0 & 0 & 0 & 0 & 0 \\
    BPB4 & $609.79 \pm 6.25$ & $960.73 \pm 7.84$ & $32.25 \pm 1.44$ & $28.86 \pm 1.36$ & $280.80 \pm 4.24$ & $53.95 \pm 1.86$ & $51.26 \pm 1.81$ \\
    BPB5 & $1487.13 \pm 13.80$ & $1931.73 \pm 15.72$ & $53.50 \pm 2.62$ & $44.03 \pm 2.37$ & $345.03 \pm 6.65$ & $47.48 \pm 2.47$ & $25.34 \pm 1.80$ \\
    BPB6 & $48.60 \pm 1.15$ & $149.89 \pm 2.02$ & $9.26 \pm 0.50$ & $8.77 \pm 0.49$ & $143.75 \pm 1.98$ & $79.34 \pm 1.47$ & $402.70 \pm 3.31$ \\
    BPB7 & $5119.20 \pm 36.19$ & 0 & 0 & 0 & 0 & 0 & 0 \\
    BPB8 & $2555.38 \pm 18.08$ & $247.51 \pm 5.63$ & 0 & 0 & 0 & 0 & 0 \\
    \hline\hline
    \end{tabular}}
    \caption{Number of LLP decays (with statistical uncertainties) in various components of the IDEA detector for the chosen benchmark points from \textbf{Case B}, assuming 100\% efficiency.}
    \label{tab:decay_events_b}
\end{table}

\begin{figure}[hbt!]
    \centering
    \includegraphics[width=0.5\linewidth]{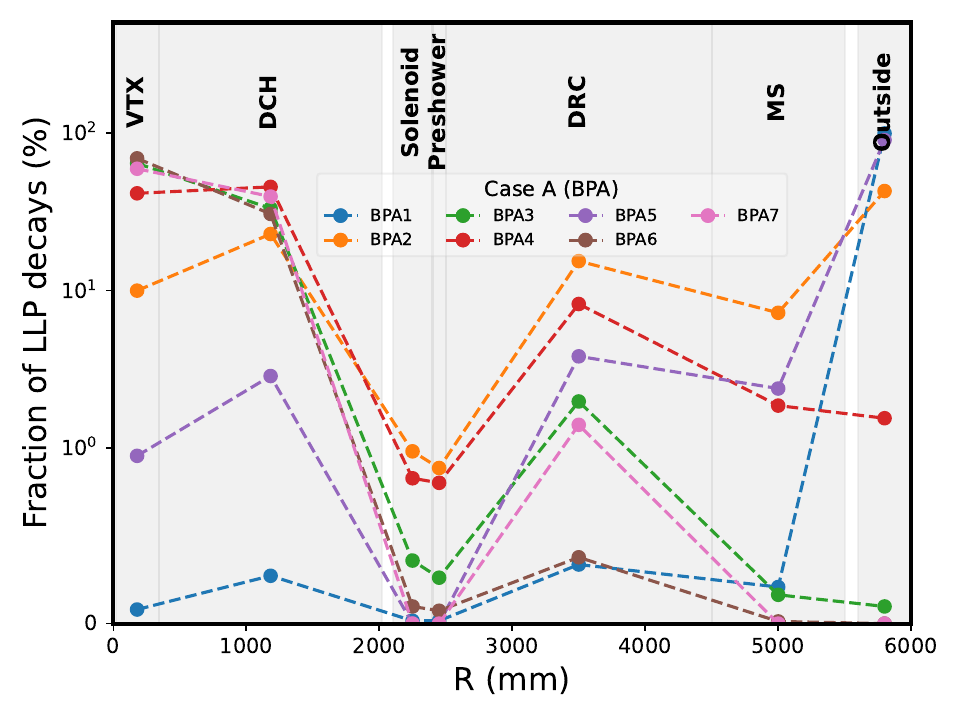}~
    \includegraphics[width=0.5\linewidth]{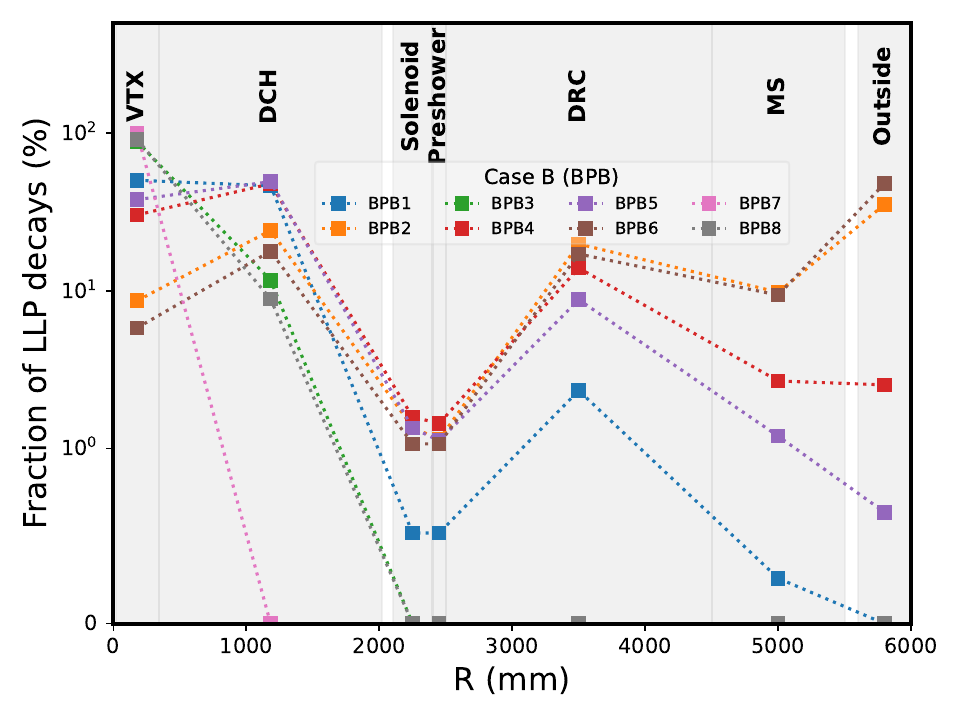}
    \caption{Fraction of LLP decays in \% within each detector component for benchmarks in \textbf{Case A} $(left)$ and \textbf{Case B} $(right)$. The $x$-axis represents the radial direction $R$ of the IDEA detector.}
    \label{fig:BP_plot}
\end{figure}

In Tables~\ref{tab:decay_events_a} and \ref{tab:decay_events_b}, we list the 
number of observed events that are expected in various parts of the IDEA 
detector assuming 100\% efficiency for our chosen benchmark points of the dark 
scalar model, where the dark Higgs bosons are produced from the decay of $B$ 
mesons and Higgs bosons, respectively. Figure~\ref{fig:BP_plot} shows the percentage of LLP decays in each detector component, for \textbf{Case A} ({\it left panel}) and \textbf{Case B} ({\it right panel}) benchmarks. This suggests, for each benchmark, the detector elements that will be the most useful for analyzing these 
benchmarks. For instance, BPA1 and BPA5 mostly decay outside the detectors, 
whereas BPA3, BPA4, and BPA6 have a significant number of events within the VTX 
and DCH detector elements. Benchmarks BPA1, BPA2, BPA3, BPA4, and BPA5 also 
have a considerable number of decays within the MS, in addition to the inner 
detectors. 
In \textbf{Case B}, BPB3 primarily decays within the VTX component, while BPB6 predominantly decays outside the IDEA detector.
This is important since the collider analysis of the benchmarks 
depends on where the dark 
Higgs boson decays, in addition to its decay mode.

\begin{table}[htb!]
\centering
\resizebox{\textwidth}{!}{
\begin{tabular}{|c | c | c | c|}
\hline
Detector component & Particles & Energy threshold (GeV) & Efficiency \\ \hline\hline
\multirow{2}{*}{Vertex detector (VTX)} & \multirow{3}{*}{charged particles} & $0.1 < E < 0.3$ & 0.06 \\ 
\multirow{2}{*}{Drift Chamber (DCH)} &  & $0.3 < E < 0.5$ &  0.65 \\
& & $E > 0.5$ & 0.997 \\
\hline
Solenoid & $e^\pm$, $\mu^\pm$, charged hadrons & $E > 0.5$ & \multirow{2}{*}{0.98}\\
Preshower & $\gamma$ & $E > 0.10$ & \\
\hline
DR calorimeter & $e^\pm$, $\gamma$, hadrons & $E > 0.5$ & 0.98 \\
\hline
Muon system & $\mu^\pm$ & $E > 0.10$ & 0.98 \\
\hline\hline
\end{tabular}
}
\caption{Fixed efficiencies of particle selection in different components of the FCC-ee IDEA detector.}
\label{tab:fixed_eff}
\end{table}

We have assumed 100\% detection efficiency of the LLP decay 
products until now. However, the number of signal events we observe in each 
detector element will depend on the decay mode, the position of the decay 
vertex, and the efficiency of observing the particles from the LLP decay in 
the subsequent detector elements. Based on estimates from Refs.~\cite{IDEAvtx}, 
\cite{Akchurin:2014zna}, and \cite{Bencivenni_2020} for the tracker, 
calorimeter, and muon spectrometer efficiencies, respectively, we present in
Table\,\ref{tab:fixed_eff} the detection efficiencies we have assumed for various particles within each component of the IDEA detector. For 
the solenoid and preshower regions, we have employed 
similar efficiencies. We perform our analyses with these fixed efficiencies. 
However, the efficiencies will also depend on the position of the decay 
vertex of the LLP. The farther it decays from the IP in a particular detector 
component, the smaller the expected detection efficiency. The parametrization 
of this effect is still unavailable in fast detector simulations, like
\texttt{Delphes}\,\cite{deFavereau:2013fsa}. We later comment on the results 
assuming that the efficiencies drop linearly with increasing displacement,
\textit{i.e.}, they depend on the $d_T$ of the LLP decay, and the $R_{in}$ and 
$R_{out}$ of the detector element. Next, we discuss the possible backgrounds 
from the SM in the analysis of a light, long-lived particle decaying into 
muons, pions, kaons, and light jets. 

\subsection{Backgrounds}
\label{ssec:bkgs}

The dominant backgrounds for light, long-lived particles are the SM long-lived 
mesons and baryons. In this section, we enumerate all the possible SM long-
lived hadrons with varying masses and decay lengths. Looking at the lab frame 
decay length distributions of these particles and knowing their possible decay 
modes are crucial in order to recognize the relevant backgrounds in light 
LLP searches. Some of these particles might have a smaller decay length than 
the signal LLP. Despite that, the high production cross-section of these 
SM particles and the exponential nature of the decay length distribution 
can lead to a significantly long tail, contributing to backgrounds in 
our analysis. 
Moreover, some of these particles can come from the decay of another long-lived 
SM particle, which increases the displacement of the decay.

\begin{table}[hbt!]
\centering
\begin{tabular}{|c|cc||c|cc|}
\hline
Mesons & m (GeV) & $c\tau$ (mm) & Baryons & m (GeV) & $c\tau$ (mm) \\ \hline
\hline
$K_L$ & 0.498 & 15330 & $\Xi^0$ & 1.315 & 87.1 \\
$K_S$ & 0.498 & 26.84 & $\Lambda$ & 1.116 & 78.9 \\
$B^\pm$ & 5.279 & 0.491 & $\Xi^-$ & 1.322 & 49.1 \\
$B^0$ & 5.279 & 0.459 & $\Sigma^-$ & 1.197 & 44.34 \\
$B_S^0$ & 5.367 & 0.439 & $\Omega^-$ & 1.672 & 24.61 \\
$D^\pm$ & 1.869 & 0.312 & $\Sigma^+$ & 1.189 & 24.04 \\
$D_S^\pm$ & 1.968 & 0.150 & $\Lambda_b^0$ & 5.619 & 0.369 \\
$D^0$ & 1.865 & 0.123 & $\Xi_b^-$ & 5.791 & 0.364 \\
 &  &  & $\Xi_b^0$ & 5.788 & 0.364 \\
 &  &  & $\Xi_c^+$ & 2.468 & 0.132 \\
 &  &  & $\Lambda_c^+$ & 2.286 & 0.06 \\
 &  &  & $\Xi_c^0$ & 2.471 & 0.0336 \\ \hline\hline
\end{tabular}
\caption{List of SM long-lived mesons and baryons, along with their mass 
($m$) and decay length ($c\tau$) in decreasing order of $c\tau$. The 
values of mass and $c\tau$ are obtained from \texttt{PYTHIA 8}.}
\label{tab:bkg_list}
\end{table}

In order to obtain a list of these long-lived hadrons, we generate $Z$ 
bosons at $\sqrt{s}=91.2$\,GeV using 
\texttt{PYTHIA\,8}\,\cite{Sjostrand:2014zea,Bierlich:2022pfr}, with the $Z$ 
bosons decaying to quarks. These hadronize to produce a variety of mesons and 
baryons, which can have displaced decays. We generate $10^9$ 
events,\!\footnote{FCC-ee shall generate $10^{12}$ $Z$ bosons; however, due to 
limited computing resources, we generate $10^9$.} which is still a factor of 
$\sim 8900$ smaller than the expected number of $Z$ bosons produced at 
the FCC-ee $Z$-pole. This demonstrates how large these 
backgrounds from displaced hadron decays can be, with a significant 
number of events populating even the tails of the $d_T$ distributions.
Table\,\ref{tab:bkg_list} lists the mesons and baryons with long decay 
lengths in decreasing order of their $c\tau$. Although $J/\Psi$ and $\Psi(2S)$ 
are not shown in Table\,\ref{tab:bkg_list} because they are not long-lived, it 
is important to note that they still travel a few millimeters before decaying, 
originating from the displaced decay of a boosted $B$ meson. Many of these 
particles have the same final states as the decay products of the 
light dark Higgs bosons. For example, $J/\psi$ can decay to two muons or $K_S$ 
can decay to two pions. This would make the search difficult for LLPs having 
masses and lifetimes closer to these SM hadrons. We discuss them individually 
while discussing the analyses for the different decay modes of $\phi$. But 
before that, we want to obtain a generic $d_T$ threshold that can suppress most 
SM backgrounds.

\begin{figure}[hbt!]
    \centering
    \includegraphics[width=0.5\textwidth]{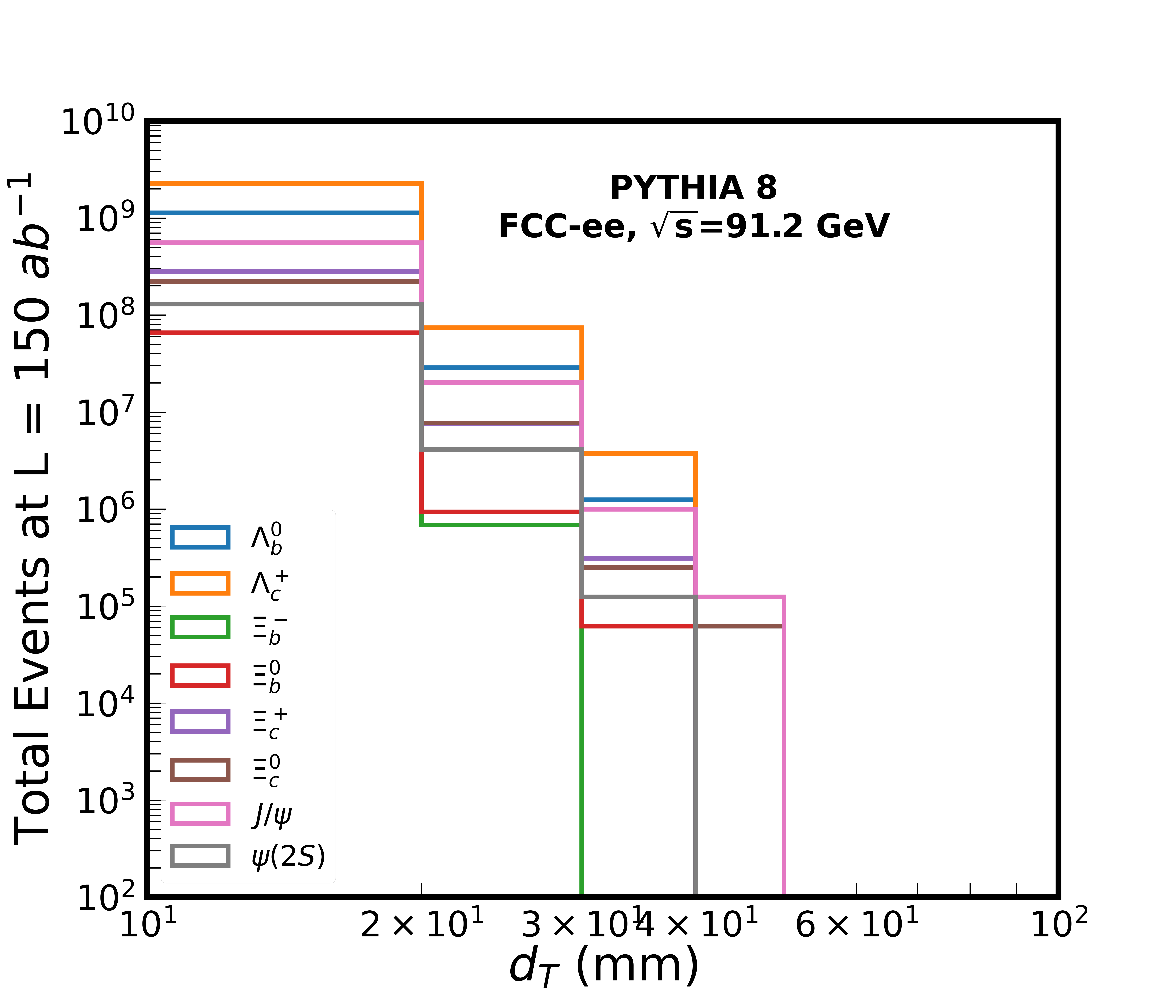}~
    \includegraphics[width=0.5\textwidth]{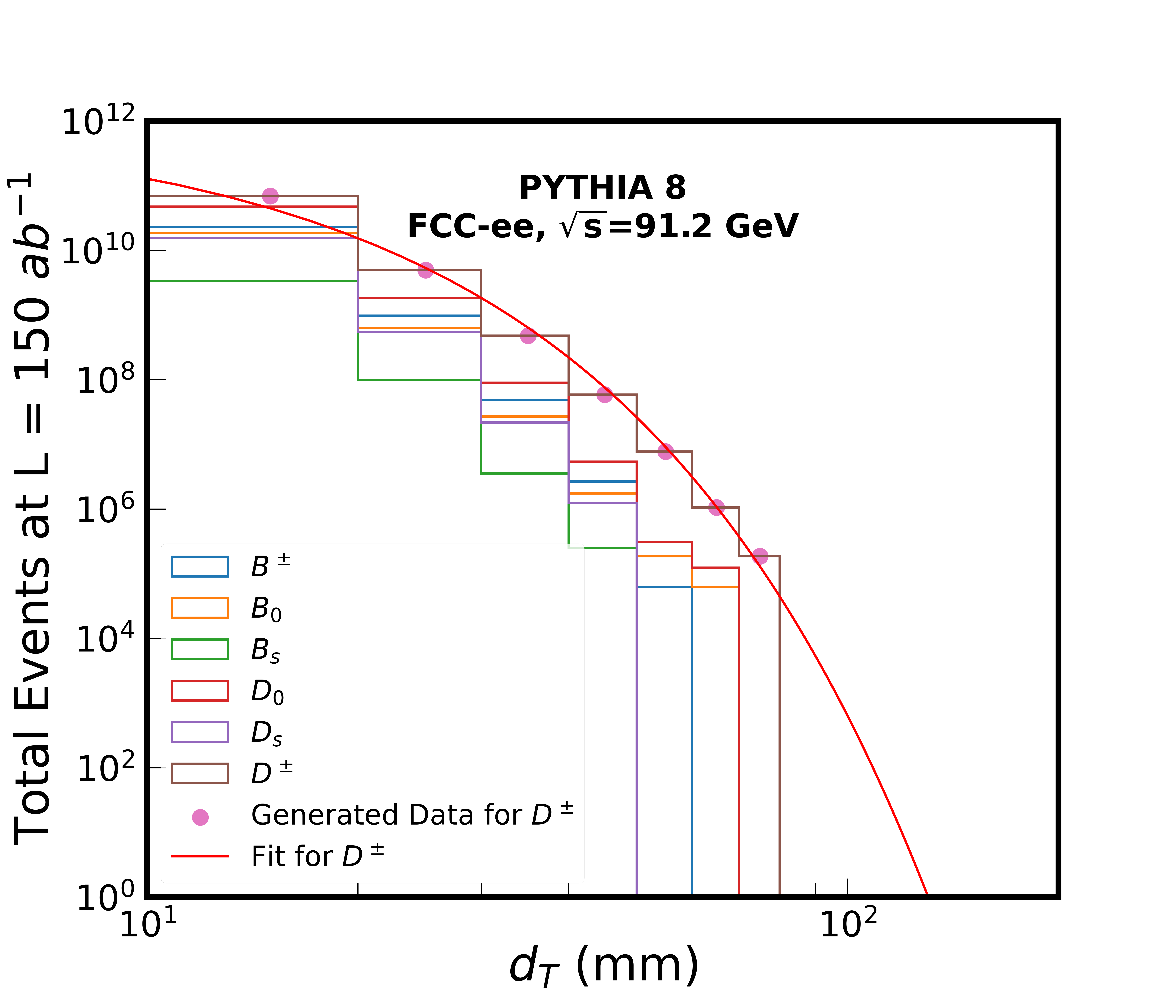}\\
    \includegraphics[width=0.5\textwidth]{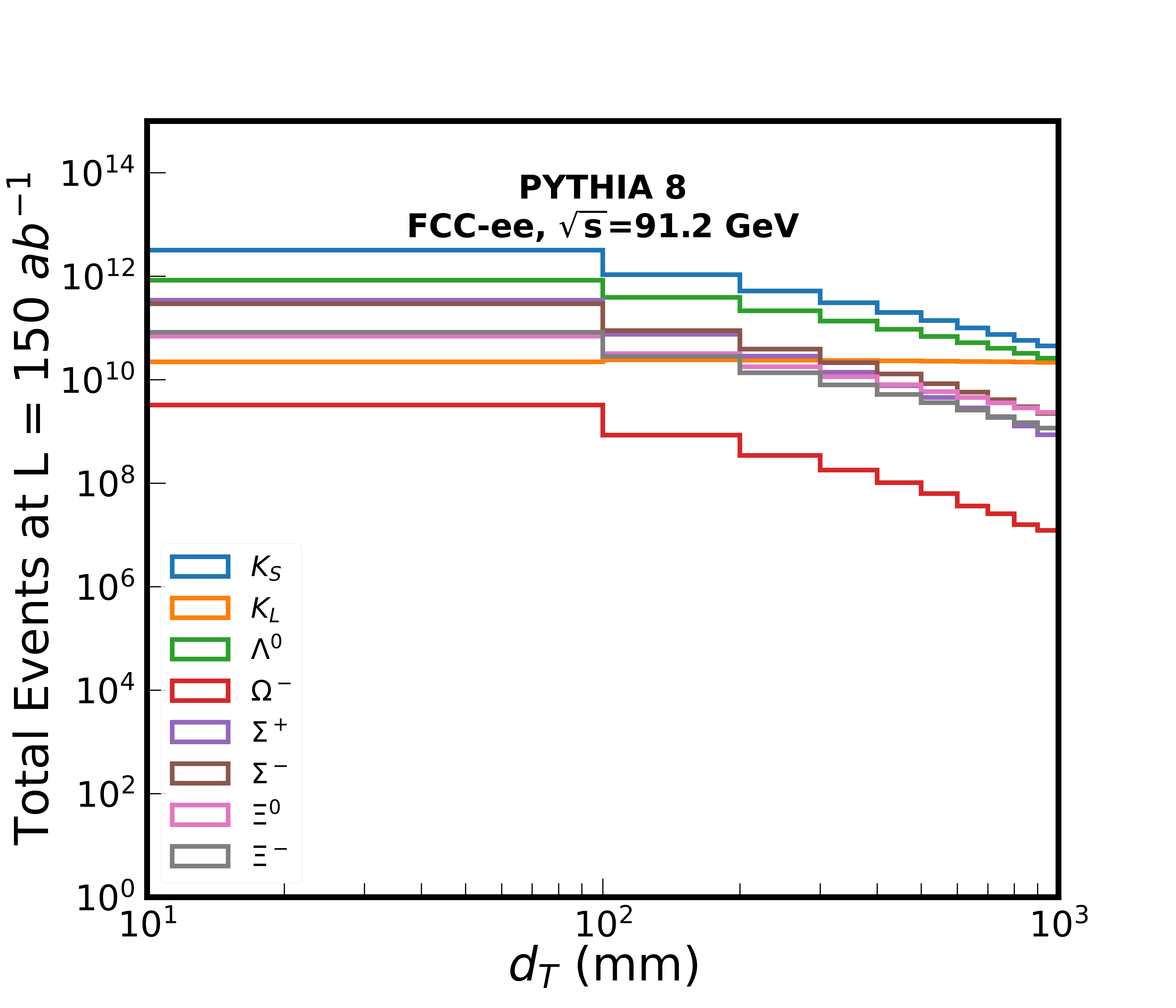}
    \caption{Number of events as a function of the transverse decay length in the lab frame, $d_T$, for various SM long-lived mesons and baryons.}
    \label{fig:bkg_dist_BKphi}
\end{figure}

We divide the SM long-lived particles into three categories based on 
their decay lengths and show their $d_T$ distributions at the $Z$ factory 
($\sqrt{s}=91.2$\,GeV) in Fig.\,\ref{fig:bkg_dist_BKphi}, where the number of
events is calculated for an integrated luminosity of 150\,ab$^{-1}$. The {\it 
top left}, {\it top right}, and {\it bottom} panels respectively show the 
distributions of particles as a function of $d_T$. There 
is a significant number of events extending up to $\sim 50$\,mm 
(\textit{e.g.}, $J/\psi$), $\sim 100$\,mm (\textit{e.g.}, $D^\pm$) and going 
beyond $1$\,m ($K_S$). We observe that the bottom and charm mesons belong to 
the second category and travel a transverse distance up to $\sim$ 100\,mm. 
Among these, the number of $D^\pm$ 
meson decays extends the farthest in $d_T$, since the $D^\pm$ 
itself has a long decay length, and additionally, it can come from the decay of 
a $B^\pm$. We fit the number of $D^\pm$ decay 
events as a function of $d_T$ to estimate the minimum $d_T$ threshold for 
which the SM $D^\pm$ background becomes zero, or negligible. We obtain a 
value of $d_T$ close to 130\,mm. Taking a conservative approach, we shall use a 
cut of $d_T>150~\text{mm}$ to reduce all the SM backgrounds from the decay of 
charm and bottom mesons.

Note that this does not eliminate the third category of hadrons, which have 
much longer tails in $d_T$. However, fitting won't provide the correct estimate 
of the maximum value of $d_T$ for these hadrons, as they would populate 
displacements up to the calorimeters, where they deposit energy. We discuss 
them in detail in the subsequent sections, depending on the decay modes of the 
LLP and these hadrons, and highlight the specific challenges and ways to deal 
with them.

In this work, we do not consider the background contributions from material interactions. The high transparency design of the drift chamber (DCH) is expected to minimize particle scattering to an extent where the contribution from multiple scattering to the momentum resolution of the tracking system is below 0.0003 for the entire momentum range from 0 to 100 GeV~\cite{IDEAStudyGroup:2025gbt}. A full detector simulation, including the material effects, is beyond the scope of this work.
Furthermore, when the material map of the VTX detector of IDEA is available, similar to the one available for LHCb\,\cite{Alexander:2018png}, it can be used to effectively veto displaced vertices that might come from material interactions.

\subsection{Analysis Strategy}

\begin{figure}[hbt!]
    \centering
    \includegraphics[width=\linewidth]{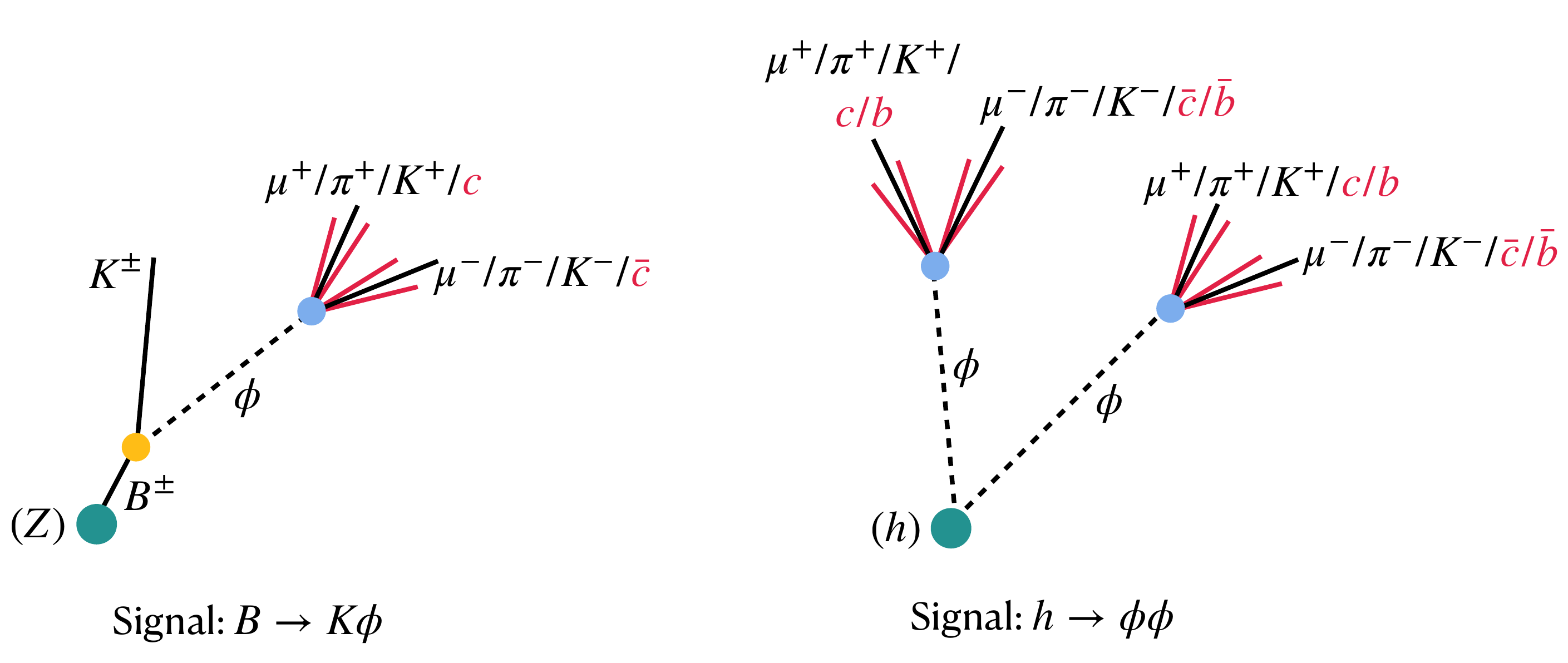}\\
    \includegraphics[width=\linewidth]{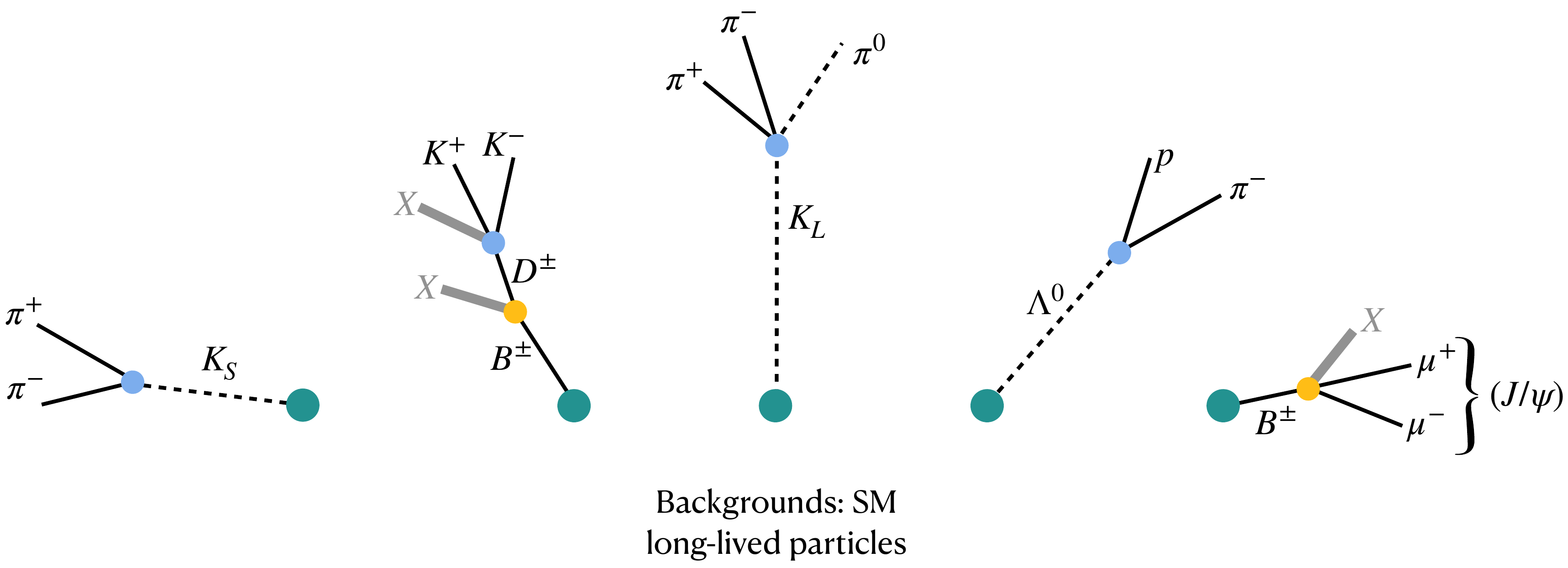}
    \caption{Illustrations of the two signal processes at the FCC-ee ({\it top panel}) and the SM backgrounds ({\it bottom panel}).}
    \label{fig:sig_bkg}
\end{figure}

In this section, we briefly discuss the analysis strategy used in this study. 
Before doing so, we illustrate in Fig.\,\ref{fig:sig_bkg} the signal processes 
we are examining, which include both the decays of $B$ mesons at the $Z$ factory and the decays of Higgs bosons during the $HZ$ production peak run at FCC-ee. We also highlight some dominant SM backgrounds discussed in the 
previous section in Fig.\,\ref{fig:sig_bkg}.

\begin{figure}[hbt!]
    \centering
    \includegraphics[width=\linewidth]{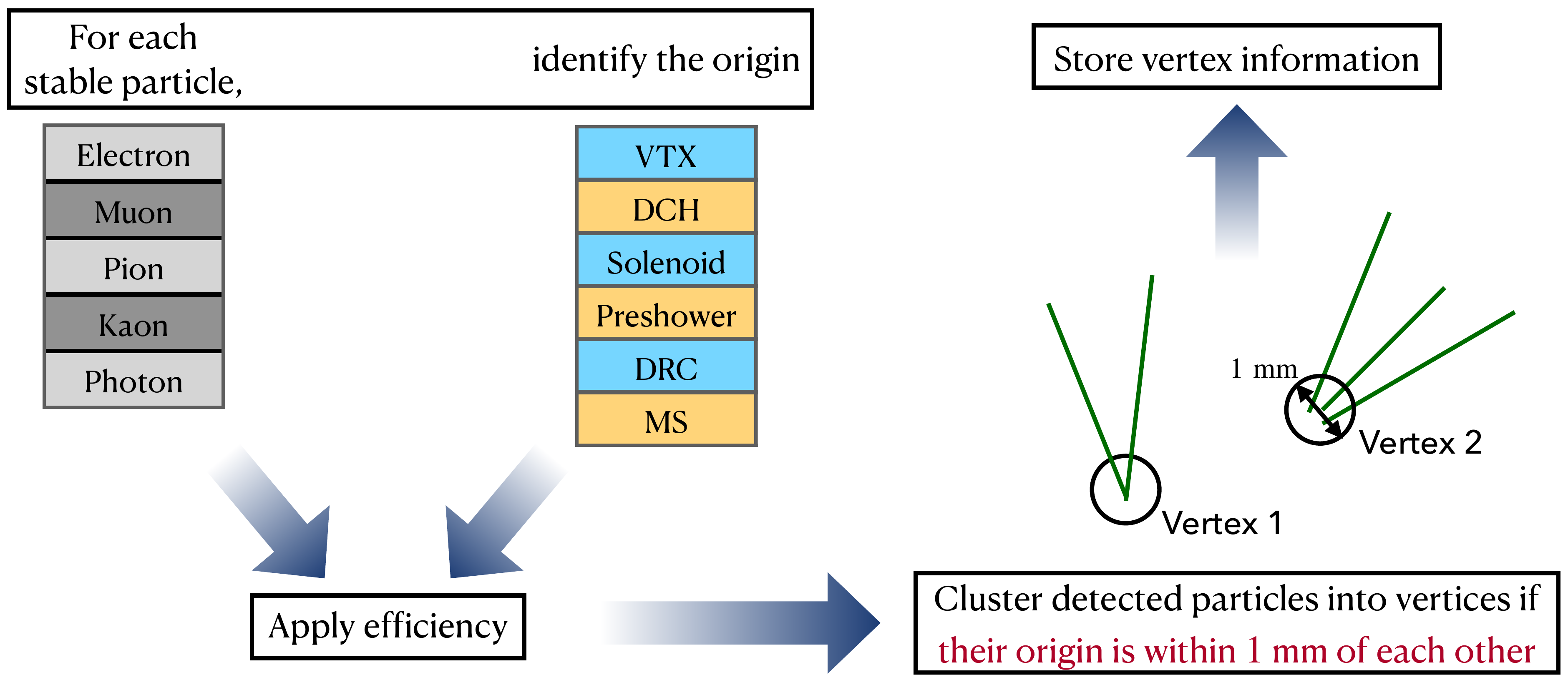}
    \caption{Schematic outline of our analysis strategy.}
    \label{fig:analysis}
\end{figure}

For our analysis, we first identify all the stable particles and the 
coordinates of the decay vertex where they originate 
in the IDEA detector. This information helps us to apply the appropriate 
efficiency factors based on Table\,\ref{tab:fixed_eff}. Next, the particles 
that meet the efficiency criteria are clustered into vertices with other 
detected particles, given their origins are within 1\ mm of each other. 
Finally, we have a list of vertices and store variables  
describing these vertices as listed below. 
This strategy is illustrated in Fig.\,\ref{fig:analysis}.

To analyze the signal and background, we consider the following variables to describe the features of a displaced vertex in an event:
\begin{itemize}
   \item detector element based on the location of the vertex, $D_{\rm vtx}$\,,
   \item number of charged particles created at the vertex, 
    $N_{\rm ch}$\,,
   \item total energy of the particles produced at the vertex, $E$\,,
   \item transverse displacement of the vertex from the origin, $d_T$\,,
   \item number of muons ($N_{\mu}$), pions ($N_{\pi}$), kaons ($N_{K}$), and
   electrons ($N_{e}$) originating at the vertex\,,
   \item invariant mass of the final state particles associated with the vertex, $m_{\rm vtx}$\,,
   \item the impact parameter of the vertex, $d_0$\,.
\end{itemize}

Based on the final state, we apply different cuts based on these variables to minimize the SM background as much as 
possible. In the following sections, we will categorize our benchmarks 
according to their final states and discuss the specific analyses and results 
for each of them.

\subsection{LLPs with Negligible Trilinear Coupling}

In this section, we explore the prospects of LLPs with negligible trilinear 
coupling at the FCC-ee, \textit{i.e.} \textbf{Case A}. The signal benchmarks  are listed in 
Table\,\ref{tab:bpchoice_a}. 
We generate the signal and the SM background 
events using \texttt{PYTHIA\,8} 
for an $e^+e^-$ beam at a center of mass energy $\sqrt{s}$ = 91.2\,GeV. The 
background is due to $Z$ bosons produced at 
$\sqrt{s}$ = 91.2\,GeV, that decay either hadronically to quarks or 
leptonically.
The hadronic branching of the $Z$ boson is 69.2\%, out 
of which the $Z\to b\bar{b}$ branching is 15.6\%. The majority of SM long-lived 
particles arise from the hadronization of quarks into various mesons and 
baryons, as discussed in detail in Section\,\ref{ssec:bkgs}. We generate $10^9$ 
events for the SM background from inclusive $Z$ decays and $10^5$ events 
for each signal benchmark, where we only simulate the process $Z\to b\bar{b}$ 
for the signal. 

\subsubsection{Analysis in the Di-muon Final State}
\label{sssec:A_mumu}

From Table\,\ref{tab:bpchoice_a}, we observe that benchmarks BPA1, BPA4, BPA5, 
and BPA6 respectively have 10\%, 12\%, 5\%, and 8\% branching ratio of
the $\phi$ to $\mu^+\mu^-$. We generate $10^5$ events for these LLP signal 
benchmarks however, with a 100\% branching ratio for $\phi\to\mu^+\mu^-$.
The SM background $Z\to \mu^+\mu^-$ is a prompt decay and, therefore, does not 
contribute at larger values of $d_T$. Displaced muons can also be produced from 
the decays of $J/\psi (1S)$ and $\psi (2S)$ into $\mu^+\mu^-$. As discussed in the previous section, long-lived $B$ mesons decay into these particles after a few millimeters. Therefore, even the prompt decays of $J/\psi$ and $\psi$ lead to displaced muons.

\begin{figure}[htb!]
\centering
\includegraphics[width=0.6\textwidth]{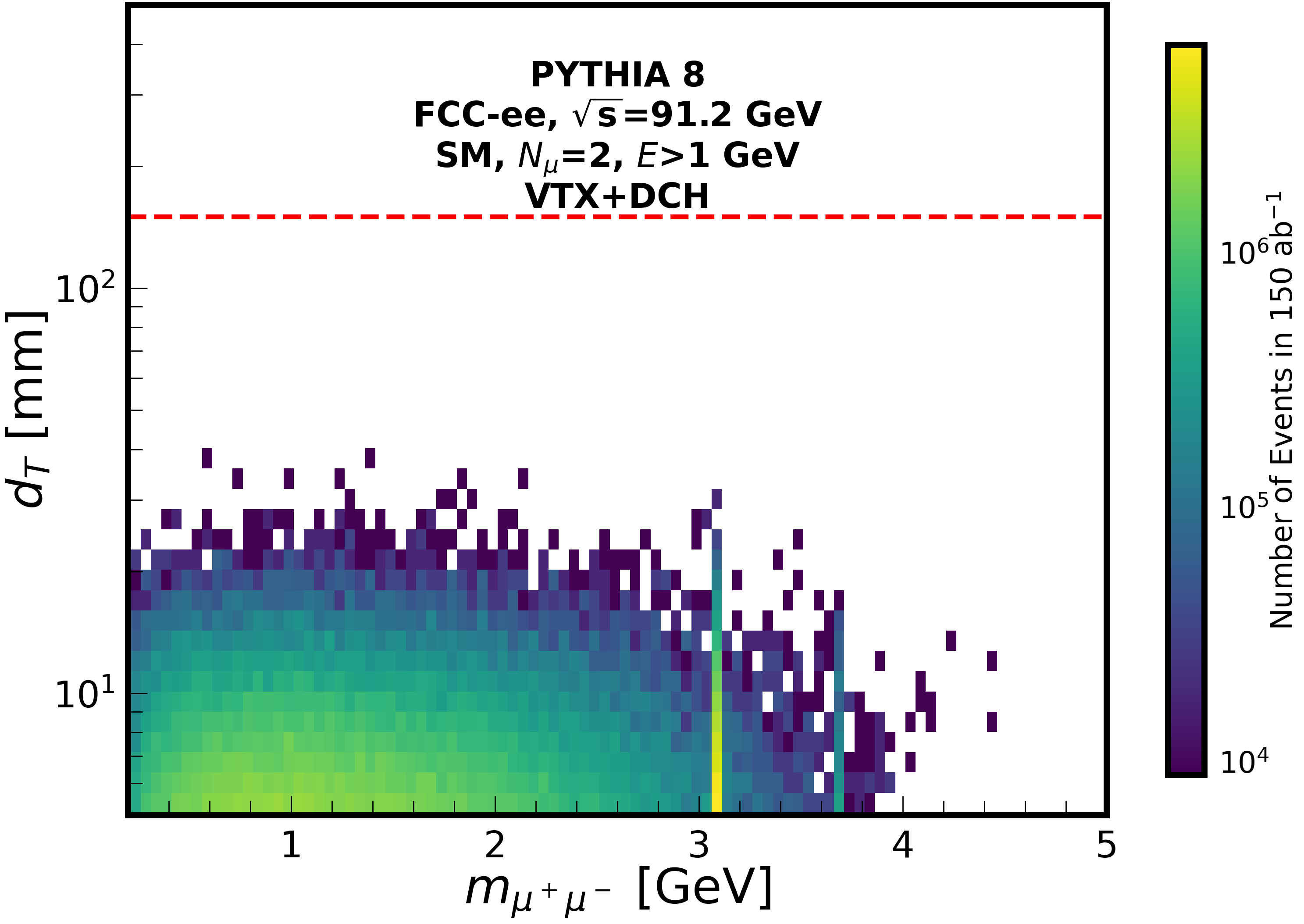}
\caption{Two-dimensional distribution of the SM background as a 
function of the invariant mass of the dimuon vertex and the $d_T$. The two distinct regions at $m_{\mu^+\mu^-}\sim 3.1~\text{GeV}$ and $m_{\mu^+\mu^-}\sim 3.7~\text{GeV}$ correspond to the $J/\psi$ and $\psi(2S)$ resonances respectively. The $d_T$ threshold motivated from fitting the SM background is shown as a {\it red dashed line} at $d_T=150$\,mm.}
\label{fig:dimuon_a}
\end{figure}

To identify the contribution of the SM backgrounds for the dimuon 
final state, we apply a set of cuts to select events with a reconstructed 
displaced vertex having exactly 
two muons in the vertex. We also demand that the vertex location is within the 
VTX or the DCH detectors, since SM hadrons travel further to the calorimeters 
and deposit their energy. But the dimuon signal can be observed even when the 
$\phi$ travels beyond the DCH all the way to the MS. These 
cuts are summarized in Table\,\ref{tab:bkg_mumu_cuts}. We calculate the 
invariant mass of the particles produced at the vertex,  
$m_{\mu^+\mu^-}$. The two-dimensional distribution of the SM background of inclusive $Z$ decays as a 
function of $m_{\mu^+\mu^-}$ and $d_T$ is shown in Fig.\,\ref{fig:dimuon_a}. 
We observe a continuous background populating the low invariant mass and 
$d_T$ region, coming from displaced muons produced from various $B$ meson decay 
chains. We also observe two resonances in $m_{\mu^+\mu^-}$ at $m_{J/\psi}$ 
$\sim$ 3.1\,GeV and $m_{\psi(2S)}$ $\sim$ 3.7\,GeV, corresponding to the 
displaced $J/\psi$ and $\psi$, coming from $B$ meson decays, with their $d_T$ 
values extending up to 40\,mm.

\begin{table}[hbt!]
    \centering
    \begin{tabular}{c}
    \hline
        Selection cuts to study the SM dimuon background \\ \hline\hline
        $D_{\rm vtx}$ $\in$ VTX or DCH\\
        $N_{\rm ch}$ = 2 \\
        $N_{\mu}$ = 2 \\
        $E > 1$\,GeV \\
        $d_T > 5$\,mm \\ 
        \hline\hline
    \end{tabular}
    \caption{Selection cuts for the dimuon final state to study the SM background.}
    \label{tab:bkg_mumu_cuts}
\end{table}

\begin{table}[hbt!]
    \centering
    \begin{tabular}{c}
    \hline
        Final cuts for the dimuon final state \\ 
        \hline\hline
        $D_{\rm vtx}$ $\equiv$ out to the MS \\
        $N_{\rm chg}$ = 2 \\
        $N_{\mu}$ = 2 \\
        $E > 2$\,GeV \\
        $d_T > 150$\,mm  \\
        \hline\hline
    \end{tabular}
    \caption{Final cuts applied for the dimuon final state for the signal benchmarks to reduce the SM backgrounds to zero.}
    \label{tab:sig_mumu_cuts}
\end{table}

From Section\,\ref{ssec:bkgs}, we choose the cut $d_T>150$\,mm  
for a conservative analysis, 
which can reduce the background from all the displaced charm and bottom mesons 
to zero, as can be seen from Fig.\,\ref{fig:dimuon_a}. Given that the 
dominant background for the dimuon final state comes 
from displaced $B$ meson decays, we apply this conservative $d_T$ threshold for 
all the benchmarks and calculate the resulting number of signal events. 
Since the final state contains muons, we extend the detection of signal events 
all the way to the MS detector component. The final cuts applied on the signal benchmarks are listed in Table\,\ref{tab:sig_mumu_cuts}.
We apply these cuts on the signal events for benchmarks BPA1, BPA4, BPA5, and 
BPA6. 
The number of signal events obtained from the analysis 
for each benchmark, including the respective branching to muons, is listed in 
Table\,\ref{tab:BM_mumu}. We observe that BPA4 and BPA6 can be easily probed in 
the dimuon channel. BPA1 has a very large decay length, causing most of the 
decays to lie beyond the MS. Still, three events with zero background can be 
observed with our analysis. We recommend that the LLP search be carried 
out in this benchmark with increased luminosity. BPA5 has the weakest prospect, where we expect to observe only a single event. Due to low sin\,$\theta$, the $\phi$-production rate of BPA5 is very low, and a high $c\tau$ causes most of the LLPs to decay outside the detector.
We shall revisit these two benchmarks when we study the prospect of dedicated detectors for FCC-ee in Section\,\ref{sec:dedicated}.

\begin{table}[hbt!]
\centering
\begin{tabular}{|c|c|c|c|}
\hline
\multicolumn{1}{|c|}{\multirow{2}{*}{Benchmarks}} & \multicolumn{3}{c|}{ 
Dimuon analysis for decays out to the MS} \\ \cline{2-4} 
\multicolumn{1}{|c|}{} & $m_{\phi}$ (GeV) & $c\tau$ (mm) & Number of events \\ \hline\hline
BPA1 & 0.4 & 39666.6 & 2.99 $\pm$ 0.21 \\
BPA4 & 2.0 & 135.2 & 9351.88 $\pm$ 54.34 \\
BPA5 & 3.5 & 10285.4 & 1.06 $\pm$ 0.02\\
BPA6 & 3.5 & 65.8 & 915.41 $\pm$ 6.10 \\ \hline\hline
\end{tabular}
\caption{Number of signal events (with statistical uncertainties) in BPA1, BPA4, BPA5, and BPA6 after a cut-based analysis. The number of background events in each case is expected to reduce to zero due to the $d_T$ cut.}
\label{tab:BM_mumu}
\end{table}

\begin{figure}[htb!]
    \centering
    \includegraphics[width=0.5\textwidth]{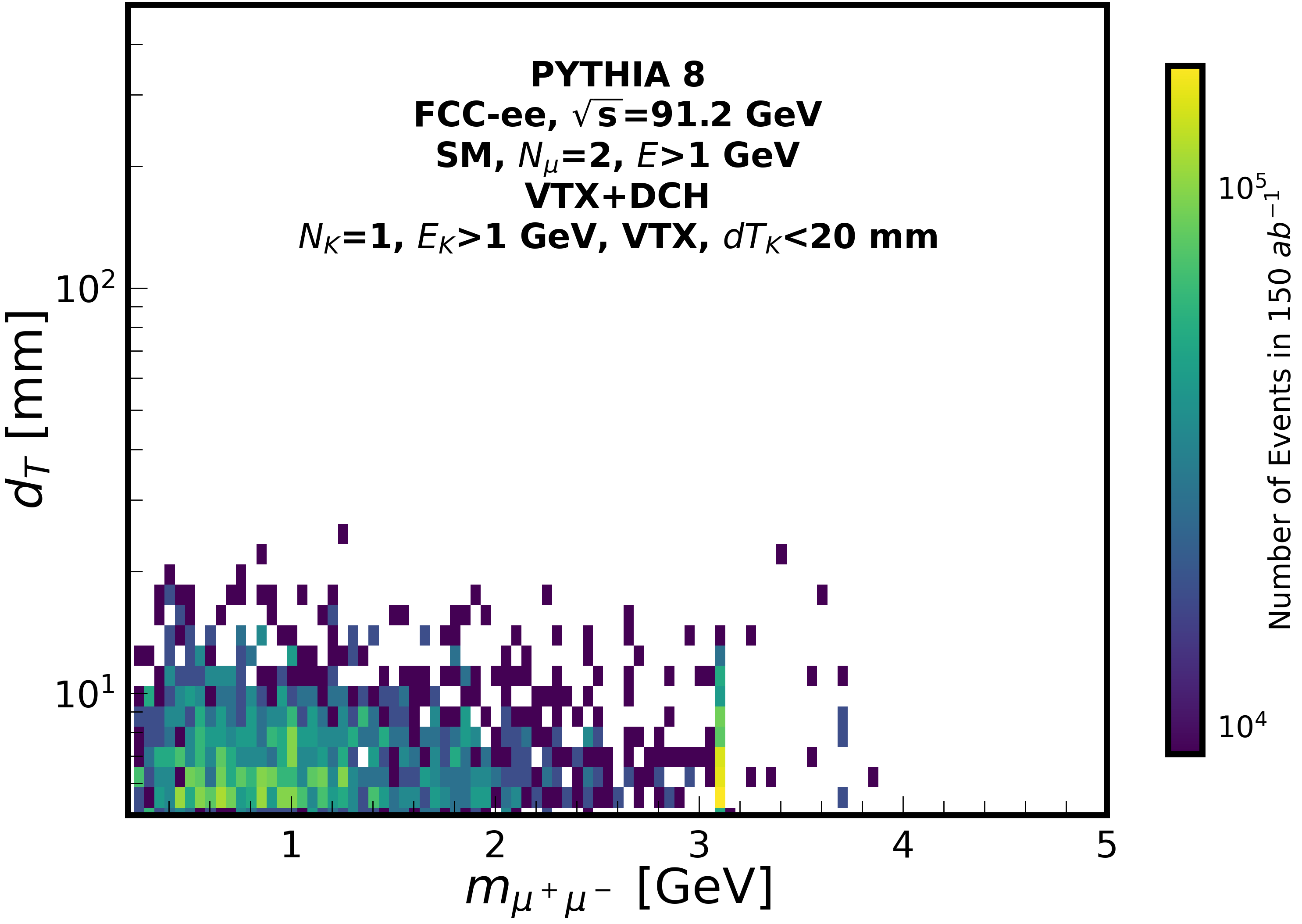}~
    \includegraphics[width=0.5\textwidth]{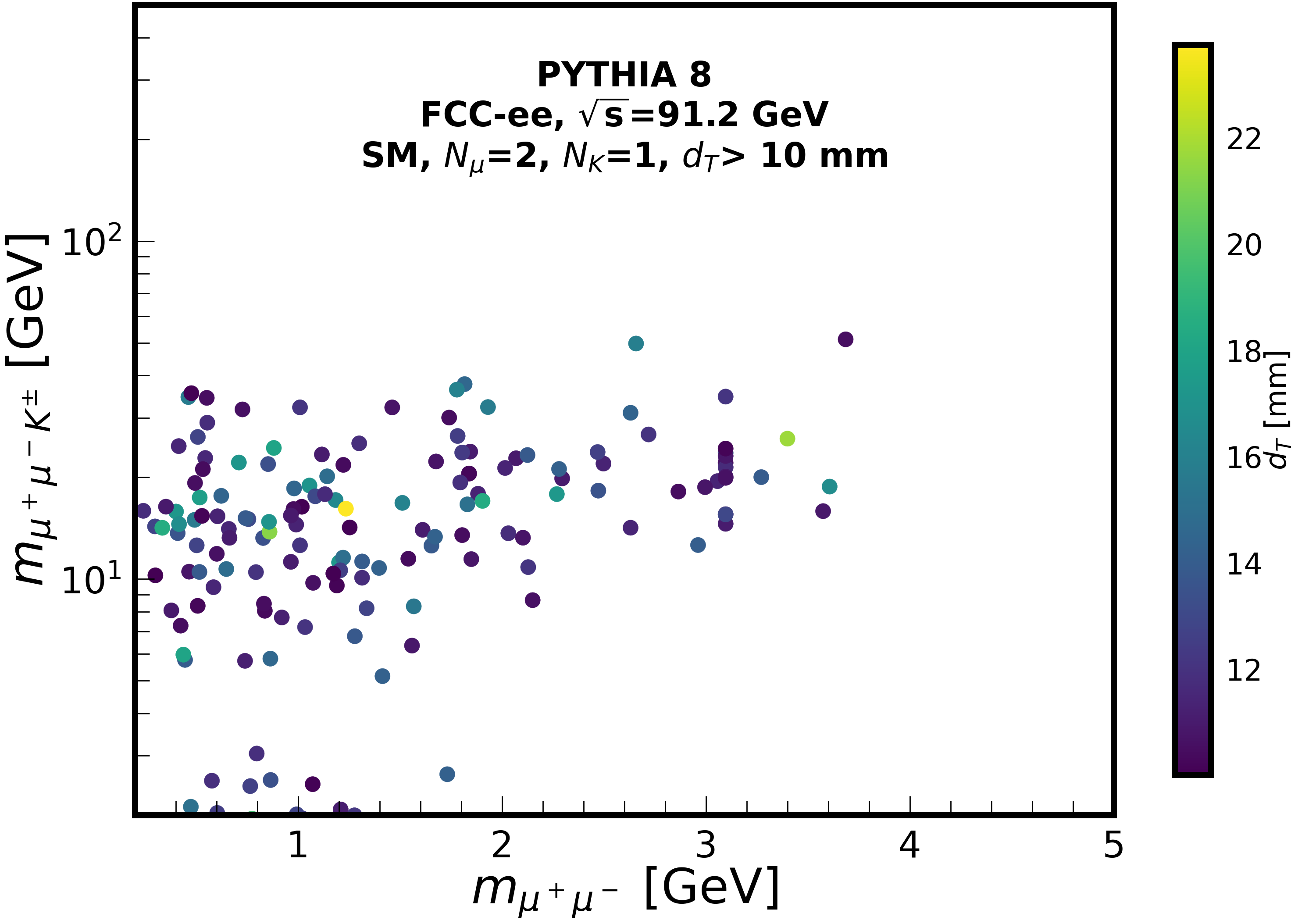}
    \caption{{\it Left:} Two-dimensional distribution of $m_{\mu^+\mu^-}$ and $d_T$ of the vertex for the SM background after demanding a prompt kaon in the event; {\it Right:} Two-dimensional distribution of $m_{\mu^+\mu^-}$ and $m_{\mu^+\mu^-K^\pm}$ for the SM background after demanding a prompt kaon in the event and $d_T>10$\,mm for the dimuon vertex.}
    \label{fig:dimuon_kaoncut}
\end{figure}

We have achieved zero background with the above analysis strategy. However, we 
study whether there are ways to suppress the background 
using some other features of the signal process, while relaxing the $d_T$ 
cut. In the signal process, $B\to K\phi$, a kaon is produced along with the $
\phi$ from the decay of the $B$ meson. Thereby, in addition to our list of 
selection cuts in Table\,\ref{tab:bkg_mumu_cuts} for the SM background events, 
we demand that there must be an isolated prompt kaon in the VTX detector 
element with $E > 1$\,GeV, but $d_T < 20$\,mm. The $d_T$ cut ensures that the 
$K^\pm$ from the $B$ decay in the signal is not displaced very much. The
{\it left} panel of Fig.\,\ref{fig:dimuon_kaoncut} shows the distributions of 
the SM background in the $m_{\mu^+\mu^-}-d_T$ plane, with the $N_{K^\pm}=1$ 
condition. The {\it right} panel of Fig.\,\ref{fig:dimuon_kaoncut} presents the 
background events that remain after applying the kaon cut, with the dimuon 
vertex having a transverse distance $d_T > 10$\,mm. The panel shows the $(m_{\mu^+\mu^-},m_{\mu^+\mu^-K^\pm})$ 
plane. Here $m_{\mu^+\mu^-K^\pm}$ represents the combined invariant 
mass of the two identified muons and the isolated charged kaon, respectively. 
The color bar indicates the values of $d_T$ for the dimuon vertex. For signal 
events, the invariant mass $m_{\mu^+\mu^-K^\pm}$ should be close to the mass of 
the $B$ meson. 

If the additional criteria of a kaon or even a mass window cut on $m_{\mu^+
\mu^-K^\pm}$ near the $B$ meson mass is applied, the background can be
suppressed even at moderate values of $d_T$, which can  
increase the signal efficiency. The reduction in the background can be already 
seen from comparing the number of SM events shown in the {\it color bars} of 
Fig.\,\ref{fig:dimuon_a} and the {\it left} panel of 
Fig.\,\ref{fig:dimuon_kaoncut}. This provides an alternate strategy to reduce 
the SM background to the LLPs decaying to two muons. This would only be 
possible for charged kaons coming from $B^\pm$ decays. For the neutral $B^0$ 
decays, the neutral kaon will have displaced decays in various detector 
elements, which is difficult to identify and associate with the $B$ meson decay.
Since we use an inclusive branching ratio of $B$ mesons to the dark Higgs 
boson, we cannot separate the charged $B$ meson decays from the neutral ones to 
quote the corresponding signal efficiencies here. This would require estimating 
the individual branching fractions, which have large uncertainties from form 
factors.

\begin{figure}[htb!]
    \centering
    \includegraphics[width=0.5\textwidth]{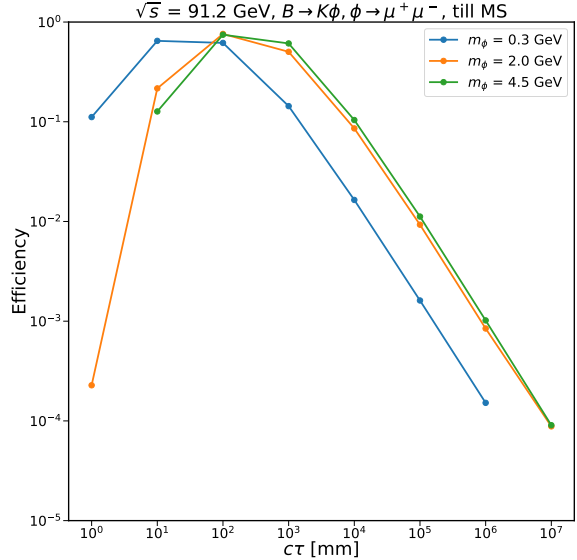}
    \caption{Efficiency for the di-muon decay mode of the LLP $\phi$ with the variation of $c\tau$ for three masses $m_\phi$. The generation of the LLP involved the $Z\to b\bar{b}$ process where one $B$ meson decays via the
    BSM process $B\to K \phi$, and the other $B$ decays to SM particles. The final state muons were detected while requiring the cut: 150\,mm $ < d_T 
    <$  5500\,mm.}
    \label{fig:dimuon_eff}
\end{figure}

Beyond the benchmark studies, we generate LLPs decaying to two muons for various $m_{\phi}$ and $c\tau$ values. From the $Z\to b\bar{b}$ process, we require one of the $B$ mesons to have a non-standard decay to LLP, while the other $B$ decays to SM particles. We follow the selection criteria in Table\,\ref{tab:sig_mumu_cuts} and calculate the signal efficiency of the analysis.
We show the efficiency for this analysis in Fig.\,\ref{fig:dimuon_eff}, which 
can be useful to translate the results for a range of masses and lifetimes, with arbitrary branching to muons. 



\subsubsection{Analysis in the Di-pion and Di-kaon Final 
States}
\label{sssec:A_pipiKK}

Almost all the benchmarks in Table\,\ref{tab:bpchoice_a} 
have a significant branching ratio of the $\phi$ to 
the di-pion final state, Br($\phi\to\pi^+\pi^-$). It  
ranges from 76\% for BPA1, 50\% for BPA2 and BPA3, 41\% for 
BPA4, and 53\% for BPA5 and BPA6. Benchmarks BPA2, BPA3, 
BPA4, BPA5, and BP6 have a Br($\phi \to K^+K^-$) of 50$\%$, 
50$\%$, 41$\%$, 21\%, and 21$\%$, respectively. Both 
the di-pion and di-kaon analyses proceed similarly. We first select the appropriate 
vertices inside the vertex detector and the drift chamber. 
These selection cuts for the background events are listed in 
Table\,\ref{tab:bkg_pik_cuts}.

\begin{table}[hbt!]
    \centering
    \begin{tabular}{c}
    \hline
        Selection cuts to study the SM di-pion/di-kaon background \\ \hline\hline
        $D_{\rm vtx}$ $\in$ VTX or DCH \\
        $N_{\rm ch}$ = 2 \\
        $N_{\pi/K}$ = 2 \\
        $E > 1$\,GeV \\
        $d_T > 5$\,mm \\ \hline\hline
    \end{tabular}
    \caption{Selection cuts applied to study the SM 
    background events when the final state contains two 
    charged pions or two kaons.}
    \label{tab:bkg_pik_cuts}
\end{table}

\begin{figure}[htb!]
    \centering
    \includegraphics[width=0.5\textwidth]{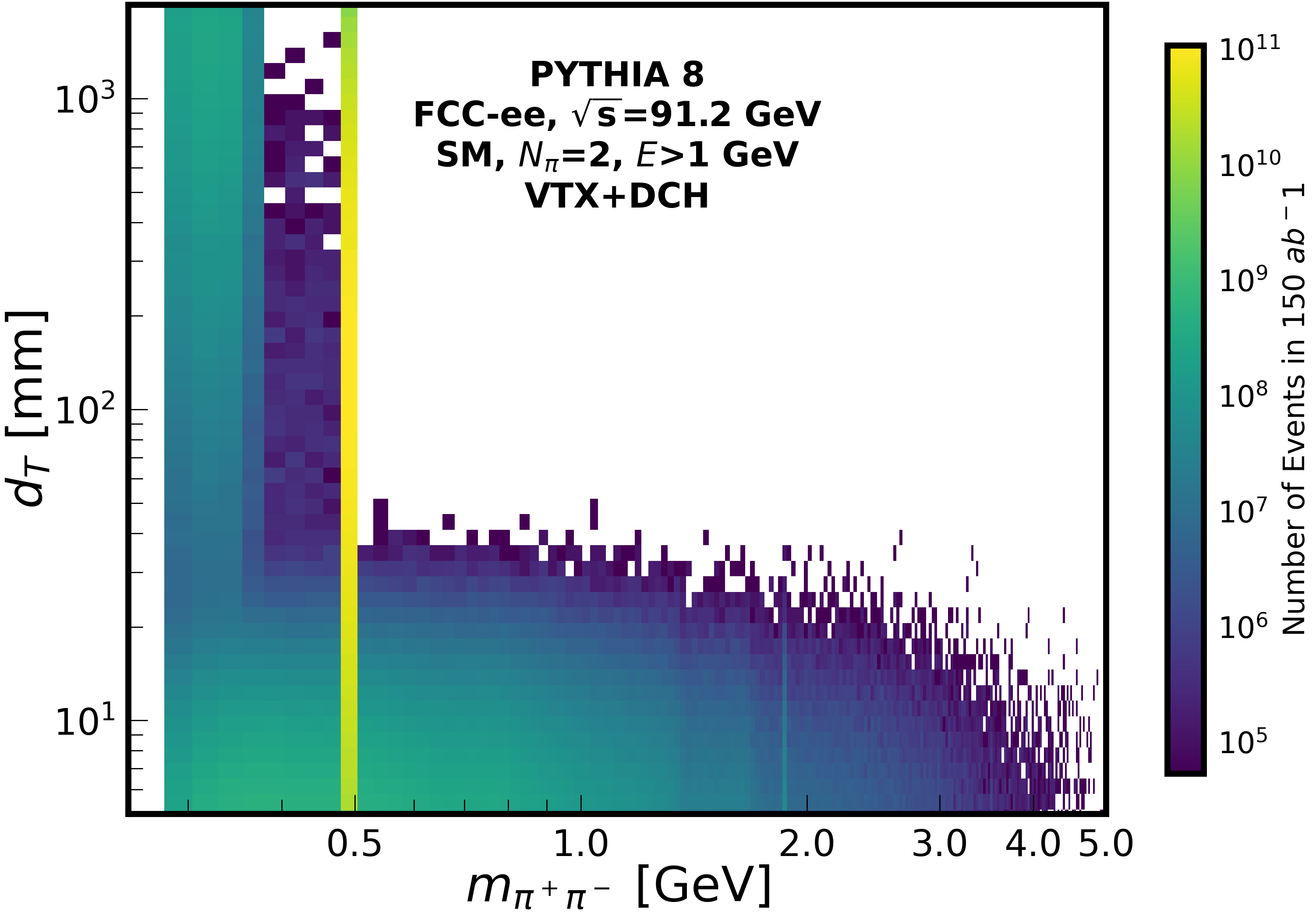}~
    \includegraphics[width=0.5\textwidth]{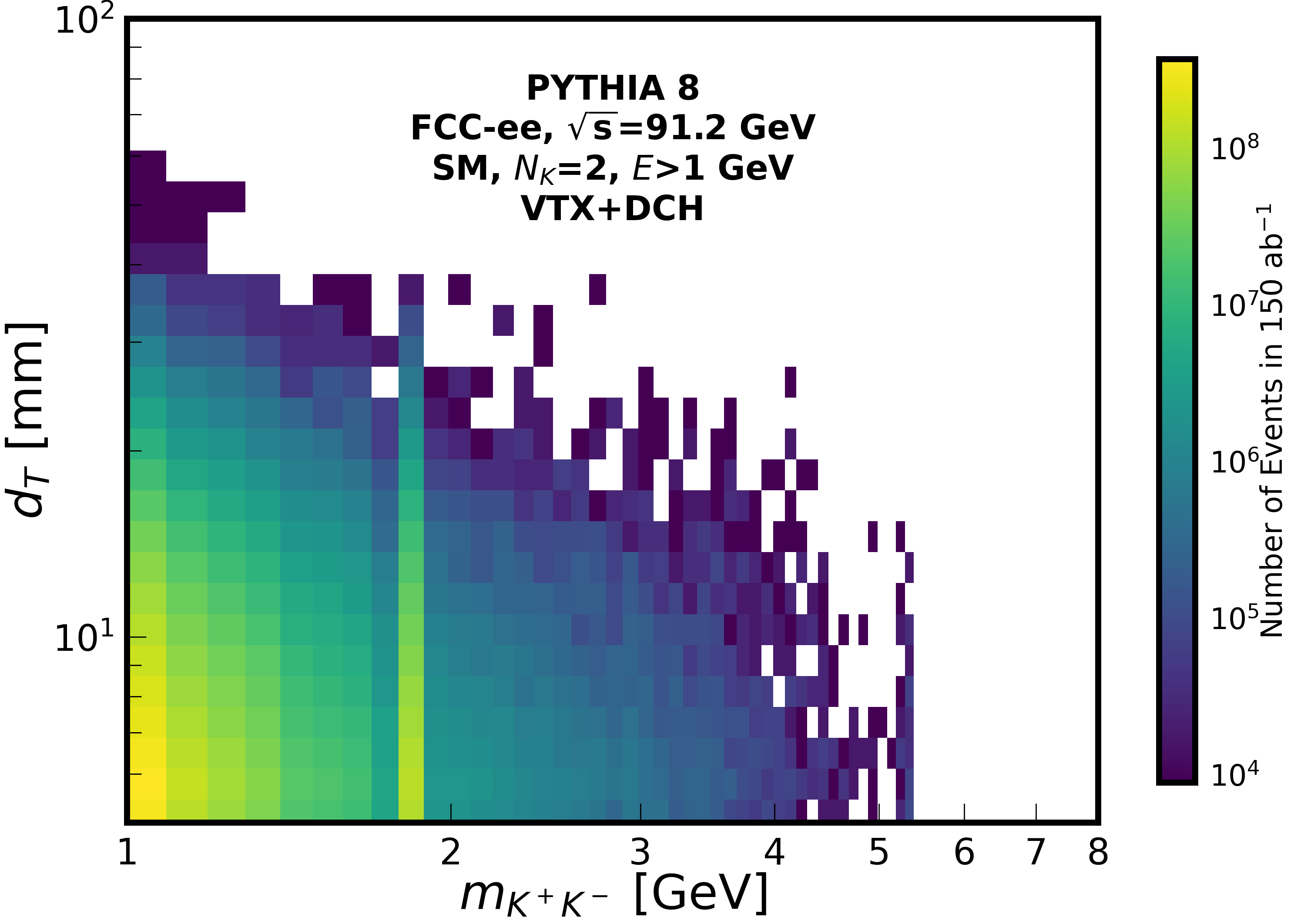}
      \caption{\textit{Left:} Invariant mass of the di-pion decay with respect to the $d_T$ of the vertex for the SM background. The distinct region at $m_{\pi^+\pi^-} \sim 0.5$\,GeV corresponds to the resonance decay $K_S\to\pi^+\pi^-$. The wide band extending from 0.28\,GeV (2$m_\pi^\pm$) $\lesssim m_{\pi^+\pi^-} \lesssim$ 0.36\,GeV ($m_{K_L^0} - m_{\pi^0}$) corresponds to the $K_L\to \pi^+\pi^-\pi^0$ decay, where the $\pi^0$ remains undetected. The $0.36\,{\rm GeV} \lesssim m_{\pi^+\pi^-} \lesssim 0.5$\,GeV region is populated by $K_S^0 \to \pi^+\pi^- e^+ e^-$ decays, where missing low-energy electrons lead to improper reconstruction of the $K_S^0$. \textit{Right:} Invariant mass of the di-kaon decay with respect to the $d_T$ of the vertex for the SM background. The distinct line at $m_{K^+K^-}\sim 1.87~\text{GeV}$ corresponds to the decay $D^0\to K^+K^-$.}
    \label{fig:dipik_mdT1}
\end{figure}

The two-dimensional distribution of the background events in 
di-pion and di-kaon final states in the
di-meson invariant mass - $d_T$ plane after applying the 
selection cuts is shown in Fig.\,\ref{fig:dipik_mdT1}. The 
mass of the charged pion is 0.139\,GeV. Hence, the events start populating from $m_{\pi^+\pi^-} \gtrsim$ 0.28\,GeV. 
The SM events with pions coming from the decay of a $K_S$ are observed at $m_{\pi^+\pi^-}= m_{K_S} \sim 0.5$\,GeV, 
which corresponds to the resonance in the {\it left} panel of Fig.\,\ref{fig:dipik_mdT1}. These di-pion vertices are 
displaced, as inferred from the $d_T$ reaching up to 2\,m. 

Furthermore, we can also have displaced di-pion vertices 
from the decay $K_L^0\to \pi^+\pi^-
\pi^0$, where the VTX and DCH detectors reconstruct only the 
two charged pions and associate them to the same vertex.
The reconstructed mass of the $K_L$ decay products depends on the missing energy of the 
unassociated $\pi^0$. 
Hence, we observe the band of events with decay vertices
clustered within 0.28\,GeV (2$m_\pi^\pm$) $\lesssim m_{\pi^+\pi^-} \lesssim$ 0.36\,GeV ($m_{K_L^0}$ - $m_{\pi^0}$) in the {\it left} panel of
Fig.\,\ref{fig:dipik_mdT1}. These extend to large values of $d_T$ due to the large decay length of the $K_L$.
The SM process $K_S^0 \to \pi^+\pi^- e^+ e^-$ also contributes to the production of events observed as
di-pion vertices, when the low energy electrons miss detection, albeit with a small decay branching ratio for the $K_S^0$ ($5\times10^{-5}$).
This populates the region $0.36\,{\rm GeV} \lesssim m_{\pi^+\pi^-} \lesssim 0.5$\,GeV, extending again to large values of $d_T$.

\begin{figure}[htb!]
    \centering
    \includegraphics[width=0.5\textwidth]{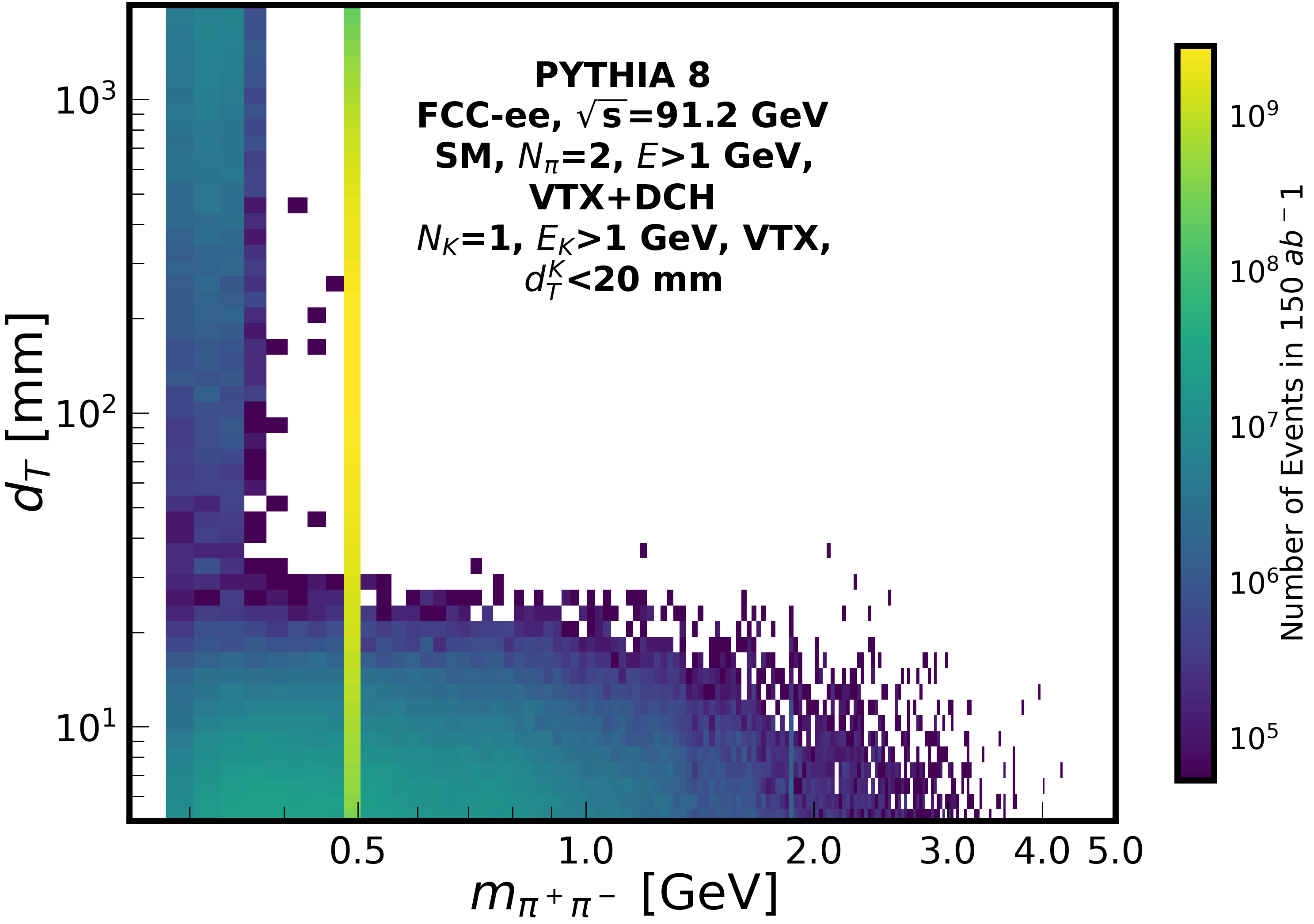}~
    \includegraphics[width=0.5\textwidth]{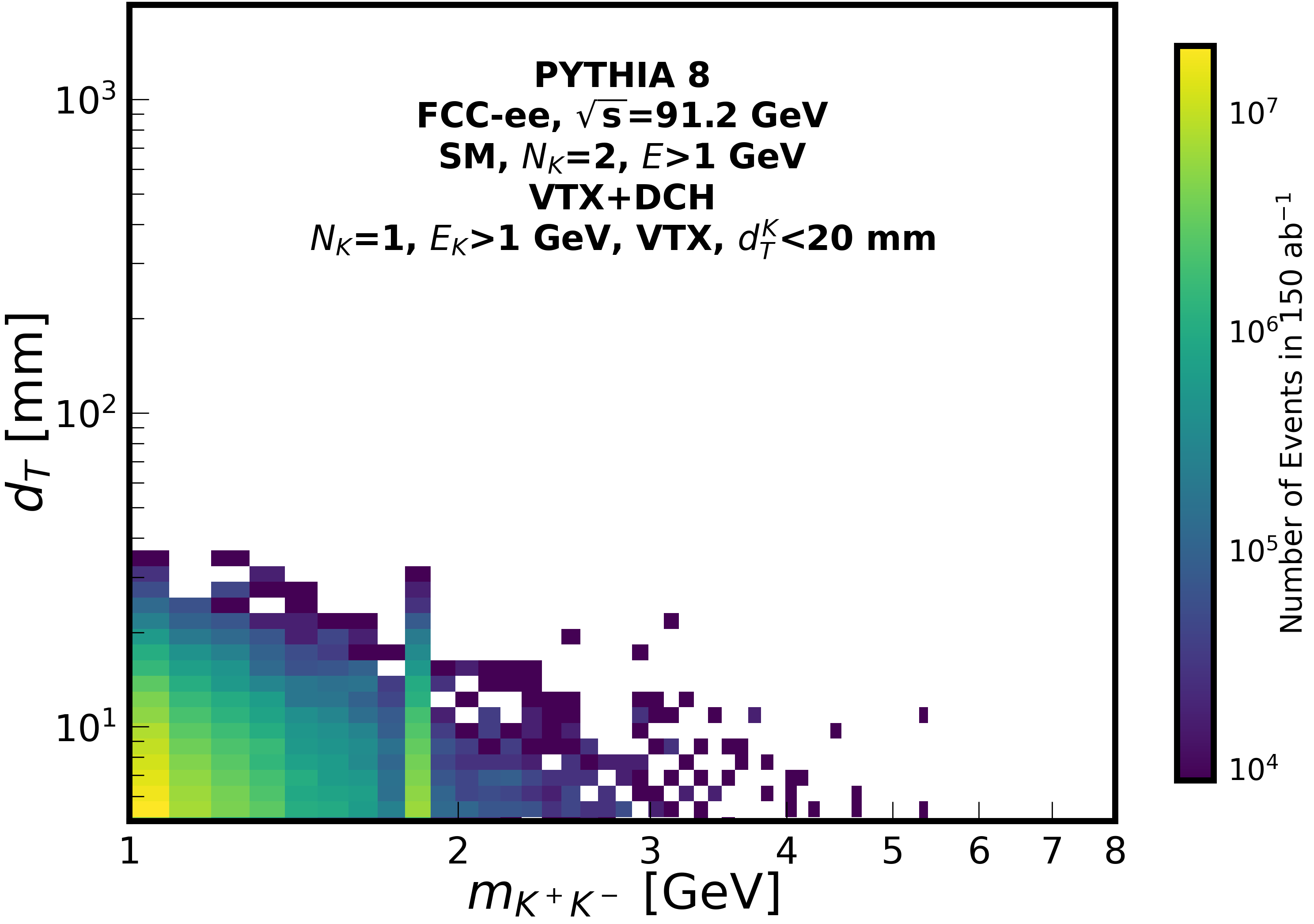}\\
    \includegraphics[width=0.5\textwidth]{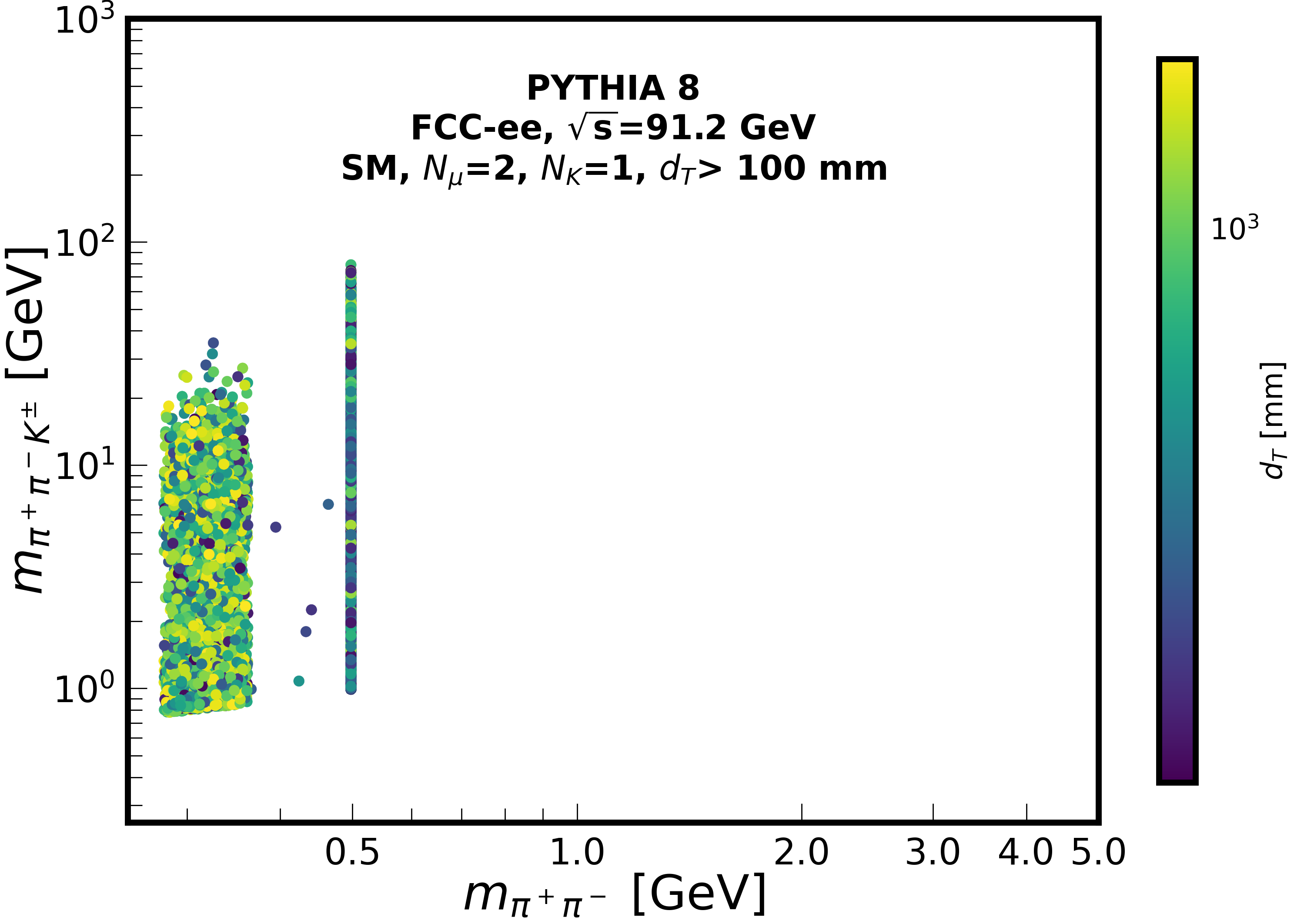}~
    \includegraphics[width=0.5\textwidth]{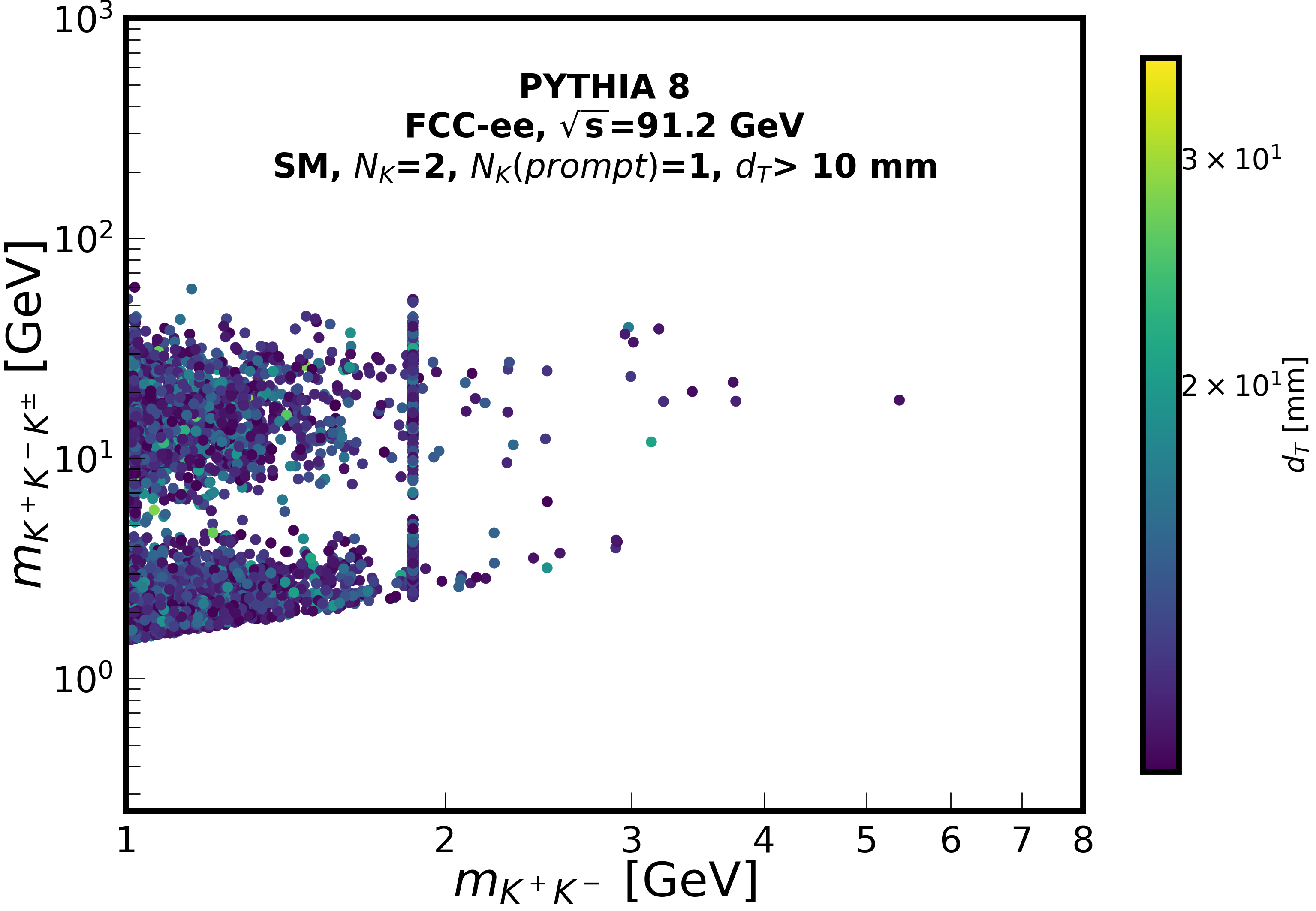}
    \caption{\textit{Top row:} Invariant mass of the di-pion \textit{(left)} and the di-kaon final state \textit{(right)} with respect to the $d_T$ of the vertex for the SM background. \textit{Bottom row:} The left panel shows the invariant mass of the di-pion final state with respect to the invariant mass of the di-pion and kaon system; the right panel shows the invariant mass of the di-kaon final state with respect to the invariant mass of the di-kaon and kaon system.}
    \label{fig:dipik_mdT2}
\end{figure}

From the distribution for the di-pion case, it is clear that benchmarks with masses less than 0.5\,GeV decaying to two pions are incredibly difficult to probe, at least with the use of the invariant mass and $d_T$ variables. The long lifetimes of the $K_{L/S}$ mesons result in a huge background in the low mass region. 
Even with an additional prompt kaon, the background still populates the $m_{\pi^+\pi^-K^\pm} \stackrel{\sim}{=} m_{B^\pm} = 5\,\text{GeV}$ region for $m_{\pi^+\pi^-} = 0.4$\,GeV. This is highlighted for the di-pion final state in the {\it bottom left} panel of Fig.\,\ref{fig:dipik_mdT2}.
Therefore, it is not possible to reduce the SM background for these low-mass benchmarks.

For the di-kaon final state, we observe from the {\it right} 
panel of Fig.\,\ref{fig:dipik_mdT1}, the mass of the di-kaon 
final state starts from $m_{K^+K^-} 
\gtrsim 1$\,GeV. The major background events originating 
here come from $D$ meson decays, populating the distribution 
below $m_{K^+K^-} \sim 1.87$\,GeV. The resonance at 
1.87\,GeV corresponds to the mass of the $D^0$ meson. We 
find that the $d_T$ values of the di-kaon final states 
are not as large as in the di-pion case, as we had already seen from the $d_T$ distribution of charm mesons in 
Fig.\,\ref{fig:bkg_dist_BKphi}.
This implies that a $d_T$ cut of 150\,mm might get rid of all the SM backgrounds for the di-kaon final state. Again, the background populates the $m_{K^+K^-K^\pm} \sim m_{B^\pm} = 5\,\text{GeV}$ region for the 1\,GeV benchmark.

\begin{figure}[htb!]
    \centering
    \includegraphics[width=0.6\textwidth]{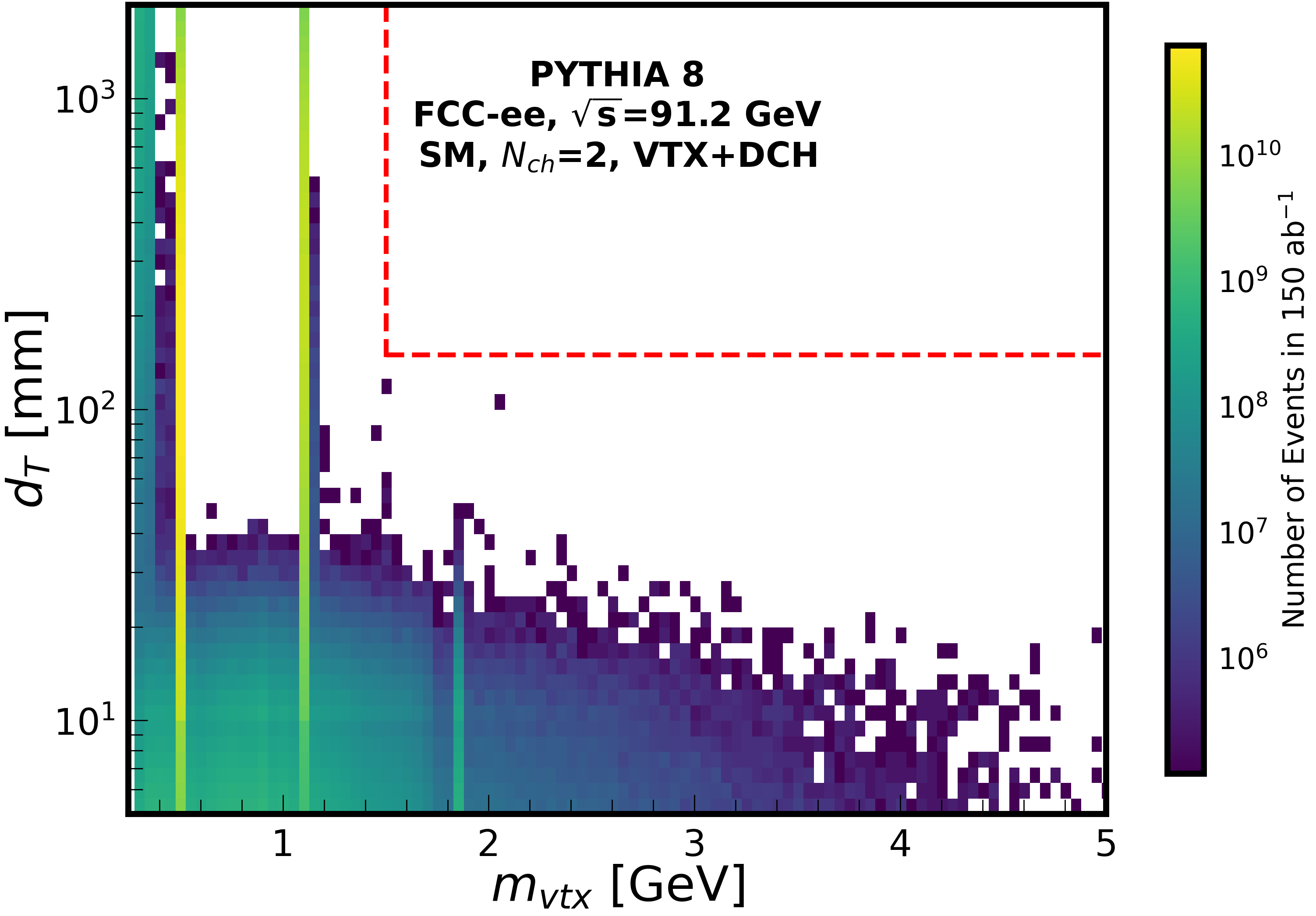}
    \caption{Invariant mass of the final state decay products with two charged particles that include pions, kaons, and protons, with respect to the $d_T$ of the vertex for the SM background. In addition to the features of Fig.~\ref{fig:dipik_mdT1}, the plot also contains a densely populated region at $m_{\rm vtx}\sim1.1\,\text{GeV}$, that extends to large $d_T$. The line corresponds to the decay $\Lambda^{0}\to p^+\pi^-$.
    The {\it red dashed lines} mark the region in the $m_{\rm vtx}-d_T$ plane which is free from the SM background, motivating the cuts on these variables.}
    \label{fig:pi_K_p}
\end{figure}

Until now, we have only considered the final states containing two pions and/or two kaons as SM backgrounds. However, we have to consider the possibility of a proton faking as a $\pi^+$ or a $K^+$. Protons can be produced from the decay of a $\Lambda^0$ baryon: $\Lambda^0\to p^+ \pi^-$. 
The misidentified protons coming from the $\Lambda^0$ decay are an important background since the $\Lambda^0$ baryon is 
long-lived, having $d_T$ values beyond 1\,m, similar to $K_{S/L}$ as shown in Fig.\,\ref{fig:bkg_dist_BKphi}. It has 
a mass of 1.1\,GeV, making it a major background for benchmarks such as BPA2 and BPA3, which have a mass of 1\,GeV. Fig.\,\ref{fig:pi_K_p} shows the distribution of the 
final state invariant mass and $d_T$, taking into account the events 
containing a proton too. Even if the excellent charge 
identification of the DCH achieves a proton rejection 
efficiency of $99.999\%$, 
we still have $\mathcal{O}(10^4)$ background events from the $\Lambda^0$ decay.
Since this affects both the di-pion and di-kaon final states equally, we cannot simply use a large $d_T$ cut even for the di-kaon signal.

\begin{table}[hbt!]
    \centering
    \begin{tabular}{c | c}
    \hline
        \multicolumn{2}{c}{Final cuts for the di-pion/di-kaon final state} \\ \hline
        $m_\phi\gtrsim 1.5$\,GeV  & All benchmarks \\ \hline\hline
        $D_{\rm vtx}$ $\in$ VTX or DCH & \multirow{2}{*}{$D_{\rm vtx}$ $\in$ MS} \\
        $N_{\rm ch}$ = 2 & \\
        $N_{\pi/K}$ = 2 & $N_{\rm ch}$ = 2 \\
        $E > 1$\,GeV & \multirow{2}{*}{$E>5$\,GeV} \\
        $d_T > 150$\,mm & \\ \hline\hline
    \end{tabular}
    \caption{Final selection cuts applied on the signal and SM background events when the final state contains two pions or two kaons.}
    \label{tab:sig_pik_cuts}
\end{table}

For the benchmarks BPA1, BPA2, and BPA3, if we were to use 
the variable $m_{\pi^+\pi^-/K^+K^-}$ for selecting the 
signal events, the analysis would entirely depend on the 
resolution of the reconstructed 
invariant mass of the final state decay products. 
Despite an excellent mass resolution, the background statistics 
play a crucial role because the SM background populates this region even for large $d_T$. 
The overlap of the signal with $\mathcal{O}(10^{10})$ background events (see Fig.~\ref{fig:pi_K_p}) shall render it impossible to detect such LLPs. 
The possible strategy to search for these benchmarks is to carry out the analysis in the muon system.
For the rest of the benchmarks, we apply a $d_T > 150$\,mm cut on the backgrounds to reduce its contribution to zero. We tabulate the final cuts used for the analyses of benchmarks with $m_\phi \gtrsim 1.5$\,GeV 
in the {\it left} column of Table\,\ref{tab:sig_pik_cuts}.
After applying these cuts, we show the number of signal events, both for $\phi\to\pi^+\pi^-$ and $\phi\to K^+K^-$ in Table\,\ref{tab:pik_vtx_dch}.

\begin{table}[htb!]
\centering
\resizebox{0.9\textwidth}{!}{
\begin{tabular}{|c|c|c|c|c|}
\hline
\multirow{3}{*}{Benchmarks} & \multicolumn{4}{c|}{VTX + DCH analysis} \\ \cline{2-5} 
 & \multirow{2}{*}{$m_{\phi}$ (GeV)} & \multirow{2}{*}{$c\tau$ (mm)} & \multicolumn{2}{c|}{Number of events} \\ \cline{4-5} 
 &  &  & $\phi\to \pi^+\pi^-$ & $\phi\to K^+K^-$ \\ \hline\hline
BPA4 & 2.0 & 135.2 & 26201.89 $\pm$ 168.14 & 26970.33 $\pm$ 170.57 \\
BPA5 & 3.5 & 10385.4 & 3.84 $\pm$ 0.10 & 1.54 $\pm$ 0.04\\
BPA6 & 3.5 & 65.8 & 9494.65 $\pm$ 63.93  & 3822.23 $\pm$ 25.53 \\ \hline\hline
\end{tabular}
}
\caption{Number of signal events (with statistical uncertainties) in BPA4, BPA5, and BPA6 after a cut-based analysis in the vertex detector (VTX) and drift chamber (DCH). The number of background events in each case is null.}
\label{tab:pik_vtx_dch}
\end{table}

The benchmarks with $m_\phi \lesssim 1.5$\,GeV, like BPA1, BPA2, and BPA3, along with the other benchmarks, are then analyzed in the MS. 
The MS provides a relatively cleaner environment since all the long-lived hadrons and their decay products, apart from muons, deposit their energy in the calorimeter. 
We select events that contain one displaced vertex detected within the MS, with two charged particles, and $E>5$\,GeV. We tabulate the cuts for the MS analysis in the {\it right} column of Table\,\ref{tab:sig_pik_cuts}.
Table\,\ref{tab:pik_ms} shows the number of signal events for all the benchmarks for the MS analysis. 
The di-pion and di-kaon final states are combined because pions and kaons might not be distinguished in the MS. 

The background is reduced to zero in this case. 
We observe 4 events for BPA1 in the MS. 
Increasing the luminosity can help in observing more events for this 
benchmark. BPA2 and BPA3 
are observed to have a sufficient number of events. BPA5, 
with a $c\tau$ of $\sim10$\,m is expected to have 4 observed 
events in the MS in addition to the 4 (2) events in the 
VTX+DCH analysis with pions (kaons) in the final state. 
The benchmark BPA6 is expected to have a significant number of events in the VTX+DCH analysis, while BPA4 has a good prospect with both the VTX+DCH and MS analyses.

\begin{table}[htb!]
\centering
\resizebox{0.7\textwidth}{!}{
\begin{tabular}{|c|c|c|c|}
\hline
\multirow{2}{*}{Benchmarks} & \multicolumn{3}{c|}{Muon Spectrometer analysis} \\ \cline{2-4} 
 & $m_{\phi}$ (GeV) & $c\tau$ (mm) & Number of events \\ \hline\hline
BPA1 & 0.4 & 39666.6 & 3.61 $\pm$ 0.65 \\
BPA2 & 1.0 & 554.3 & 206.46 $\pm$ 4.06 \\
BPA3 & 1.0 & 34.6 & 77.99 $\pm$ 9.98 \\
BPA4 & 2.0 & 135.2 & 1515.04 $\pm$ 40.38 \\
BPA5 & 3.5 & 10285.4 & 3.59 $\pm$ 0.08 \\
BPA6 & 3.5 & 65.8 & 3.01 $\pm$ 0.95 \\ \hline\hline
\end{tabular}
}
\caption{Number of signal events (with statistical uncertainties) in benchmarks BPA1 to BPA6  
after a cut-based analysis in the Muon System (MS). The 
number of background events in each case is null.}
\label{tab:pik_ms}
\end{table}

\begin{figure}[htb!]
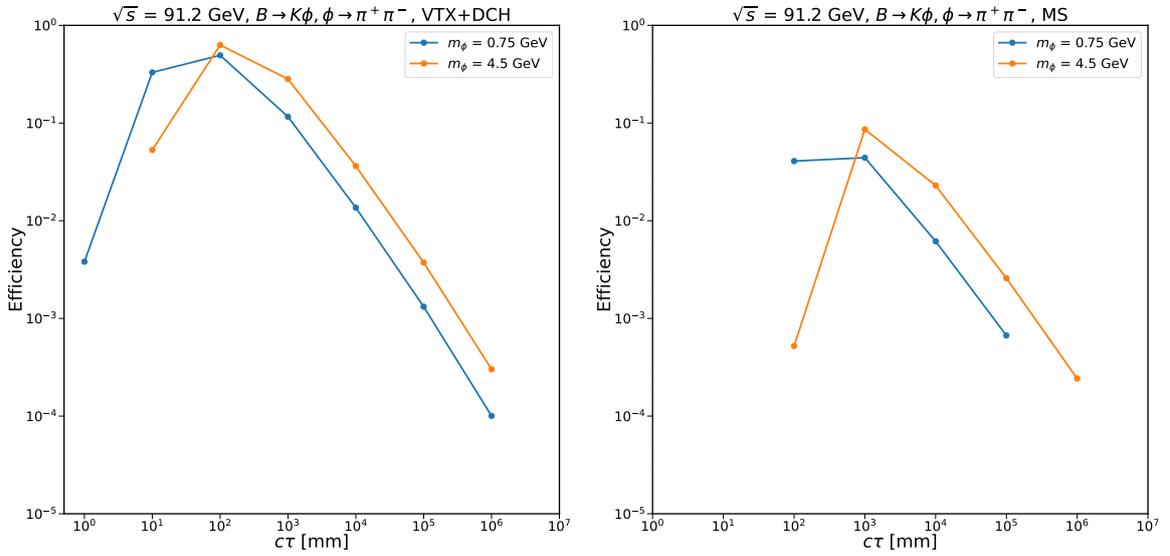

    \centering
    \includegraphics[width=0.5\textwidth]{Eff_dipion_plot.pdf}~
    \includegraphics[width=0.5\textwidth]{Eff_dipion_MS_plot.pdf}
    \caption{Efficiency for the di-pion decay mode of the LLP $\phi$ for varying decay lengths and masses. The generation of LLP involved the $Z\to b\bar{b}$ process where one $B$ meson decays via BSM process $B\to K \phi$, and the other $B$ decays to SM particles. The final state pions were detected with a cut: $d_T > 150$ mm within VTX and DCH \textit{(left)}, and in the muon system (MS) \textit{(right)}.}
    \label{fig:dipion_eff}
\end{figure}

Using the two kinds of search techniques presented above, we 
perform the analysis beyond our chosen LLP benchmarks, 
assuming 100\% decay to di-pion final states. The muon 
system analysis combines both di-pion and di-kaon events. 
The efficiencies of the VTX+DCH analysis and the MS analysis 
for the di-pion final state are presented in 
Fig.\,\ref{fig:dipion_eff}. The efficiencies of the di-kaon 
final state are similar to the di-pion efficiencies. 

\subsubsection{Analysis in the $c\bar{c}$ Final State}
\label{sssec:A_ccbar}

One of our chosen benchmarks for \textbf{Case A}, BPA7, has 
a significant branching ratio to the $c\bar{c}$ final 
state. After the decay $\phi\to 
c\bar{c}$, the charm quarks hadronize to form mesons, 
particularly the $D$ mesons: $D^0, D^\pm$. Because $\phi$ is 
long-lived, the $D^0,D^\pm$ and particles formed after the hadronization are detected at a 
large $d_T$ depending on the $c\tau$ of $\phi$. The 
signature for this LLP benchmark is the presence of multiple 
displaced vertices and/or a large number of charged 
particles connected to a displaced vertex. 
The potential SM background consists of prompt bottom and 
charm production. The $B$ hadrons lead to displaced 
vertices because of their long 
lifetimes. The $D$ mesons produced from the decay of the $B$
are thus displaced. Other long-lived mesons like $K_S$ and 
$K_L$ can also mimic the signal by decaying into multiple 
charged particles at large $d_T$. The SM particles produced 
from the hadronization of $b\bar{b}$ and $c\bar{c}$ states 
may also include some baryons, such as the $\Xi$, $\Sigma$, 
or $\Lambda_c$ that have longer decay lengths, as 
discussed in Section\,\ref{ssec:bkgs}. 

\begin{table}[hbt!]
    \centering
    \begin{tabular}{c}
    \hline
        Selection cuts to study the SM $c\bar{c}$ background \\ \hline\hline
        $D_{\rm vtx} \in$ VTX or DCH \\
        $N_{\rm ch} \ge 3$ \\
        $E > 5$\,GeV \\
        $d_T > 5$\,mm \\ \hline\hline
    \end{tabular}
    \caption{Selection cuts applied to study the SM background for the $c\bar{c}$ final state.}
    \label{tab:bkg_cc_cuts}
\end{table}

\begin{figure}[htb!]
    \centering
    \includegraphics[height=5cm]{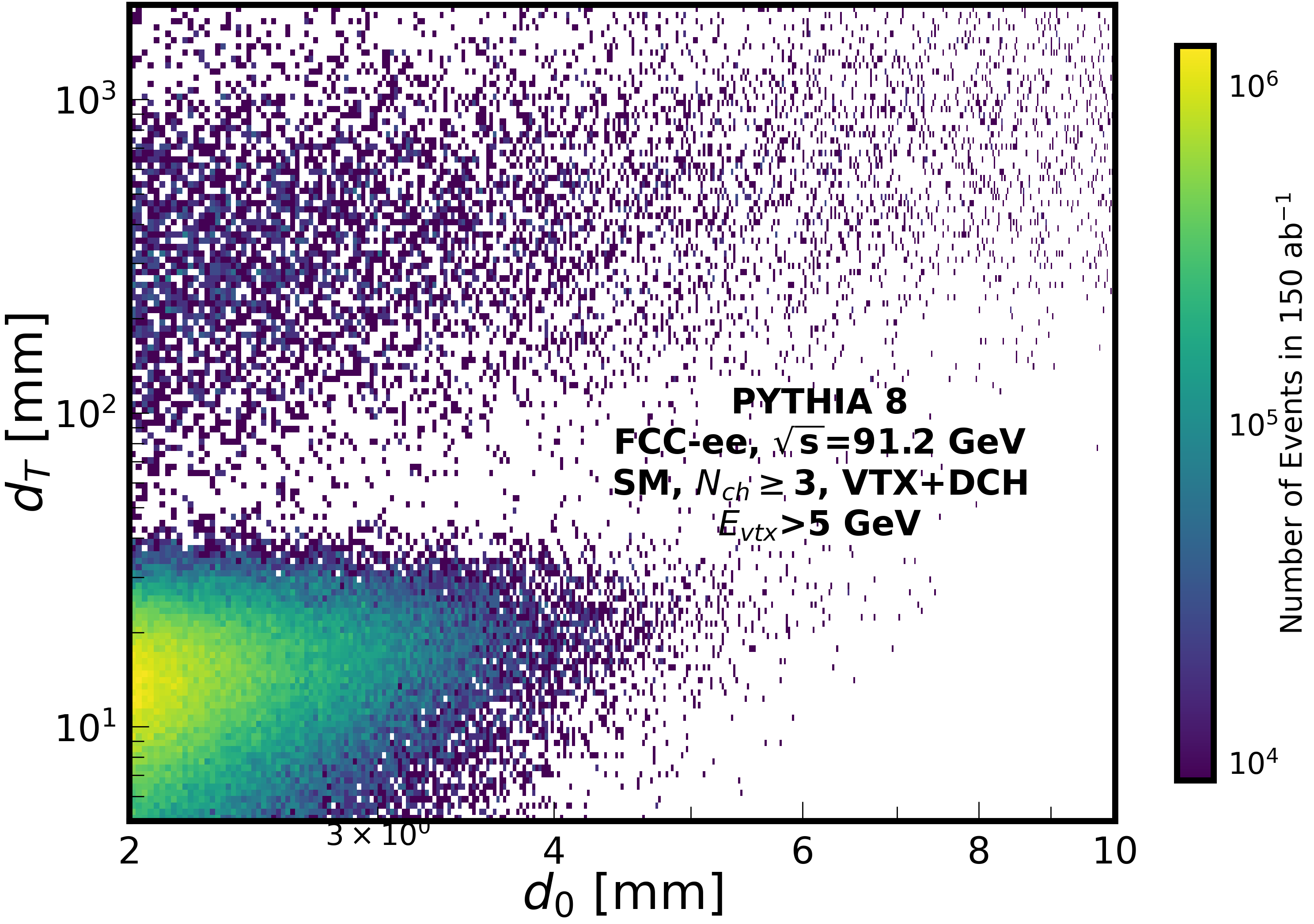}~
    \includegraphics[height=5cm]{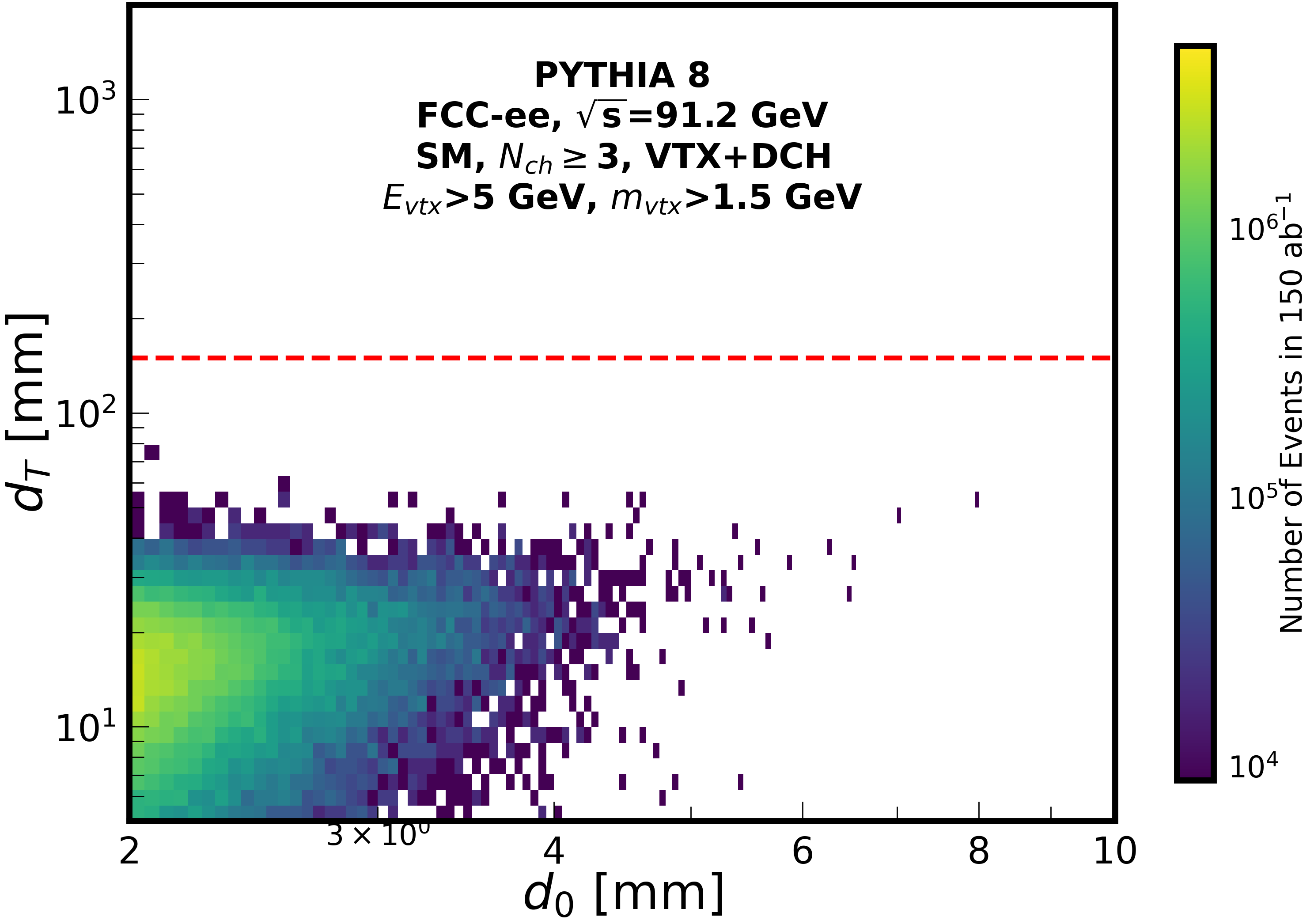}
    \caption{Distribution of the SM background events from $Z$ production in the $d_0-d_T$ plane before ({\it left}) and after ({\it right}) applying a cut on the invariant mass of the final state decay products. 
    The {\it red dashed line} marks the $d_T$ threshold which reduces the SM backgrounds. The $m_{\rm inv}>1.5\,\text{GeV}$ cut removes the background SM long-lived baryons and mesons (see Fig.~\ref{fig:pi_K_p}).
    }
    \label{fig:ccbar_d0dT}
\end{figure}

Table\,\ref{tab:bkg_cc_cuts} lists the selection cuts 
applied to study the background and design cuts to suppress 
the backgrounds completely. We require the vertex to have at 
least 3 outgoing charged particles. The total energy of the 
decay particles from a vertex must be greater than 
5\,GeV, and $d_T$ of the vertex must be at least 5\,mm. After applying 
these selection cuts, we plot the distribution of the SM 
background events in the plane of $d_T$ and the impact 
parameter, $d_0$, in the {\it left} panel of 
Fig.\,\ref{fig:ccbar_d0dT}. We observe numerous displaced vertices with $d_T$ values as high 
as a few meters, with impact parameters extending beyond 10\,mm.

\begin{table}[hbt!]
\centering
\resizebox{\textwidth}{!}{
\begin{tabular}{|c c c|c|c|}
\hline
\multicolumn{3}{|c|}{Benchmark} & $m_{\phi}$ (GeV) & $c\tau$ (mm) \\ \cline{4-5}
\multicolumn{3}{|c|}{BPA7} & 4.4 & 95.0 \\ \hline
\multicolumn{3}{|c|}{Final cuts for the $c\bar{c}$ final state} & \multicolumn{2}{c|}{Number of events} \\ \hline\hline
$D_{\rm vtx}\in$ VTX or DCH & \multirow{3}{*}{\Bigg\} +} & $d_T >$ 100\,mm,  $d_0 >$ 10\,mm & \multicolumn{2}{c|}{6.29 $\pm$ 0.08} \\
$N_{\rm ch}\ge 3$, $E>5$\,GeV & & $d_T >$ 150\,mm,   $d_0 >$ 10\,mm & \multicolumn{2}{c|}{5.96 $\pm$ 0.08} \\
$m_{\rm vtx}>1.5$\,GeV & & $d_T >$ 250\,mm, $d_0 >$ 20\,mm & \multicolumn{2}{c|}{3.19 $\pm$ 0.06} \\ \hline\hline
\end{tabular}
}
\caption{Number of signal events (with statistical uncertainties) for BPA7, with $c\bar{c}$ final state, with cuts on $d_T$ and $d_0$, where the first set of cuts are applied in all three cases. The background is expected to be zero in each case. }
\label{tab:BPA7_cut}
\end{table}

\begin{figure}[htb!]
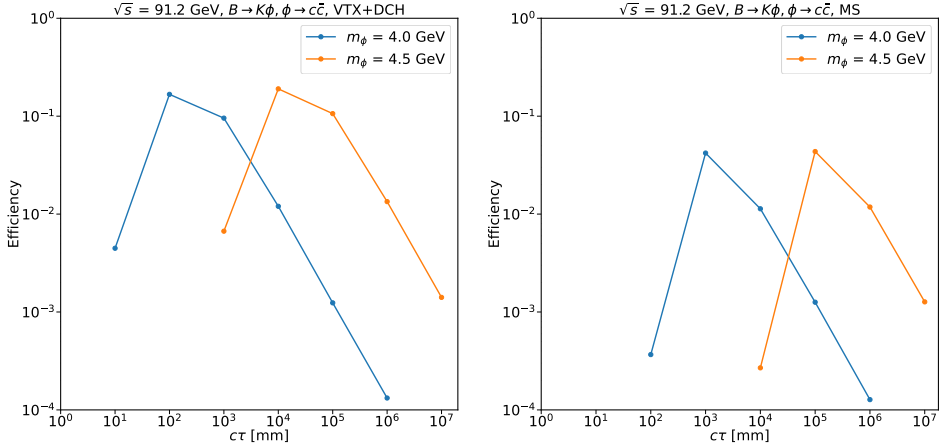

    \centering
    \includegraphics[width=0.4\textwidth]{Eff_ccbar.pdf}~
    \includegraphics[width=0.4\textwidth]{Eff_ccbar_MS.pdf}
    \caption{Efficiency for the $c\bar{c}$ decay mode of LLP $\phi$ for varying decay lengths and masses of $\phi$. The generation of LLP involved the $Z\to b\bar{b}$ process where one $B$ meson decays via BSM process $B\to K \phi$, and the other $B$ decays to SM particles. The final state pions were detected with a cut: $d_T >$ 150 mm within VTX and DCH \textit{(left)}, and in the Muon System (MS) \textit{(right)}.}
    \label{fig:ccbar_eff}
\end{figure}

If we apply a cut on the invariant mass
of 1.5\,GeV, the events with large $d_0$ and $d_T$ are 
removed, rendering the background to be constrained to $d_T 
< 100$\,mm, as shown in the {\it right} panel of 
Fig.\,\ref{fig:ccbar_d0dT}.
This is because the cut removes the major source of 
displaced background, {\it i.e.,} long-lived baryons with mass 
$\lesssim$ 1.5\,GeV (\textit{cf.} 
Table\,\ref{tab:bkg_list}). We then apply several cuts on 
$d_T$ and $d_0$ to extract signal events with zero
background.  
The final set of cuts along with the results are 
presented in Table\,\ref{tab:BPA7_cut}. Note that we don't 
perform the MS-only analysis for this benchmark since 
the events have a negligible probability of 
reaching the MS, as shown in Table\,\ref{tab:decay_events_a}.

Similar to the previous analyses, we extend our analysis for LLPs decaying to $c\bar{c}$ for various decay lengths and masses. The efficiencies for two such mass points with varying decay lengths are shown in Fig.\,\ref{fig:ccbar_eff}. 
The efficiencies for both the VTX+DCH analysis and the MS analysis are calculated over a range of mass and lifetime of $\phi$. 
For the former, the signal events must have a minimum $d_0$ of 10\,mm and a $d_T$ of 150\,mm. In the latter case, a sufficient condition is to detect displaced activities in the MS, with $N_{\rm ch}\ge 3$ and $E>5$\,GeV.

\subsubsection{Summary for {\bf Case A}}

We present in Table~\ref{tab:summary_phi_total_all} a summary of all the benchmark points, listing the total number of signal events obtained by combining the three decay modes across the VTX+DCH and MS analyses. The analyses discussed above are background-free, with the caveat that material interactions have not been considered in this study. We further evaluate the signal yield using a more conservative selection on $d_T$. Keeping all other selection criteria unchanged, the $d_T > 150~\text{mm}$ cut is replaced by a tighter cut of $d_T > 340~\text{mm}$, effectively omitting the VTX region (refer to Table~\ref{tab:IDEA_det} for the dimensions). The DCH region, characterized by a very low material budget, is expected to have negligible contamination from material-induced backgrounds~\cite{IDEAStudyGroup:2025gbt}.

\begin{table}[htb!]
\centering
\resizebox{0.95\textwidth}{!}{
\begin{tabular}{|c|c|c|c|c|}
\hline
\multirow{2}{*}{Benchmarks} & \multirow{2}{*}{$m_{\phi}$ (GeV)} & \multirow{2}{*}{$c\tau$ (mm)} &
\multicolumn{2}{c|}{Total signal events (all channels combined)} \\ \cline{4-5}
 & & & $d_T > 150$\,mm & $d_T > 340$\,mm \\ \hline\hline

BPA1 & 0.4 & 39666.6 &$6.60 \pm 0.68$ & $6.46 \pm 0.68$ \\

BPA2 & 1.0 & 554.3 & $206.46 \pm 4.06$ & $206.46 \pm 4.06$ \\

BPA3 & 1.0 & 34.6 & $77.99 \pm 9.98$ & $77.99 \pm 9.98$ \\

BPA4 & 2.0 & 135.2 & $64043.14 \pm 248.89$ & $46599.71 \pm 170.28$ \\

BPA5 & 3.5 & 10285.4 & $10.04 \pm 0.14 $ & $9.28 \pm 0.13$ \\

BPA6 & 3.5 & 65.8 & $14235.26 \pm 69.11$ & $7806.75 \pm 51.16$ \\

BPA7 & 4.4 & 95.0 &$5.96 \pm 0.08$ & $4.41 \pm 0.07$ \\

\hline\hline
\end{tabular}
}
\caption{Summary of the total number of signal events (with statistical uncertainties) for benchmarks BPA1–BPA7, combining all $\phi$ decay modes ($\phi\to\mu^+\mu^-$, $\phi\to\pi^+\pi^-$, $\phi\to K^+K^-$, and $\phi\to c\bar{c}$) across all detector regions (VTX, DCH, and MS).}
\label{tab:summary_phi_total_all}
\end{table}

\begin{figure}[hbt!]
    \centering
    \includegraphics[width=0.8\textwidth]{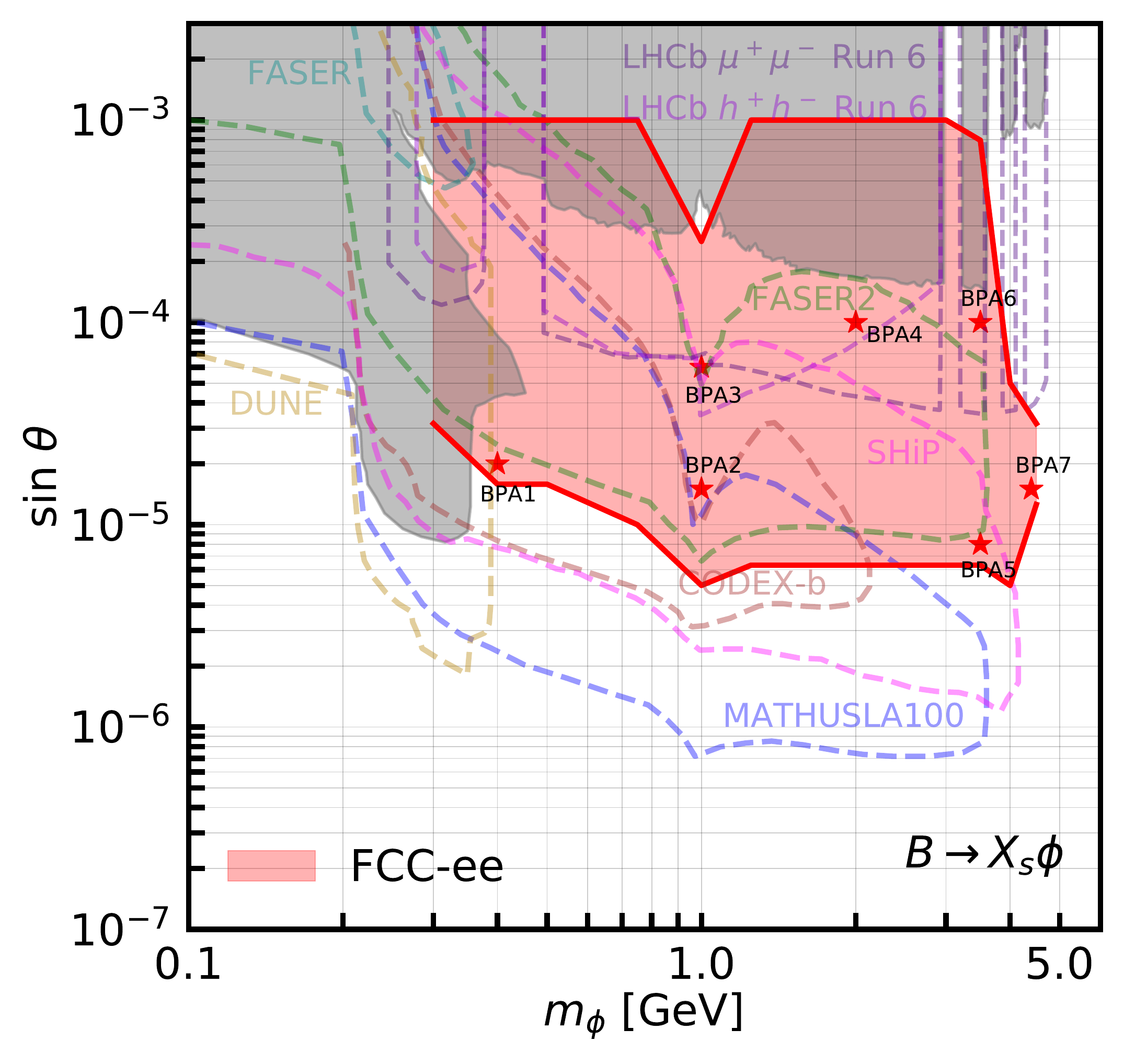}
    \caption{The FCC-ee sensitivity from our analyses with 3 observed events along with the current bounds and projected sensitivities as shown in Fig.\,\ref{fig:benchmarks_BtoKphi} for {\bf Case A}.}
    \label{fig:FCC_ee_BtoKphi}
\end{figure}

Although benchmarks like BPA4 and BPA6 undergo a drastic decrease in the number of events with a tighter $d_T$ cut, the signal yield remains sufficiently large for a discovery, especially in the null background scenario. BPA7 shows a modest yield in both cases. The rest of the benchmarks are either too long-lived or have masses that overlap with SM long-lived particles. In either case, the analysis is dominated by decays in the muon spectrometer, rendering them unaffected by the $d_T$ variable.

We now scan the $m_\phi - \sin\theta$ parameter space and combine the VTX+DCH and MS analyses in the di-muon, di-pion, di-kaon, and $c\bar{c}$ decay channels. The \textit{red} region in Fig.~\ref{fig:FCC_ee_BtoKphi} shows the bound on the $m_{\phi}-\sin\theta$ parameter space from our study, compared with the existing constraints. Our study finds that in the case of negligible trilinear coupling, FCC-ee can probe benchmarks BPA3 and BPA7, beyond the projected sensitivity of any existing proposed experiments.
This is achieved with the pion/kaon MS only analysis and the $c\bar{c}$ analysis in the VTX+DCH detectors, respectively, for BPA3 and BPA7.
For benchmarks BPA2, BPA4, and BPA6, where FASER2, LHCb, and SHiP experiments have some sensitivity, FCC-ee is expected to provide additional prospects.
While for BPA2, we expect around 200 signal events in the MS with zero backgrounds with our analysis, for BPA4 and BPA6, we expect to observe $\mathcal{O}(10000)$ ($\mathcal{O}(1000)$) signal events after the VTX+DCH (MS only) analysis.
This would help in the identification of the model and estimation of the model parameters, even when we observe a signal in the previous experiments, like SHiP.

\subsubsection{Production via $Z\to Z^*\phi$}

\begin{figure}[hbt!]
    \centering
    \includegraphics[width=0.5\linewidth]{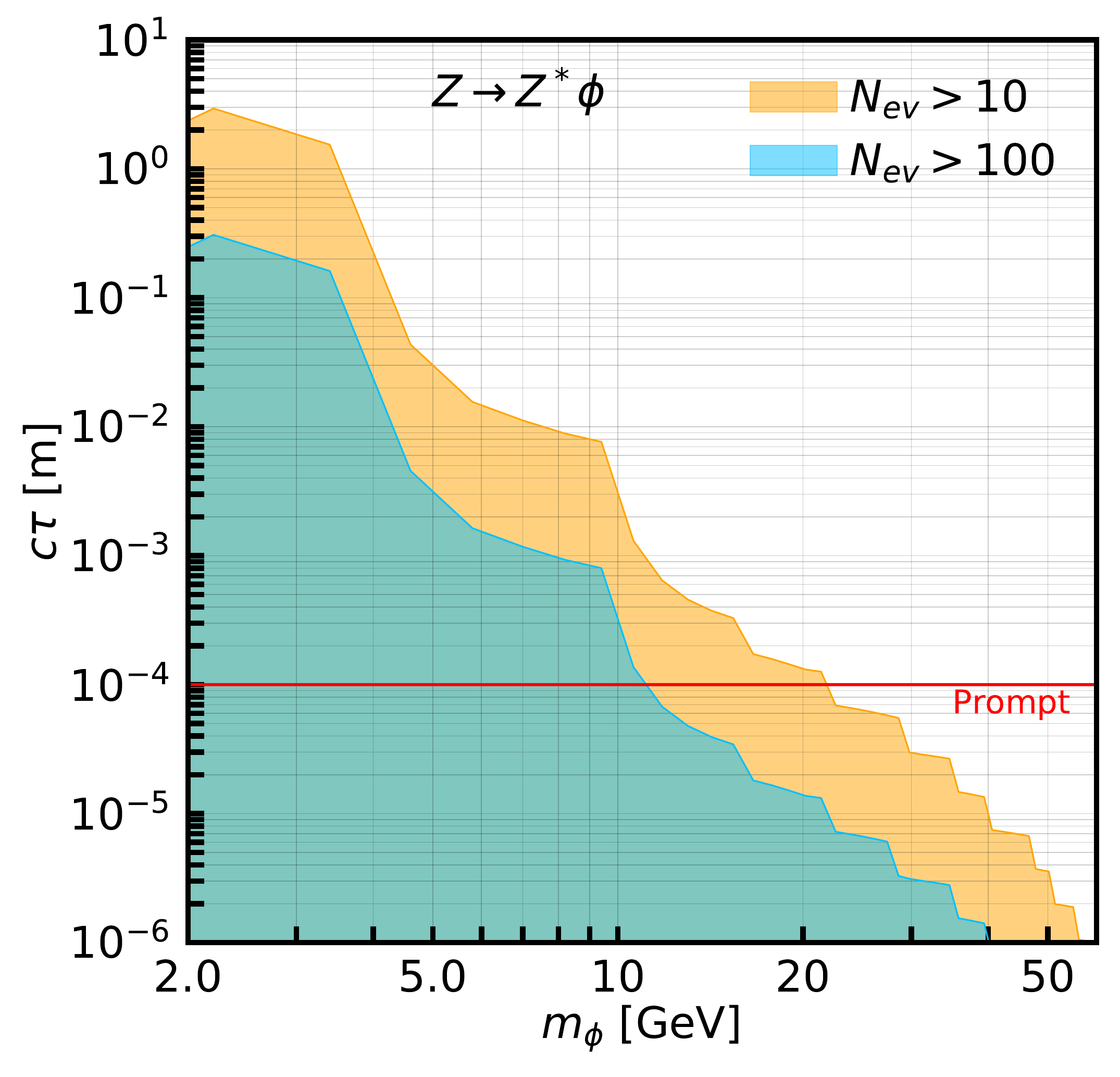}
    \caption{The dark Higgs boson mass versus decay length parameter space having at least 10 ({\it orange}) and 100 ({\it cyan}) $Z\to Z^*\phi$ events at the FCC-ee.}
    \label{fig:ztozphi}
\end{figure}

We mentioned the production of $\phi$'s from $Z$ boson 
decays in Section\,\ref{sec:model}. The partial width of the 
process $Z\rightarrow Z^*\phi$ is suppressed by the square 
of the mixing angle. However, at the $Z$ factory of FCC-ee, 
around $10^{12}$ $Z$ bosons will be produced, and it might 
be possible to observe such rare decay modes. To study the prospect of this production 
mode for $\phi$, we estimate the partial width of the 
$Z$ boson for varying masses of $\phi$ assuming 
$\sin\theta=1$ using \texttt{MadGraph5\_aMC@NLO}\,\cite{Alwall:2014hca}. We can then scale this partial decay width accordingly for any value of the mixing angle. From this, we find the branching fraction for $Z\rightarrow Z^*\phi$ and calculate the number of expected $e^+e^- \to Z^*\phi$ events at the FCC-ee running at the $Z$ pole. In Fig.\,\ref{fig:ztozphi}, we show the region of the $m_\phi-c\tau$ parameter space where we expect to have at least 10 or 100 $Z\rightarrow Z^*\phi$ events at the FCC-ee. We observe that most of this region has a prompt decay of $\phi$, which requires a different analysis strategy.

\subsection{LLPs with large Trilinear Coupling}

In this section, we study the prospects of discovering LLPs 
at FCC-ee that have a large trilinear coupling with the SM 
Higgs boson and, hence, are produced via the decay of 
the SM Higgs. The corresponding signal benchmarks are 
listed in Table\,\ref{tab:bpchoice_b}. The FCC-ee is 
expected to operate as a Higgs factory at $\sqrt{s}$ = 240\,GeV. The planned integrated luminosity achieved during this run is $\mathcal{L} = 10.8$\,ab$^{-1}$.  
We simulate this using \texttt{PYTHIA\,8}, and generate the signal via the associated production of the Higgs with 
a $Z$: $e^+e^-\to Zh,\; Z\to {\rm incl.},\;
\text{and}\;h\to \phi\phi$. Depending on the benchmark properties, 
$\phi$ decays to SM particles with the branching ratios 
listed in Table\,\ref{tab:bpchoice_b}. We observe that the 
benchmarks in this case have final states consisting of $\pi^+\pi^-$, $K^+K^-$, $c\bar{c}$, and $b\bar{b}$, depending on the mass of the LLP.

A similar study on LLPs originating from the decay of the SM 
Higgs was performed in Ref.\,\cite{Ripellino:2024tqm}. The 
analysis investigated the process $e^+e^-\to 
Zh$, followed by $Z\to \ell^+\ell^-$ and $h\to ss$, where 
$\ell^\pm$ represents a charged lepton and $s$ is 
a long-lived scalar decaying to $b\bar{b}$. 
In our study, we do not limit our analysis to the leptonic 
decay mode of the $Z$, but we consider its inclusive decay 
mode, owing to the small $Z\to \ell^+\ell^-$ branching ratio 
of $\sim$10\%.

The backgrounds from the SM that mimic the signal are the
processes $e^+e^-\to f\bar{f}$, $e^+e^-\to ZZ$, $e^+e^-\to
Zh$, and $e^+e^-\to WW$ at $\sqrt{s}=240~\text{GeV}$, where 
$f(\bar{f})$ denotes fermions (anti-fermions). Both hadronic 
and leptonic decay modes of the vector bosons are considered.
Table\,\ref{tab:ZHiggs_bkg_cs} lists the cross-section of 
these processes and the total number of such events at $\sqrt{s}$ = 240\,GeV and $\mathcal{L} = 10.8$\,ab$^{-1}$.
We generate $5\times10^8$ single $Z\to f\bar{f}$, $4\times 10^7$ $ZZ$, $5\times10^8$ $WW$, and $4\times 10^7$ $Zh$ events.
For the signal benchmarks, we simulate 20,000 events for the process $Zh$, $h\to\phi\phi$.

\begin{table}[hbt!]
    \small
    \centering  
    \begin{tabular}{|c|c|c|c|}
    \hline
    SM Backgrounds & Cross-section (pb) & Generated events & Expected events \\ \hline\hline
    $e^+e^-\to f\bar{f}$ & 21.43 & $5\times10^8$ & $2.3\times 10^8 \pm 1.03\times 10^4$\\
    $e^+e^-\to W^+W^-$ & 16.84 & $5\times10^8$ & $1.8\times 10^8 \pm 8049.84$\\
    $e^+e^-\to ZZ$ & 1.4 & $4\times 10^7$ & $1.5\times 10^7 \pm 2371.70$ \\
    $e^+e^-\to Zh$ & 0.237 & $4\times 10^7$ & $2.5\times 10^6 \pm 395.28$\\ \hline\hline
    \end{tabular}
    \caption{SM backgrounds for \textbf{Case B} with their cross-sections, generated number of events, and expected number of events,at $\sqrt{s}$ = 240 GeV, $\mathcal{L} = 10.8$\,ab$^{-1}$.}
    \label{tab:ZHiggs_bkg_cs}
    \end{table}

\subsubsection{Analysis in the Di-pion and Di-kaon Final 
States}
\label{sssec:B_pipiKK}

Among the benchmarks listed in Table\,\ref{tab:bpchoice_b}, both BPB1 and BPB2 decay into \(\pi^+\pi^-\) and \(K^+K^-\) with equal branching ratios of 50\% each. 
The strategy for selecting signal and rejecting background events involves identifying displaced charged pions and charged kaons within the detector and then reconstructing the LLPs and the Higgs boson.

\begin{figure}[htb!]
    \centering
    \includegraphics[width=0.5\textwidth]{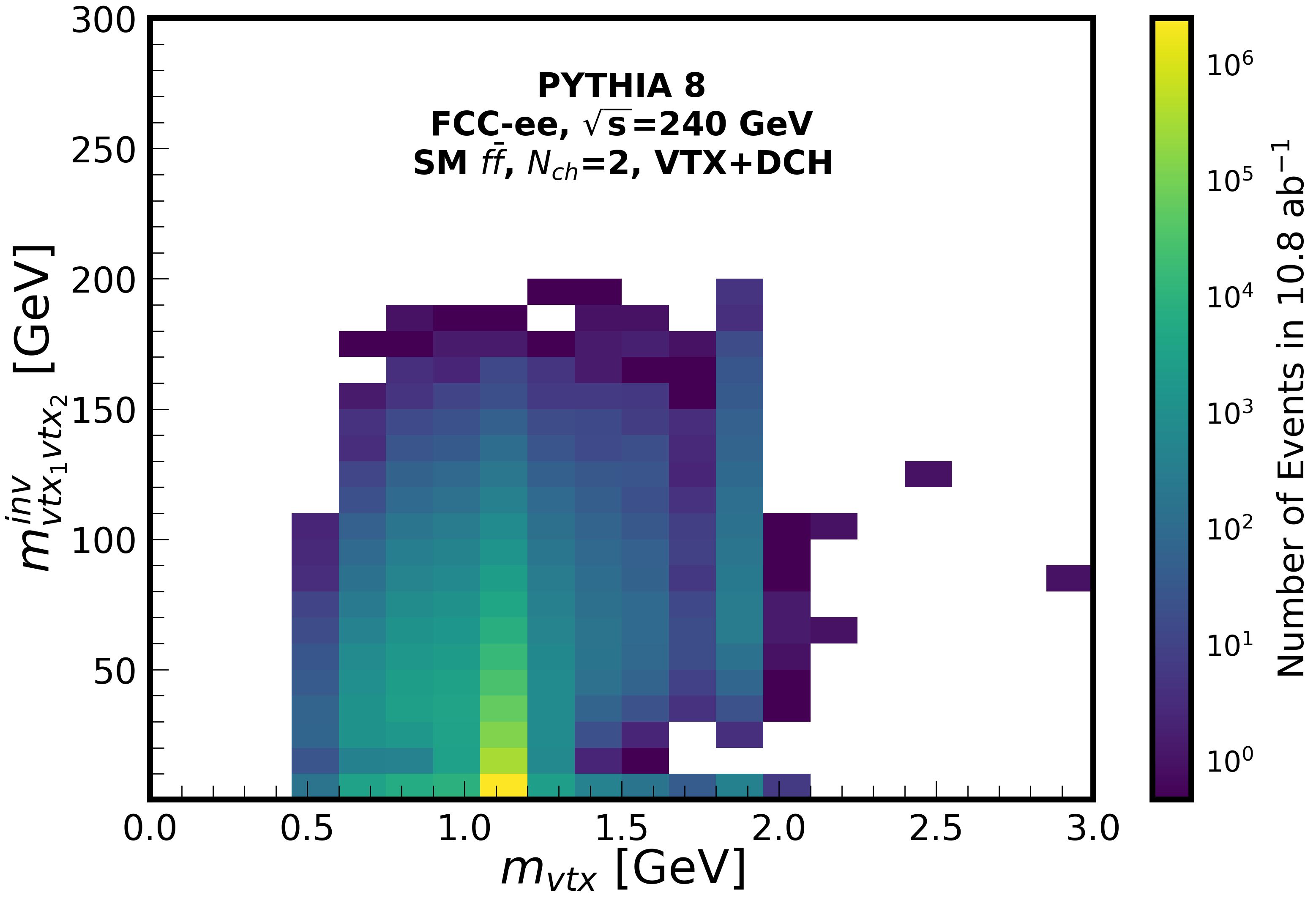}~
    \includegraphics[width=0.5\textwidth]{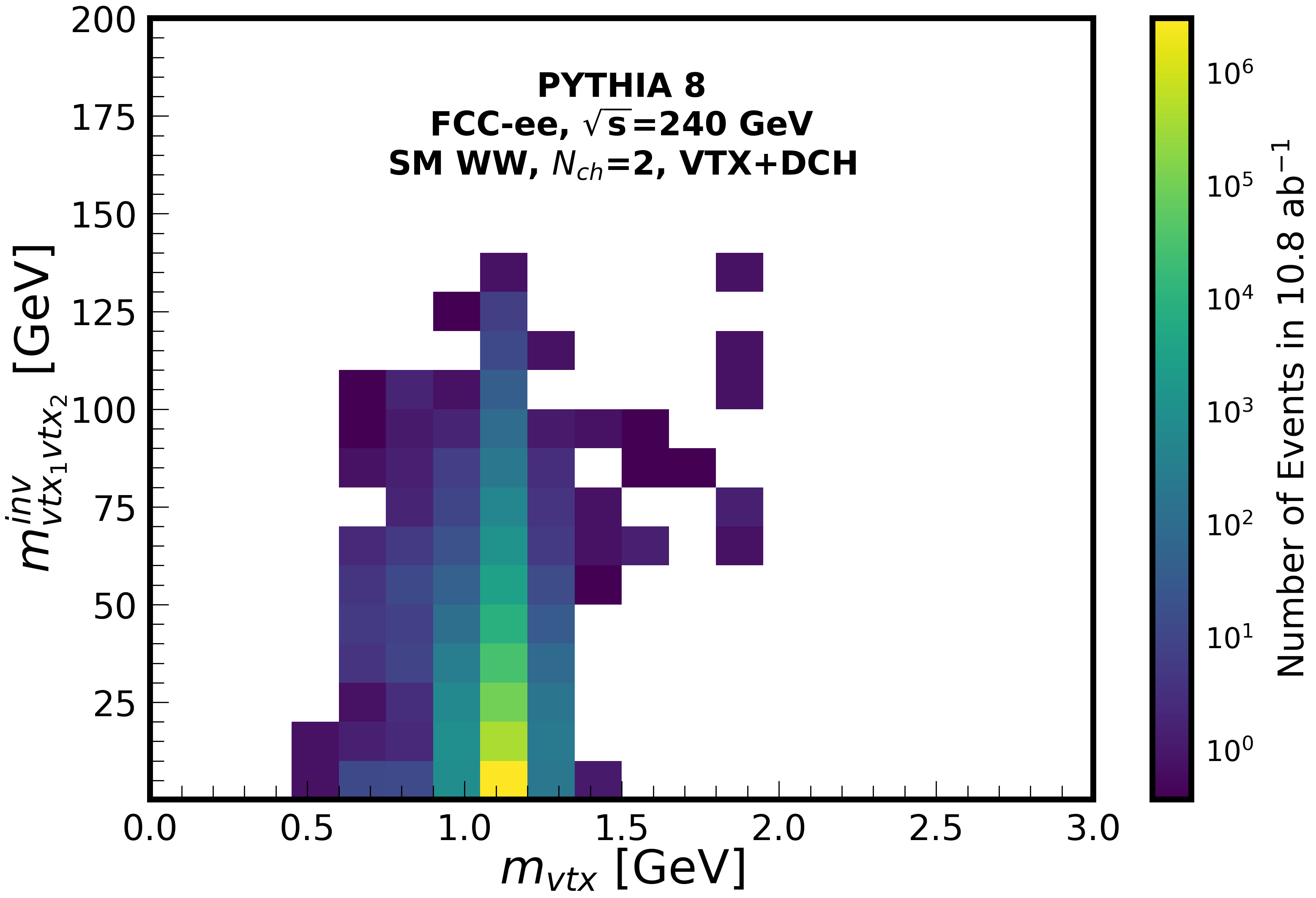}\\
    \includegraphics[width=0.5\textwidth]{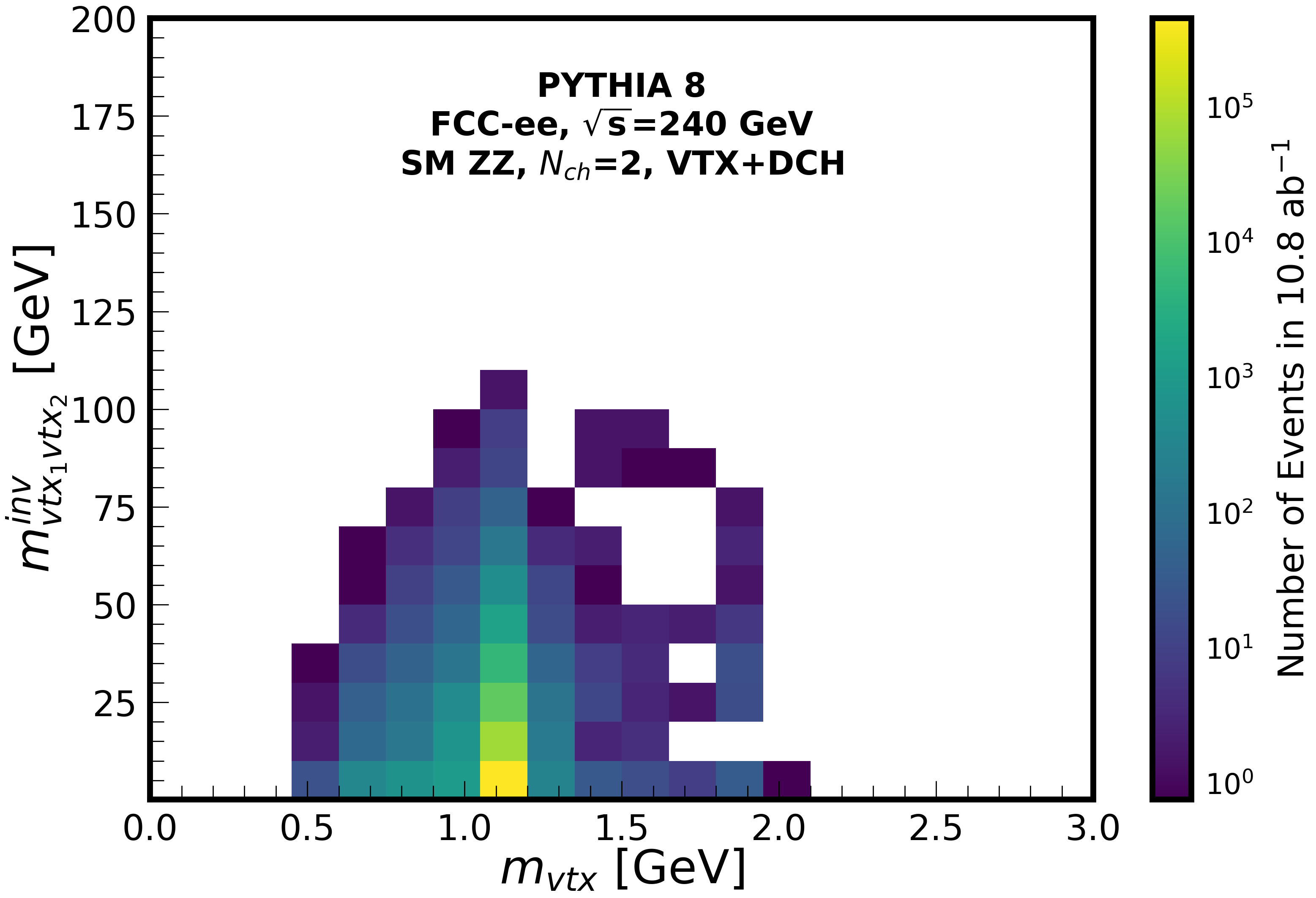}~
    \includegraphics[width=0.5\textwidth]{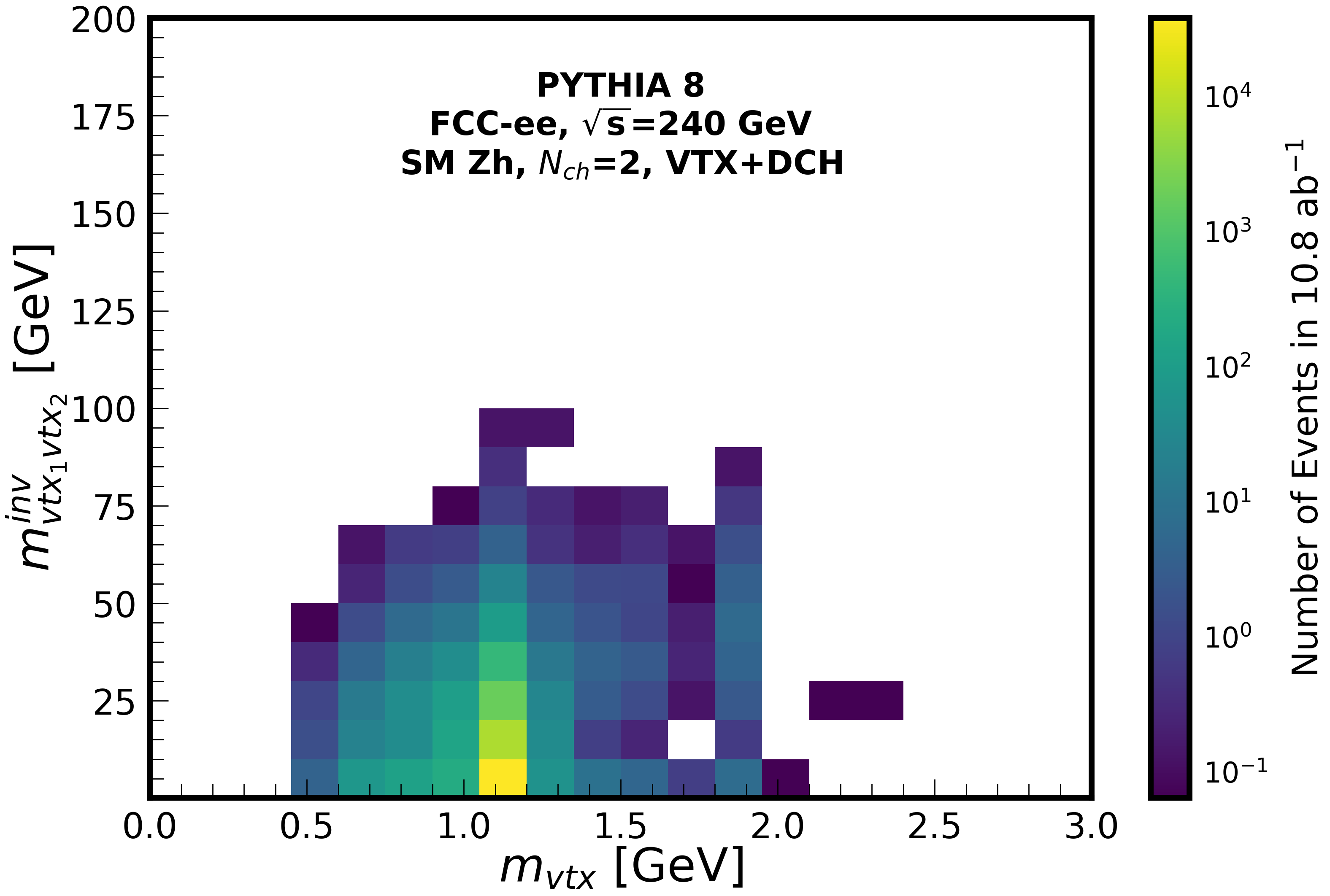}
    \caption{Distribution of the invariant mass of all the final state particles associated with the vertex with respect to the invariant mass of all the particles from two vertices in an event, for the four major backgrounds of the $e^+e^-\to Z h,h\to \phi\phi$ channel, namely ({\it top left} to {\it bottom right}) $f\bar{f},\,W^+W^-,\,ZZ,$ and $Zh$ production, respectively.} 
    \label{fig:higgs_bkg_inv}
\end{figure}

We first define the cuts a vertex and the resulting 
decay products need to satisfy to be selected. The cuts are 
similar to those in the previous analyses. The vertex must be 
detected inside the VTX or DCH. The number of charged 
particles $(N_{\rm ch}$) must be exactly equal to 2. The 
total energy of the decay particles from the vertex must 
be greater than 1\,GeV, and the $d_T$ of the vertex must be greater than 
5\,mm. If the above conditions are satisfied, we select the vertex. Now, there must be exactly two 
such vertices in the event. We compute the invariant mass of each of the 
decays vertices $(m_{{\rm vtx}_1},~m_{{\rm vtx}
_2})$. For the LLP signal, since both the displaced vertices 
come from the decay of two identical LLPs ($\phi$), we 
select events where the invariant masses of the two 
decay vertices do not differ by more than 
100\,MeV. Finally, we reduce the huge background coming from 
long-lived mesons and $K_L/K_S$ decays by selecting the 
events with invariant mass above the 
kaon threshold $(m_{{\rm vtx}_1}, ~m_{{\rm vtx}_2}) > 0.6 
~\text{GeV}$. Table\,\ref{tab:bkg_pik_higgs_cuts} summarizes 
all the selection cuts we apply to the events.

\begin{table}[hbt!]
    \centering
    \begin{tabular}{c}
    \hline
        Cuts on background in the di-pion and di-kaon final state \\ \hline\hline
        $D_{\rm vtx} \in$ VTX or DCH \\
        $N_{\rm tot}$ = 2, ~~$N_{\rm ch}$ = 2 \\
        $E > 1$\,GeV, ~~$d_T > 5$\,mm \\ 
        $N_{\rm vtx}$ = 2, ~~$m_{{\rm vtx}_{1/2}} > 0.6$\,GeV\\
        $|m_{{\rm vtx}_1} - m_{{\rm vtx}_2}| < 0.1$\,GeV \\
        \hline\hline
    \end{tabular}
    \caption{Selection cuts applied to the signal $e^+e^-\to Zh$, where $Z\to f\bar{f},
    \;\text{and}\; h\to \phi\phi$ and the SM 
    background when analysis is done in the $\pi^+\pi^-$ and $K^+K^-$ final states.}
    \label{tab:bkg_pik_higgs_cuts}
\end{table}

We reconstruct the Higgs boson by summing the four momenta of the final 
states of the two selected vertices. The invariant mass of the two should lie 
within a 3\,GeV mass window around the Higgs mass, \textit{i.e.}, $122\,\text{GeV} < m^{\rm inv}_{\rm vtx_1 vtx_2} < 128$\,GeV. The number of
signal events for BPB1 and BPB2, and background events, is listed in 
Table\,\ref{tab:BPB1_cuts}.

\begin{table}[hbt!]
\centering
\resizebox{\textwidth}{!}{\begin{tabular}{|c|c|c|c|c|c|c|}
\hline
\multicolumn{1}{|c|}{Benchmarks} & \multicolumn{2}{c|}{$m_{\phi}$ (GeV)} & \multicolumn{4}{|c|}{$c\tau$ (mm)} \\ \hline
\multicolumn{1}{|c|}{BPB1} & \multicolumn{2}{c|}{1} & \multicolumn{4}{c|}{10} \\
\multicolumn{1}{|c|}{BPB2} & \multicolumn{2}{c|}{1} & \multicolumn{4}{c|}{100} \\ \hline
\multirow{3}{*}{Cuts} & \multicolumn{6}{c|}{Number of events} \\ \cline{2-7} 
 & \multicolumn{2}{c|}{Signal} & \multicolumn{4}{c|}{Backgrounds} \\ \cline{2-7} 
 & BPB1 & BPB2 & $f\bar{f}$ & $WW$ & $ZZ$ & $Zh$ \\ \hline\hline
\multicolumn{1}{|c|}{$122 ~\text{GeV} < m^{\rm inv}_{{\rm vtx}_1 {\rm vtx}_2} < 128 ~\text{GeV}$} & 222.35 $\pm$ 1.69 & 21.27 $\pm$ 0.52 & 171.27 $\pm$ 8.90 & 1.82 $\pm$ 0.81 & 0 & 0 \\
\multicolumn{1}{|c|}{In each vertex, $N_p$ = 0} & 222.35 $\pm$ 1.69 & 21.27 $\pm$ 0.52 & 115.26 $\pm$ 7.30 & 0 & 0 & 0 \\
\multicolumn{1}{|c|}{In each vertex, $N_{\pi^\pm} =2~\text{or}~N_{K^\pm} = 2$} & 222.35 $\pm$ 1.69 & 21.27 $\pm$ 0.52 & 7.41 $\pm$ 1.85 & 0 & 0 & 0 \\
\multicolumn{1}{|c|}{$d_T > 100 ~\text{mm}$} & 143.56 $\pm$ 1.36 & 18.24 $\pm$ 0.48 & 0 & 0 & 0 & 0 \\ \hline\hline
\end{tabular}}
\caption{Number of signal and background events after applying the 
displayed cuts.}
\label{tab:BPB1_cuts}
\end{table}

The backgrounds from $e^+e^-\to f\bar{f}$ and $e^+e^-\to W^+W^-$ are 
required to satisfy the Higgs mass criterion as well. The four panels in 
Fig.\,\ref{fig:higgs_bkg_inv} show the distributions of the four types of 
background events in the $m_{\rm vtx}-m_{{\rm vtx}_1 {\rm vtx}_2}^{\rm inv}$ 
plane, respectively. The SM $f\bar{f}$  
background ({\it top 
left}) populates very high values of the 
invariant mass compared to the other three backgrounds.  
This is reflected in the results in Table\,\ref{tab:BPB1_cuts} where the $Z$ events easily satisfy the Higgs boson mass cut. The huge background at $m_{\rm vtx}$ = 1\,GeV results 
from the formation of long-lived $\Lambda$ baryons, which decay into a proton and a pion. To mitigate this effect, we apply an additional proton veto on the 
signal and backgrounds, $N_p = 0$. The background after the use of a proton veto reduces drastically. 

Until now, we have required that there are exactly two charged particles
emanating from the displaced vertex. We now 
explicitly specify that these two charged particles must be either a pair of 
pions ($\pi^+\pi^-$) or kaons ($K^+K^-$). This means that, at a given vertex, $N_{\pi^\pm}=2$ or $N_{K^\pm}=2$. It is important to note that all of the aforementioned criteria heavily depend on the accurate identification of charged particles, particularly in correctly identifying and vetoing protons. As discussed in Section~\ref{sssec:A_pipiKK}, even a very small mistag rate can lead to a large background in the $m_{\rm vtx}=1.0~\text{GeV}$ region.
As compared to the di-pion and di-kaon final states in \textbf{Case A}, the 
number of background events in this case is significantly smaller. As a result, 
the proton veto is expected to be efficient in reducing the SM background here.
Additionally, to further suppress the SM $f\bar{f}$ background, we require that the vertex must have a transverse displacement $d_T > 100$\,mm. By applying these criteria, it is 
possible to achieve a zero-background scenario. BPB1 and BPB2 still have, respectively, 143 
and 18 signal events to be observed. 

\subsubsection{Analysis in the $c\bar{c}$ and $b\bar{b}$ 
Final States}
\label{sssec:B_ccbb}

Benchmarks BPB3, BPB4, BPB5, and BPB6 decay to $c\bar{c}$ with 65\%, 65\%, 
49\%, and 49\% branching fractions, respectively, while BPB6 and BPB7 decay 
dominantly to $b\bar{b}$. To select the signal events, we require a 
decay vertex to be detected in the VTX or the DCH, similar to the 
previous analyses. The vertex must involve more than two 
charged particles in the final state. The $c\bar{c}/b\bar{b}$ partons 
hadronize and therefore, the vertices from $\phi$ decays have multiple charged particles. Their total energy must be at least 5\,GeV at 
a vertex. The displacement of such a vertex must be greater than 5\,mm from the interaction point. Vertices satisfying all the above conditions are selected.

\begin{figure}[htb!]
    \centering
    \includegraphics[width=0.5\textwidth]{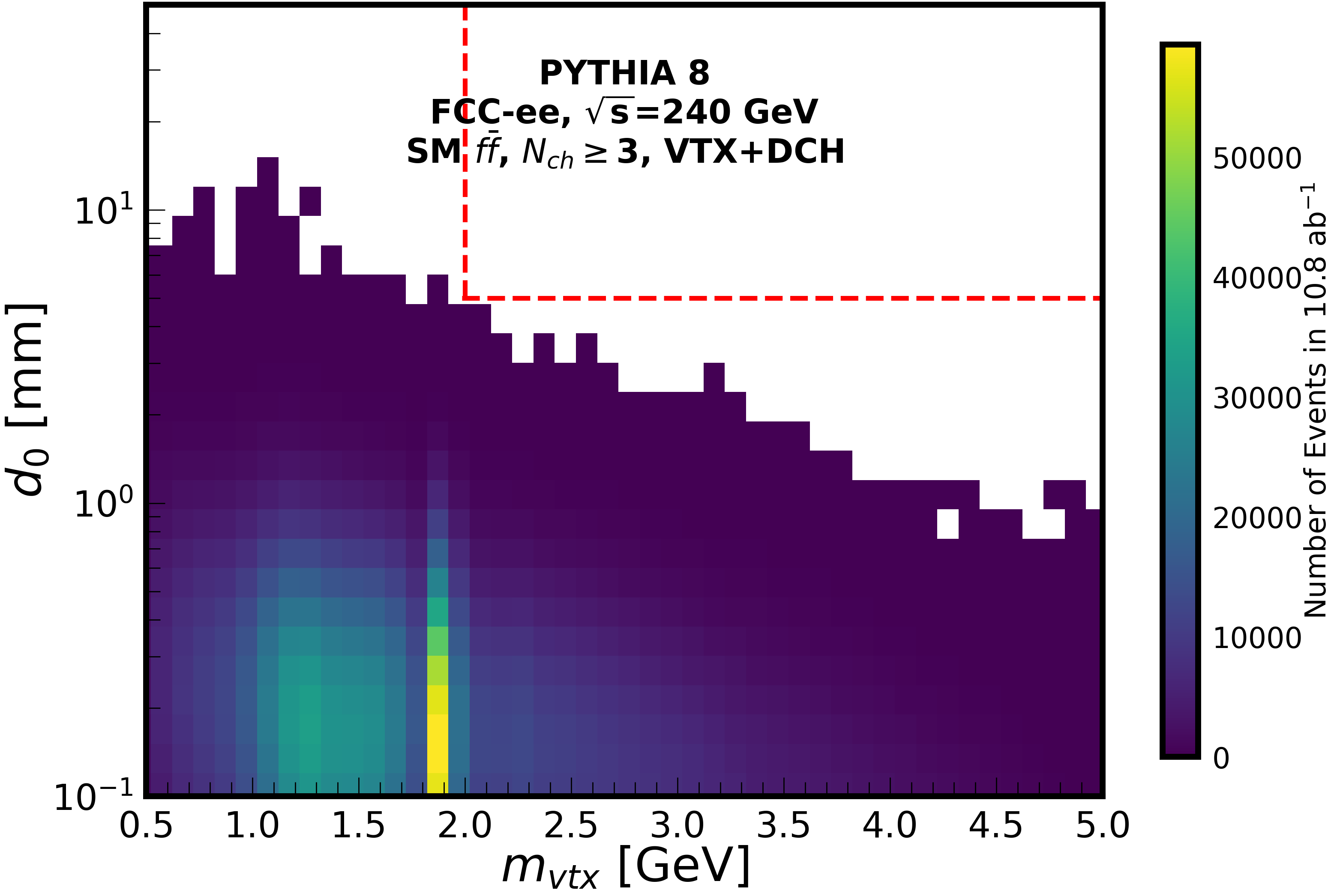}~
    \includegraphics[width=0.5\textwidth]{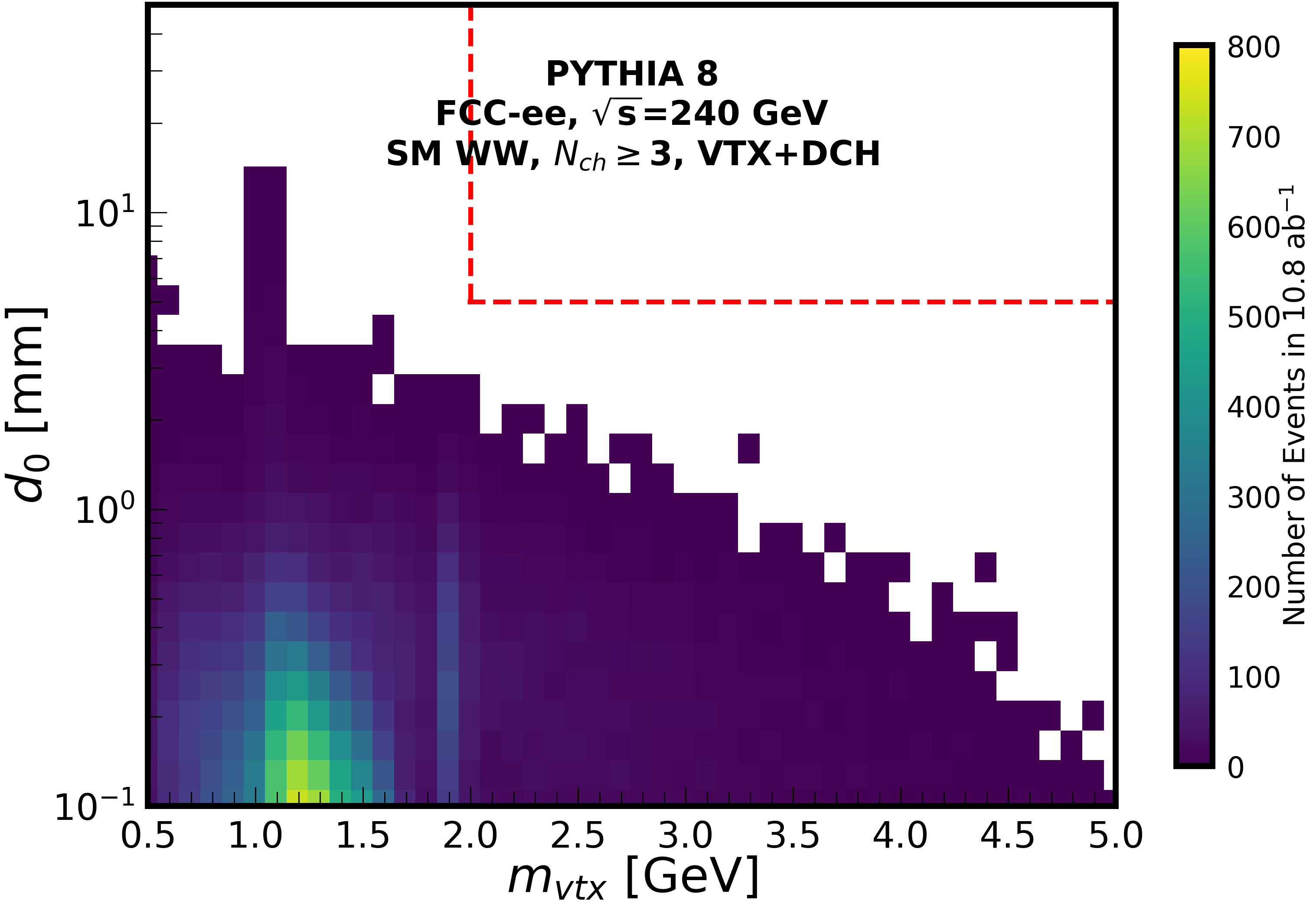}\\
    \includegraphics[width=0.5\textwidth]{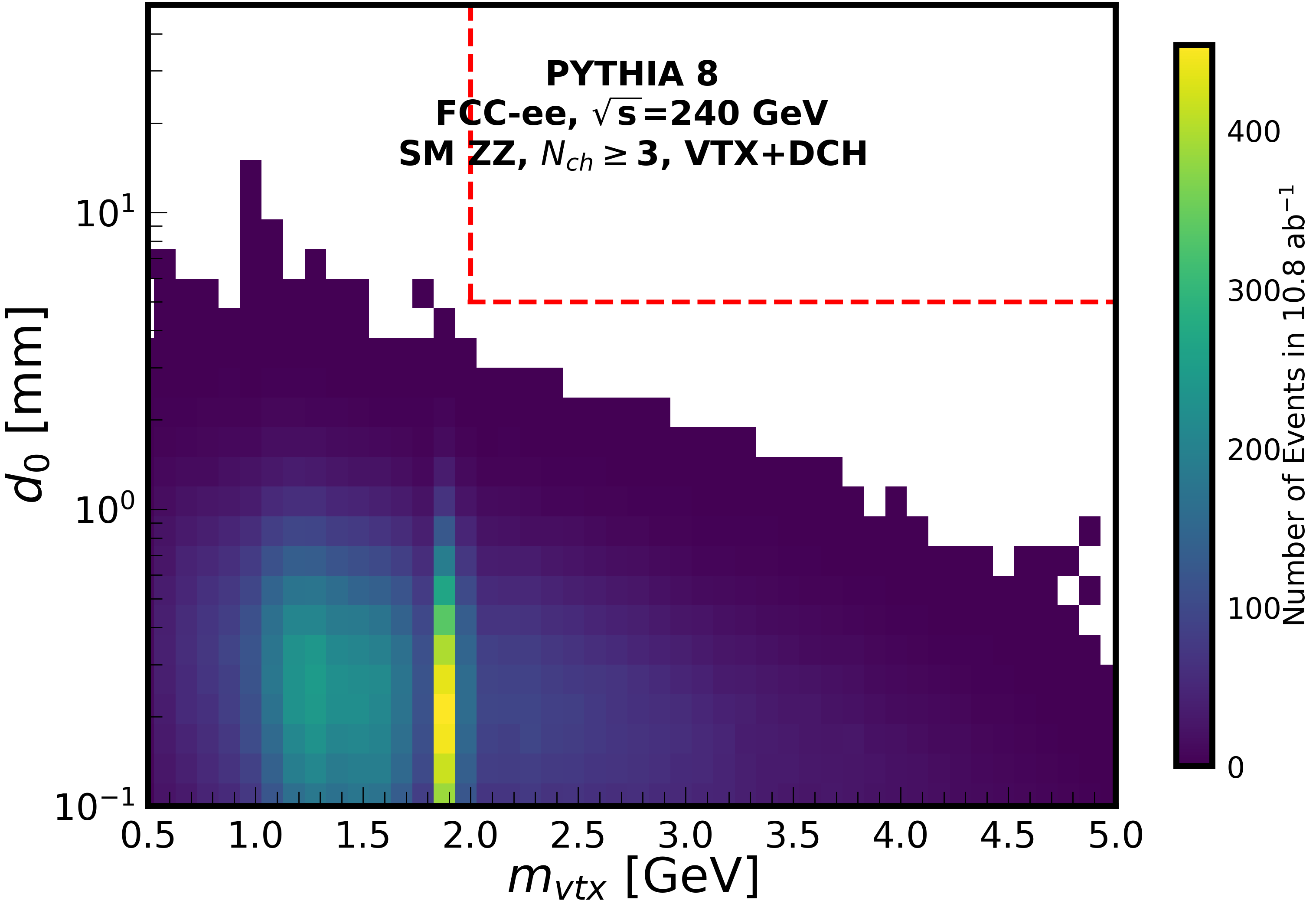}~
    \includegraphics[width=0.5\textwidth]{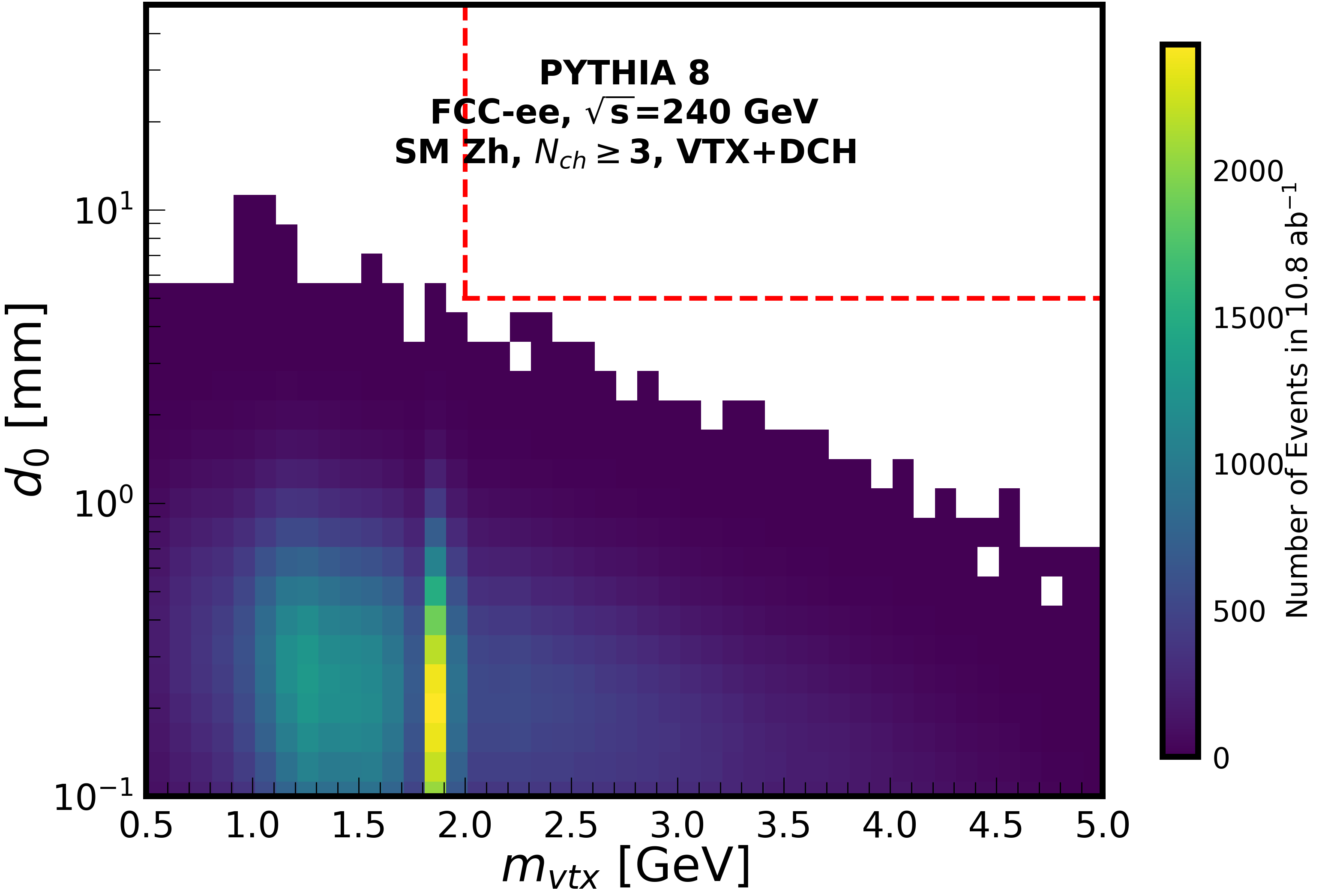}    
    \caption{Distribution of the invariant mass of the particles associated with a decay vertex with respect to the impact parameter of the vertex, $d_0$, for the four major backgrounds of the $e^+e^-\to Z h,h\to \phi\phi$ channel. 
    The {\it red dashed lines} mark the region in the $m_{\rm vtx}-d_0$ plane which is free from the SM background, motivating the cuts on these variables.
    }
    \label{fig:higgs_bkg}
\end{figure}

After applying these selection cuts on the vertices, we plot the distribution of the vertices from the background events in the $m_{\rm vtx}$ and $d_0$ plane.
Fig.\,\ref{fig:higgs_bkg} shows the two-dimensional distribution of the SM backgrounds: $f\bar{f}$, $WW$, $ZZ$, and $Zh$. 
Since there are two LLPs, we demand that there must be exactly two such vertices in the event. 
Table\,\ref{tab:bkg_ccbb_higgs_cuts} summarizes these selection cuts.

\begin{table}[hbt!]
    \centering
    \begin{tabular}{c}
    \hline
        Cuts on background in the $c\bar{c}$ and $b\bar{b}$ final states \\ \hline\hline
        $D_{\rm vtx} \in$ VTX or DCH \\
        $N_{\rm ch} \ge 3$ \\
        $E > 5$\,GeV \\
        $d_T > 5$\,mm \\ 
        $N_{\rm vtx} = 2$\\
        \hline\hline
    \end{tabular}
    \caption{Selection cuts applied on the signal $e^+e^-\to Zh, Z\to incl., h\to \phi\phi$ and the SM background when analysis is done in the $c\bar{c}$ and $b\bar{b}$ final states.}
    \label{tab:bkg_ccbb_higgs_cuts}
\end{table}

For events having two displaced vertices, we further suppress the backgrounds using two analysis methods, as described below.
 
{\begin{enumerate}
    \item \textit{Criterion 1:} Require one of the two vertices in the event to satisfy a strict cut on $m_{\rm vtx}$ and $d_0$: 
    \begin{equation}
        m_{\rm vtx} > 2 ~\text{GeV},\; d_0 > 5 ~\text{mm}    
    \end{equation}
    \item \textit{Criterion 2:} Require both the vertices in the event to satisfy a weaker cut on $d_0$: 
    \begin{equation}
        m_{\rm vtx} > 2 ~\text{GeV}, \; d_0 > 2 ~\text{mm}    
    \end{equation}   
\end{enumerate}
\captionof{table}{List describing the two criteria for the $c\bar{c}$ and $b\bar{b}$ analyses in {\bf Case B}.}
\label{tab:cc_higgs_cuts}}

\medskip

Both criteria result in a null background. This can be seen in Fig.\,\ref{fig:higgs_bkg}. The {\it red dashed lines} indicate the cuts on $m_{\rm vtx}$ and $d_0$ according to \textit{Criterion 1}. In the case of the background events, the final state particles from the vertices do not have an invariant mass above the 
$c\bar{c}$ threshold, and the $d_0$ of the vertices do not exceed $5$\,mm. There are events whose vertex has $d_0$ greater than 
2\,mm; however, no event has two such vertices. The number of signal 
events vary depending on the criteria. They are listed for each benchmark in Table\,\ref{tab:BPB3B7_cuts}.
For BPB7 and BPB8, where the mass
of the LLP is 40\,GeV, the momentum boost for such a particle is less.
Therefore, we observe that the number of displaced events increases for higher $c\tau$, as it helps the LLP satisfy the displacement criteria.

\begin{table}[hbt!]
\centering
\begin{tabular}{|c|c|c|c|c|}
\hline
\multirow{2}{*}{Benchmarks} & \multirow{2}{*}{$m_{\phi}$ (GeV)} & \multirow{2}{*}{$c\tau$ (mm)} & \multicolumn{2}{c|}{Number of signal events} \\ \cline{4-5} 
 &  &  & {\it Criterion 1} & {\it Criterion 2} \\ \hline\hline
BPB3 & 4.4 & 10 & 32.44 $\pm$ 4.19 & 2.70 $\pm$ 1.21 \\
BPB4 & 4.4 & 100 & 8.08 $\pm$ 0.47 & 1.19 $\pm$ 0.18 \\
BPB5 & 6.0 & 100 & 71.38 $\pm$ 1.48 & 18.41 $\pm$ 0.75 \\
BPB6 & 6.0 & 1000 & 0.87 $\pm$ 0.07 & 0.20 $\pm$ 0.04 \\
BPB7 & 40 & 10  & 562.27 $\pm$ 10.80 & 320.53 $\pm$ 8.15 \\
BPB8 & 40 & 100 & 1747.05 $\pm$ 13.46  & 887.15 $\pm$ 9.59  \\ \hline\hline
\end{tabular}
\caption{The number of signal events for each benchmark after applying the selection cuts of Table\,\ref{tab:bkg_ccbb_higgs_cuts} and the cuts in \textit{Criterion 1} and \textit{Criterion 2}. The number of background events is zero in both cases.}
\label{tab:BPB3B7_cuts}
\end{table}

Taking a more conservative approach, we select a region that lies farther from the background-populated region in Fig.\,\ref{fig:higgs_bkg}. We tighten \textit{Criterion 1} by imposing a stricter cut of $d_0>10~\text{mm}$. This results in signal yields of $6.49\pm1.87$, $4.76\pm0.36$, $48.49\pm1.22$, $0.64\pm0.06$, $213.96\pm6.66$, and $1393.24\pm12.02$ for benchmarks BPB3, BPB4, BPB5, BPB6, BPB7, and BPB8 respectively. For a majority of benchmark scenarios, the more conservative selection retains sufficient number of signal events, while maintaining zero background.

We note that for a lighter mass of the LLP, the higher $c\tau$ results in 
the loss of events because the search is in the VTX and/or DCH, so with a 
higher boost and longer lifetime, the particles decay after passing the 
VTX/DCH and thus escape detection. BPB6, especially, has less than 1 
remaining signal event. In that case, it shall be critical to search for 
events from such benchmarks in the muon spectrometer. An MS analysis 
would entail selection cuts similar to those listed in 
Table\,\ref{tab:bkg_ccbb_higgs_cuts}, except for the fact that the decay 
vertex is detected in the muon spectrometer ($D_{\rm vtx}\in~\text{MS}$) and 
$N_{\rm vtx}\geq1$. The last condition indicates that an event is selected even 
if only one of the two $c\bar{c}$ clusters is detected in the MS. With 
this analysis, we find that BPB6 results in 9 events, while the background 
remains zero. It is also important to note that the absence of any events in 
the VTX+DCH analysis for BPB6 is a direct consequence of the $N_{\rm vtx}
=2$ condition. To demand that both of the light LLPs with decay lengths of 1\,m 
decay within the VTX and DCH components is excessively restrictive. We weaken 
the $N_{\rm vtx}$ cut and instead impose stricter cuts on $d_T$ and 
$d_0$: $N_{\rm vtx}\geq 1,~d_T>150~ \text{mm},~d_0>10~ \text{mm},~m_{\rm vtx}
>2~\text{GeV}$. With these cuts, the VTX+DCH analysis yields 7 events for BPB6. 

\begin{figure}[htb!]
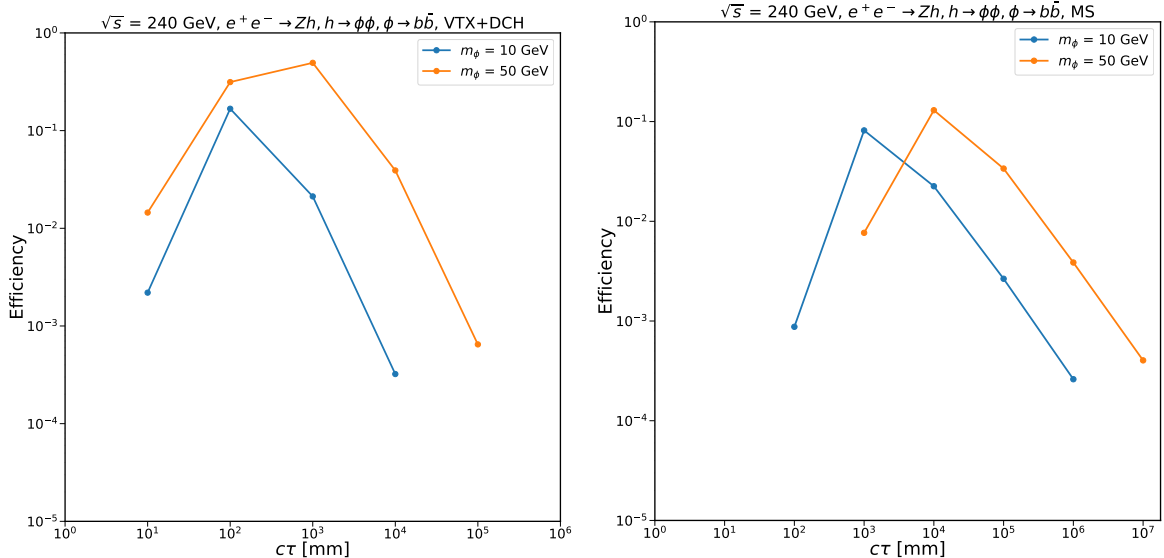

    \centering
    \includegraphics[width=0.5\textwidth]{Eff_Higgs_bb_plot.pdf}~
    \includegraphics[width=0.5\textwidth]{Eff_Higgs_bb_MS_plot.pdf}
    \caption{Efficiency plot of the $b\bar{b}$ decay mode of LLP $\phi$, in the $m_{\phi}$-$c\tau$ plane. The generation of LLP involved the $h\to \phi\phi$ process, where $\phi$ decays to $b\bar{b}$. The final state charged particles were detected within VTX and DCH \textit{(left)}, and in the Muon System (MS) \textit{(right)}.}
    \label{fig:Higgs_bb_eff}
\end{figure}

Fig.\,\ref{fig:Higgs_bb_eff} shows the signal efficiency for the two analyses: 
the VTX+DCH analysis and the MS analysis. Both use the same cuts on $E$ and 
$N_{\rm ch}$. We observe that the MS analysis can probe LLPs with much 
higher $d_T$.

\subsubsection{Summary for {\bf Case B}}

In the case of large trilinear coupling, 
we expect to observe at least $\sim 100$ and $\sim 10$ signal events, respectively, for benchmarks BPB1 and BPB2 in the pion/kaon decay modes within the VTX and DCH detectors.
For the rest of the benchmarks except BPB6, the VTX+DCH analysis yields around 10-1000 signal events, depending on the benchmark, when the backgrounds are suppressed to zero.

\begin{figure}[hbt!]
    \centering
    \includegraphics[width=0.5\textwidth]{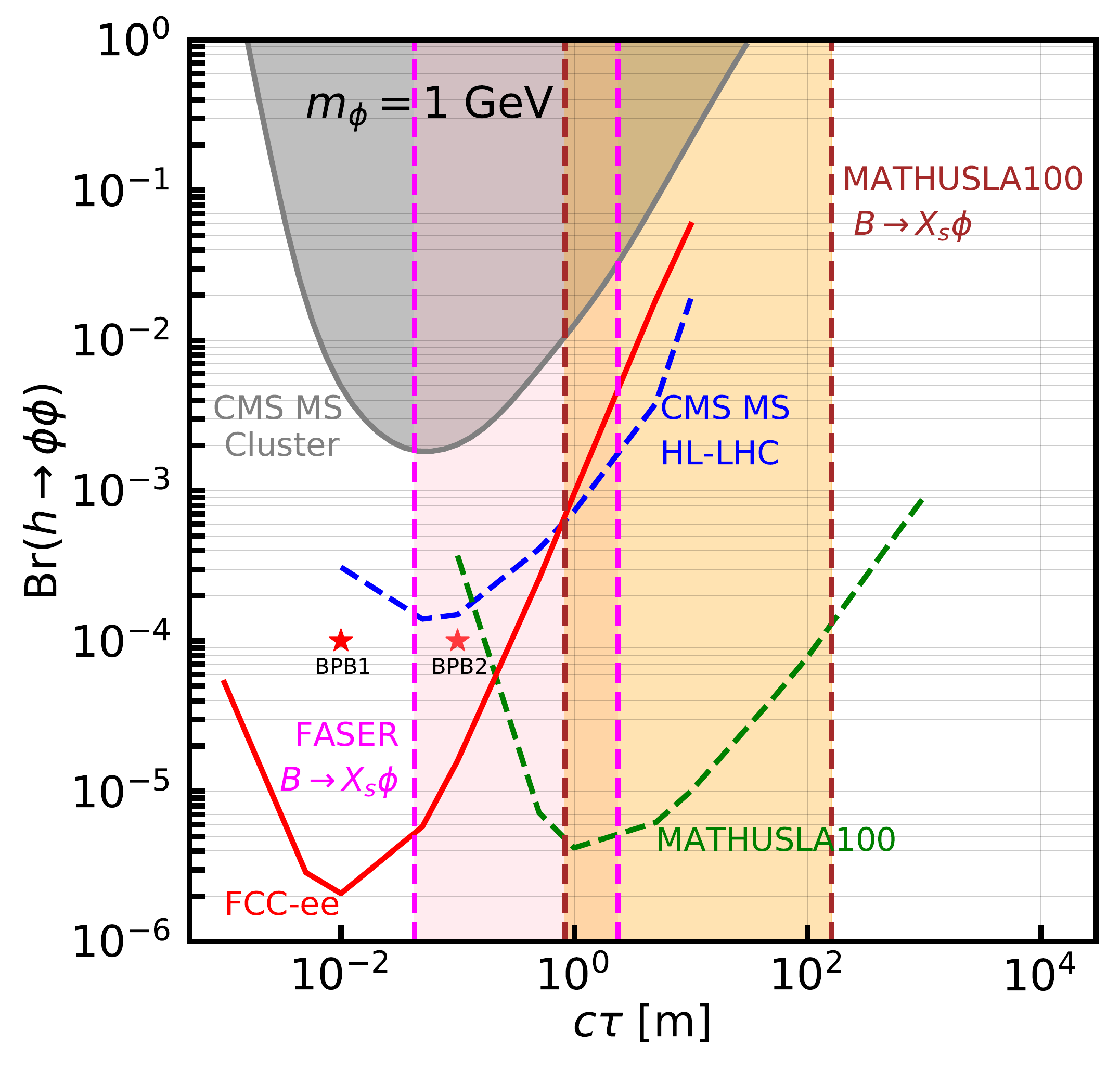}~
    \includegraphics[width=0.5\textwidth]{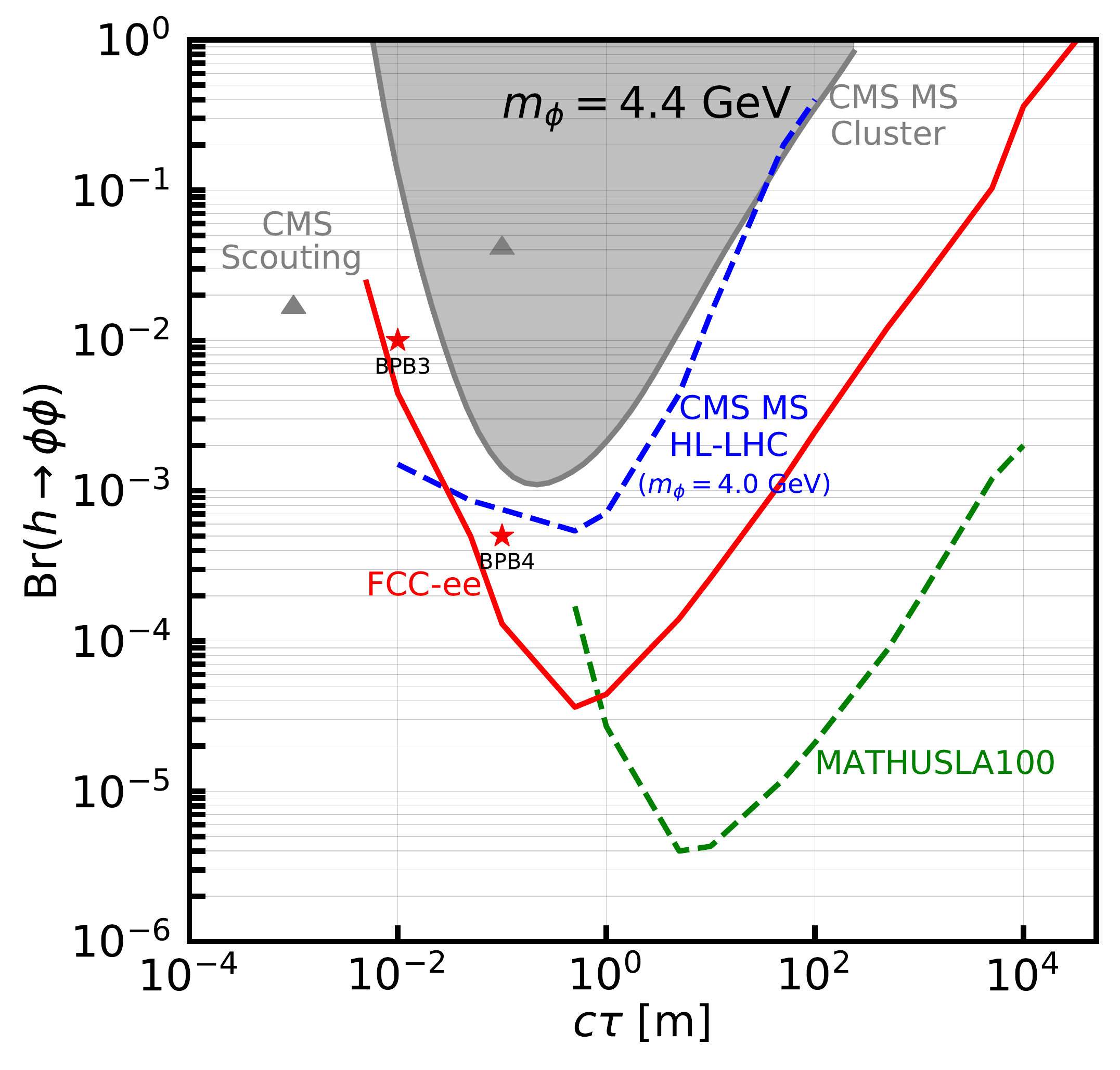}\\
    \includegraphics[width=0.5\textwidth]{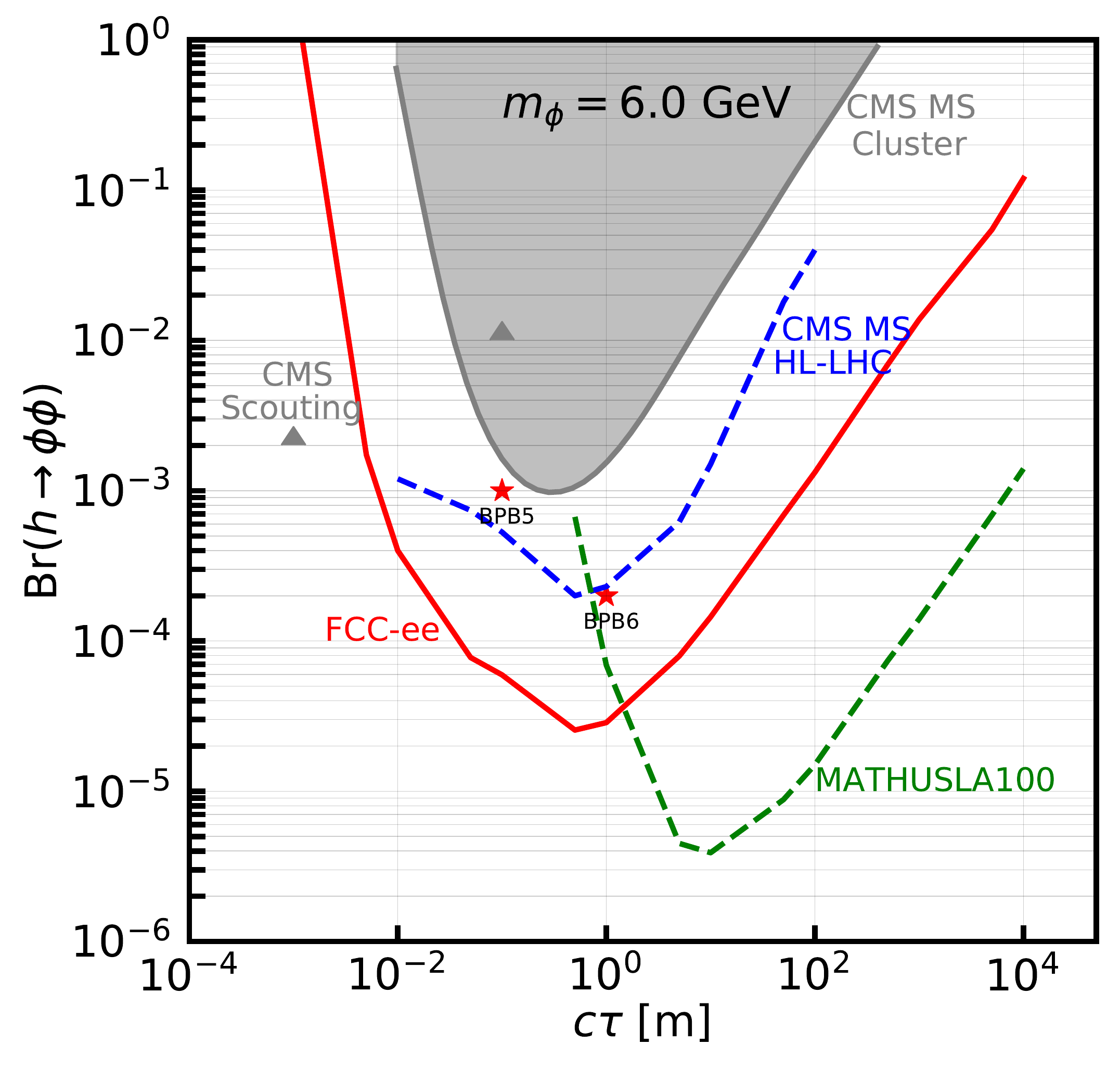}~
    \includegraphics[width=0.5\textwidth]{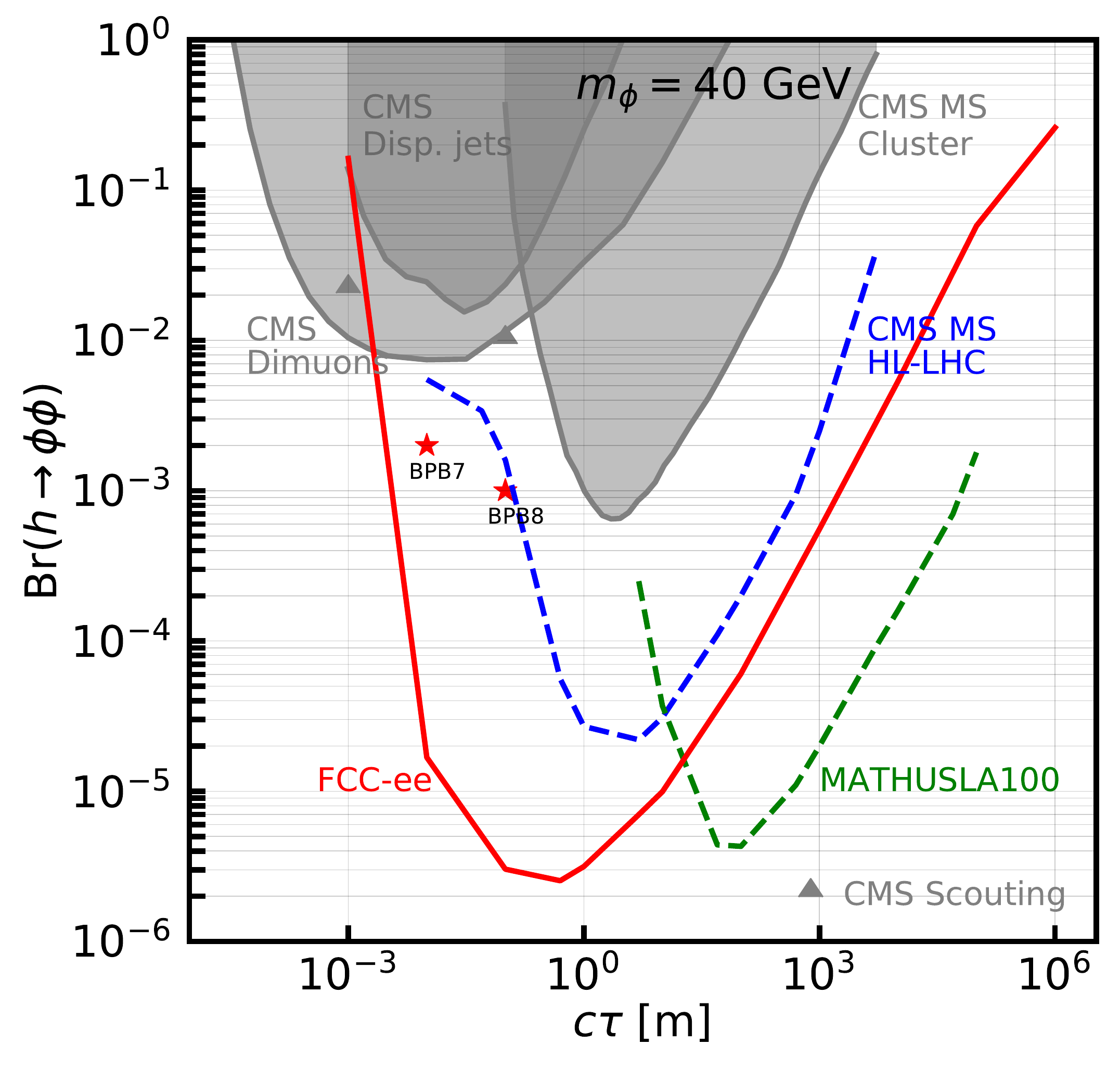}
    \caption{The FCC-ee sensitivity from our analyses with 3 observed events along with the current bounds and projected sensitivities as shown in Fig.\,\ref{fig:benchmarks_htophiphi} for {\bf Case B}.
    }
\label{fig:FCC_ee_htophiphi}
\end{figure}

Fig.~\ref{fig:FCC_ee_htophiphi} shows the upper limits on Br$(h\to\phi\phi)$ when 3 events are observed in the FCC-ee, combining the VTX+DCH and MS analysis in the di-pion, di-kaon, $c\bar{c}$, and $b\bar{b}$ decay modes. FCC-ee can probe Br$(h\to\phi\phi)$ as low as $\sim 2\times10^{-6}$ for an LLP of $m_{\phi}=1~\text{GeV}$ and $c\tau=10~\text{mm}$. In all cases, the IDEA detector of FCC-ee improves the sensitivity, at the most sensitive $c\tau$ value, by at least an order of magnitude as compared to the HL-LHC projection.
While the sensitivity of MATHUSLA seems to be comparable or better than the FCC-ee projection, we should keep in mind that this sensitivity is for the MATHUSLA100 detector. The projection weakens to Br$(h\to\phi\phi)\sim 10^{-4}$ for the MATHUSLA40 design as discussed in Ref.\,\cite{MATHUSLA:2025zyt}.

\begin{table}[hbt!]
    \centering
    \resizebox{\textwidth}{!}{
    \begin{tabular}{|c|c|c|c|c|}
    \hline
     & Final State & Backgrounds & Selection cuts & Section \\
    \hline\hline
    \multirow{3}{*}{{\bf Case A}} & $\mu^+\mu^-$ & $J/\psi$, $\psi(2S)$ & Table\,\ref{tab:sig_mumu_cuts} & \ref{sssec:A_mumu} \\\cline{2-5}
     & $\pi^+\pi^-$/$K^+K^-$ & $K_S$, $K_L$, $D^0$, $\Lambda$ & Table\,\ref{tab:sig_pik_cuts} & \ref{sssec:A_pipiKK} \\\cline{2-5}
     & $c\bar{c}$ & SM displaced vertices with $N_{\rm ch}>2$ & Table\,\ref{tab:BPA7_cut} & \ref{sssec:A_ccbar} \\
    \hline
    \multirow{2}{*}{{\bf Case B}} 
     & $\pi^+\pi^-$/$K^+K^-$ & $f\bar{f}$, $WW$, & Table\,\ref{tab:BPB1_cuts} & \ref{sssec:B_pipiKK} \\\cline{2-2}\cline{4-5}
     & $c\bar{c}$/$b\bar{b}$ & $ZZ$, $Zh$ & Tables\,\ref{tab:bkg_ccbb_higgs_cuts} and \ref{tab:cc_higgs_cuts} & \ref{sssec:B_ccbb} \\
    \hline\hline
    \end{tabular}}
    \caption{Summary of the final states, possible backgrounds, references to the selection cuts in the analyses performed and the corresponding sections where they are discussed.}
    \label{tab:summary}
\end{table}

In Table\,\ref{tab:summary}, we present a summary of all final states considered for {\bf Case A} and {\bf Case B}, including the relevant backgrounds, references to the tables listing the selection cuts, and the sections where they are discussed, for easy navigation.

\medskip

\noindent \textbf{Use of variable efficiencies}
\medskip

Our analyses involved the use of fixed efficiencies of the detector components. We now modify the efficiencies as a function of the radial distance from the IP
of the detector component and the displacement $d_T$ of the particles. The fixed values of efficiencies are listed in Table\,\ref{tab:fixed_eff}. 
For the VTX detector, the same efficiency is maintained throughout due to a large number of hits which improve the reconstruction. For the DCH, DRC, 
and MS, the efficiency values from Table\,\ref{tab:fixed_eff} are multiplied by the function $(R_{out}-d_T)/(R_{out}-R_{in})$, where $R_{in}$ and $R_{out}$ are the inner and outer radius of the detector element, as listed in Table\,\ref{tab:IDEA_det}. Thus, when $d_T = R_{in}$ \textit{i.e.}, the decay 
particle is produced at the inner edge of the detector, the probability 
of the particle being detected is the maximum, and the probability falls as the particle traverses the length of the detector component, and becomes completely zero at the outer edge of the detector when $d_T = R_{out}$.

With this setup, the di-pion analysis of the BPA1 benchmark results in 2 events in the muon system. The signal efficiency for BPA2 in the MS analysis of di-pion and di-kaon final states drops from 6.4\% to 4.1\%. There is a 25\% reduction in the signal efficiency for BPA3 in the MS analysis and a 20\% reduction for BPA6 in the VTX+DCH analysis of the di-pion final state. For BPB1, we observed 143 events after the cut-based analysis. With the modified detector efficiencies, we now observe 80 events.

Our analyses for the displaced decays of the dark Higgs boson in the IDEA detector for both \textbf{Case A} and \textbf{Case B} show promising results for most benchmarks. 
However, there are still a few benchmarks, such as BPA1, BPA5, and BPB6, for which we expect very few events to be detected in the IDEA detector after applying cuts to reduce backgrounds. 
Additionally, if the detector efficiency of various components of the IDEA detector decreases linearly with increasing displacement of the decay, the expected signal yield will be affected, as discussed above. 
In such scenarios, dedicated LLP detectors surrounding the FCC-ee collider complex provide a complementary approach. 
In the next section, we propose various dedicated detector geometries and examine their sensitivity to the dark Higgs boson.



\section{Dedicated Detectors at $e^-e^+$ Colliders}
\label{sec:dedicated}

We have discussed the detection prospects of the LLPs coming from both $B$-
mesons and Higgs boson decays at various parts of the proposed IDEA detector. 
However, for some of the benchmarks in both cases, most of the LLP decays 
escape the IDEA detector as seen from Tables\,\ref{tab:decay_events_a} and 
\ref{tab:decay_events_b}. This happens mostly for lighter LLPs due to a higher 
Lorentz boost factor and/or LLPs having larger decay 
lengths. We find that benchmarks BPA1 and BPA5 from \textbf{Case A}, and BPB6 
from \textbf{Case B} are expected to have less than 5 observable events 
after the analyses in the VTX+DCH or the MS of the IDEA detector. An additional 
approach for these benchmarks would be to employ dedicated LLP detectors 
positioned outside the main IDEA detector. There are already proposals of 
different dedicated detectors for lepton colliders, like the HErmetic Cavern TrackEr (HECATE)~\cite{Chrzaszcz:2020emg}, near and far detectors~\cite{Wang:2019xvx,Schafer:2022shi}, and the LAYered CAvern Surface Tracker (LAYCAST)~\cite{Lu:2024fxs} for FCC-ee, ILC and CEPC, specifically designed to search for LLPs.

The authors of HECATE have proposed locating the detector along the inside of the cavern walls, forming a 4$\pi$ detector. Using 
the example of heavy neutral leptons (HNL), they showed that such a detector would enhance the sensitivity to the 
squared mixing parameter by almost half an order of magnitude compared to the 
FCC-ee main detector. The far detectors proposed in Ref.\,\cite{Wang:2019xvx} 
should study Higgs decays 
into a pair of long-lived light scalars at 240\,GeV, as well as $Z$-boson 
decays into either a long-lived HNL and an active neutrino or a pair of long-
lived neutralinos at the $Z$-pole. The sensitivity of this detector 
results in a modest improvement for LLPs with high decay 
lengths if its dimensions are as large as the MATHUSLA 
detector. The LAYCAST detector is proposed to be installed on the ceiling and 
the wall of the cavern of future electron-positron colliders such as CEPC and FCC-ee. 
This detector has a cuboid shape with a size of 40\,m $\times$ 20\,m $\times$ 
30\,m, where three physics models are tested: $h\to \phi\phi$, HNLs, and 
neutralinos. 
Compared to the far detectors, LAYCAST is observed to have a better sensitivity for lighter LLPs with smaller $c\tau$ values, and larger LLP masses in the long-lifetime regime, being closer to the IP.
For the HNL case, its sensitivity is similar to that of HECATE.

For the future colliders, the biggest advantage is the possibility of 
optimizing the position and configuration of the dedicated LLP detectors for 
maximizing the sensitivity to a particular model. We 
performed such an optimization for light long-lived dark Higgs bosons at the 
FCC-hh in order to propose transverse detectors, 
called DELIGHT\,\cite{Bhattacherjee:2021rml}, and forward detectors, FOREHUNT\,\cite{Bhattacherjee:2023plj}.
While optimizing the position of the transverse and forward detectors, we respectively considered the processes $h\to\phi\phi$ and $B\to X_s\phi$. 
In this section, we study various dedicated detector configurations for the FCC-ee to determine which configuration is feasible in size and optimized for detecting LLPs.
These detectors should be capable of capturing LLPs originating from both $B$-meson decays and Higgs boson decays. 

\begin{figure}[h!]
    \centering
    \includegraphics[scale=0.45]{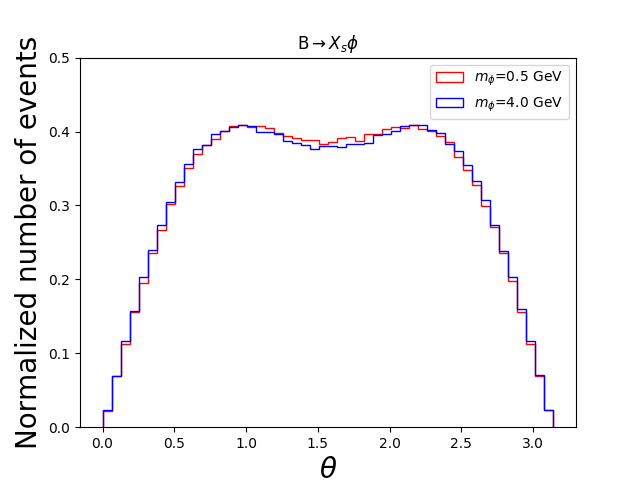} 
     \includegraphics[scale=0.45]{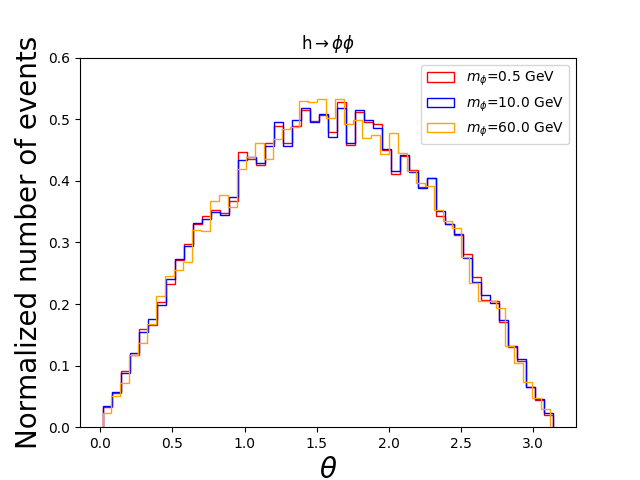} \\
\caption{Normalized distributions of the polar angle ($\theta$) of the LLPs coming from the decays of $B$-mesons ({\it left}) and the Higgs boson ({\it right}).}
    \label{fig:dedicated_dist}
    \end{figure}

The geometry and placement of these detectors are chosen based on the production direction of the LLP 
being considered. Before 
going into detail about the geometry of the detector design, we first investigate the 
optimal placement of the detector. To get an idea of the directionality of the LLP, we plot the normalized $\theta$ (angle between the direction of the LLP 
with the $z$-axis, which is the beam axis) distribution for the LLP in Fig.\,\ref{fig:dedicated_dist}.
The {\it left} panel of Fig.\,\ref{fig:dedicated_dist} presents the $\theta$ 
distribution for LLPs with masses of 0.5\,GeV and 4.0\,GeV produced from $B$-meson decays. We observe that the LLPs can move in any direction from $[0,\pi]$ 
range, having a peak at $\frac{\pi}{4}$ and $\frac{3\pi}{4}$. 
The situation is different for the LLPs coming from the Higgs boson decay, as depicted in the {\it right} panel of Fig.\,\ref{fig:dedicated_dist}, where most of the LLPs move 
centrally and peak at $\frac{\pi}{2}$\,\footnote{Since the Higgs boson is produced primarily at rest and the decays are spherically isotropic, the distribution in the $\theta$ direction is proportional to $\sin{\theta}$.}.  
This indicates that LLPs from Higgs decays are predominantly emitted perpendicular to the beam axis. By comparing the direction of the LLPs, we should keep in mind that our placement of the detector should maximize its coverage for both decay scenarios.
We deduce that a detector covering the central $\eta$ region between 0.5 and 2.5 would be optimal for both cases.

The final key consideration for the placement of the dedicated detectors is the design of the collider complex.
The FCC-ee collider is planned to be designed as a quasi-circular tunnel with a circumference of 97.75\,km and an internal diameter of 5.5\,m~\cite{FCC:2018evy}. 
The proposed tunnel layout includes approximately 8\,km of bypass tunnels, 18 shafts, 12 large caverns, and 12 new surface sites. 
The colliding electron and positron beams of the FCC-ee cross at two IPs and large caverns are required to accommodate the experiments. Although the exact construction sequence for these caverns is yet to be confirmed, large-span caverns with dimensions of 66\,m $\times$ 35\,m $\times$ 35\,m are proposed. 
A service cavern with a span of 25\,m is required at each of the 12 access points around the ring. 
A service cavern at the machine level with dimensions of 100\,m $\times$ 25\,m $\times$ 15\,m is needed near the experiment caverns. Shorter service caverns are planned at the remaining 10 access points, depending on the infrastructure equipment that needs to be housed underground. 
These caverns will be connected to the surface via shafts with diameters ranging from 10\,m to 18\,m. 
Each IP requires one shaft with a diameter of 15\,m and one with a diameter of 10\,m. 
Additionally, two shafts are proposed to be located near the existing CERN accelerators. 
The IDEA detector is designed to extend up to a half-length of approximately 6.5\,m in the $\pm$ $z$-direction and 5.5\,m in the radial direction inside the large experimental cavern. 

Based on this discussion, we propose several detector designs optimized to capture LLPs originating from both $B$-meson and Higgs decays, as discussed below. The configurations of the dedicated detectors have been planned keeping in mind the sketch of the structure of the FCC-ee experimental cavern as shown in Fig.\,7.5 of Ref.\,\cite{FCC:2018evy}. The detailed illustrations of these detectors are presented in Appendix\,\ref{app:dedi_det}.

\begin{enumerate}
    \item \textbf{A-TYPE}: This category of detectors consists of full- and half-cylindrical shells around the IDEA detector in the FCC-ee tunnel. There are six detectors in this category with varying inner radius, outer radius, and lengths. A1 is a full cylinder surrounding the IDEA detector, covering the $4\pi$ solid angle, and A2, A3, A4, A5, and A6 are half cylinders that surround the $0 < \theta < \pi$ region, where $\theta$ is the angle between the radius vector and the $X$-axis. The half-cylinders are proposed to accommodate the constraints in the experimental cavern in the case that the bottom half of the IDEA detector is inaccessible for a fully cylindrical structure. Fig.\,\ref{fig:det_A} shows the different configurations of detectors in this series. Table\,\ref{tab:det_A} shows the values of the parameters $R_1$, $R_2$, and $L$. 
    
    \item \textbf{B-TYPE}: This category consists of half-cylindrical shells like the A-type detectors; however, this time, the cylindrical shells are tilted at an angle of $45\degree$ in the $X-Y$ plane. We choose this configuration for the possibility that the IDEA detector is positioned in the FCC-ee experiment cavern such that the left wall and the floor are obstructed. There are five detectors in this category- B1, B2, B3, B4, and B5. B5 is a detector with similar coverage but of a different configuration. B5 is a combination of two box-shaped detectors of length $L$ placed at right angles. The detectors are illustrated in Fig.\,\ref{fig:det_B}, and the parameters of B1-B5 are listed in Table\,\ref{tab:det_B}.

    \item \textbf{C-TYPE}: At the time of the FCC-ee run, the space in the cavern for FCC-hh could be utilized to place dedicated detectors for LLPs. In such a scenario, we propose the C-type detectors, which are to be placed in the FCC-hh tunnel at 10 m from the IDEA detector. In this category, C1, C2, C3 are solid half-cylinders with $-\pi/2 < \theta < \pi/2$ and varying $L$. C4, C5, and C6 are box-shaped detectors of different dimensions parallel to the IDEA detector. C7 and C8 consist of multiple boxes along the $Z$-axis, parallel to the IDEA detector, and in the FCC-hh tunnel. Fig.\,\ref{fig:det_C} and the parameters in Tables\,\ref{tab:det_C_1} and \ref{tab:det_C_2} highlight the different types of detectors in the C-series.

    \item \textbf{D-TYPE}: The service cavern is 50\,m away from the FCC-ee experiment cavern. LLPs with large decay lengths can traverse these distances, making the service cavern a suitable location to place dedicated LLP detectors. The service cavern is 100 to 150\,m long, 15\,m high, 25\,m wide. The D-type detectors are to be placed in this location, at 80 m away from the IDEA detector. Six detectors are considered in this series, two of which are cylinders, and four are box-shaped detectors, as shown in Fig.\,\ref{fig:det_D}. Table\,\ref{tab:det_D} lists the various parameters used to design the D-type detectors.

    \item \textbf{E-TYPE}: We propose forward detectors for FCC-ee in this category. The E-type detectors are cylindrical shells placed along the $Z$-axis in the forward direction of the IDEA detector. There are ten detectors of this type, with varying radii, lengths, and distances from the IDEA detector. Out of these, E5 and E6 consist of forward detectors placed on both sides of the IDEA detector. Fig.\,\ref{fig:det_E} shows the configuration of six E-type detectors. Table\,\ref{tab:det_E} consists of the values of the radii and position for the six different types.

    \item \textbf{F-TYPE}: FORward Experiment for HUNdred TeV (FOREHUNT) ~\cite{Bhattacherjee:2023plj} was proposed as a forward detector for FCC-hh for the detection of light LLPs produced from the decay of $B$ mesons and Higgs bosons at FCC-hh. They can also be utilized as detectors for LLPs produced in the FCC-ee. If a dedicated detector were to serve the purpose of both the experiments at FCC-ee and FCC-hh, it would be a cost-effective solution, too. They comprise the F-type detectors in our study. The radius and position of the detectors are varied, as illustrated in Fig.\,\ref{fig:det_F} and in Table\,\ref{tab:det_F}.

    \item \textbf{G-TYPE}: Similar to FOREHUNT, another category of dedicated detectors, DELIGHT, specific to the FCC-hh experiment, was proposed in ~\cite{Bhattacherjee:2021rml}. However, in contrast to FOREHUNT, these are to be placed in the transverse direction to FCC-hh, as shown in Fig.\,\ref{fig:det_G}. In our study, we consider the DELIGHT detectors, rotated azimuthally, to be the G-type detectors to see if they are sensitive to LLPs produced at FCC-ee. As mentioned, we aim for an optimum design that is equally sensitive to the two experiments so that the dedicated detector can be reused for the FCC-hh run. Three configurations of DELIGHT were explored in Ref.\,\cite{Bhattacherjee:2021rml} $-$ DELIGHT (A), DELIGHT (B), and DELIGHT (C). Here, we rename them G1, G2, and G3. We also consider some additional dimensions and placement positions. In total, there are nine configurations of G-type detectors, as listed in Table\,\ref{tab:det_G}. 
    
    \item \textbf{H-TYPE}: A combination of near and far detectors shall capture LLPs of both smaller and larger decay lengths. Considering this, we design two cylindrical arcs (a quarter of a cylinder), one near the IDEA detector and one far from the IDEA detector. These have radii $R_1$, $R_2$ and $R_1'$, $R_2'$. They possess the same length. We consider two configurations of this type. They are described in Fig.\,\ref{fig:det_H} and Table\,\ref{tab:det_H}.
    
    \item \textbf{I-TYPE}: The simplest detector design is that of a box. Until now, we have considered many box-shaped detectors placed in various locations away from the IDEA detector. We propose placing such box-shaped detectors directly above the IDEA detector as shown in Fig.\,\ref{fig:det_I}. There are six I-type detectors with different elevations along the $Y$-axis, according to the dimensions parametrized by the values in Table\,\ref{tab:det_I}.
\end{enumerate}

For each dedicated detector, we scan over the $mc\tau-m_\phi$ plane of LLPs coming from the $B\to K\phi$ process (\textbf{Case A}) and the $h\to \phi\phi$ 
process (\textbf{Case B}). 
We calculate the signal efficiencies for each point in the parameter space. 
The results are shown in terms of efficiency maps in Appendices~\ref{app:dedi_BKphi} and \ref{app:dedi_hphiphi}. 
We present the observations in each category of dedicated detectors for 
\textbf{Case A} and \textbf{Case B}. 

In \textbf{Case A}, among the A-type detectors, A1 is the most sensitive, even for $c\tau = 10^9\,\text{mm}$ and $m_{\phi}\geq 2~\text{GeV}$.
However, the configuration for A1 is that of a cylindrical shell that completely surrounds the IDEA detector. 
If it is unfeasible to construct such a detector due to space constraints, we also consider half-cylinders. 
The tilted designs of the B-type are more likely to fit the space. B4 shows the best performance among the B-type detectors. 
The ratio of the decay volumes of A1 and B4 is exactly 0.5, and so is the ratio of the efficiencies of the two. 

The C and the D-type detectors are placed in the FCC-hh tunnel and the service cavern, respectively. LLPs from $B\to K\phi$ do not possess enough boost to reach these detectors, especially the D-type detectors; hence, they are less efficient. 
Even the LLPs with large $c\tau$ are not detected in these because of the lack of coverage in solid angle. 
Moreover, the branching fraction of $B\rightarrow K\phi$ is inversely proportional to the mixing angle. Therefore, for larger decay lengths, the production rate of $\phi$ decreases, reducing the sensitivity of dedicated detectors. While for smaller decay lengths, even when the branching fraction of $B$-mesons decaying to $\phi$ increases, the LLPs do not reach the dedicated detectors.
Although C2 is a half-cylinder and C5 is a box, they have similar decay volumes (3140 and 3000\,m$^3$, respectively), and they are comparable in terms of efficiencies.

Among the E-type detectors, E5, which is symmetrically placed on either side of the IDEA detector in the forward direction, has the highest sensitivity. The FOREHUNT detectors, named F-type detectors here, again lose much of the sensitivity for LLPs from $B\to K\phi$. The DELIGHT B configuration we proposed for FCC-hh, denoted as the G2 detector here, is the most sensitive to LLPs with high decay lengths out of all the detector types considered here.
We re-emphasize that the most optimum detector choice would be the one sensitive to LLPs from both FCC-ee and a future FCC-hh run. 
Among the I-type detectors, I4 is the most sensitive for \textbf{Case A}.    

In \textbf{Case B}, the LLPs receive a boost from the decay of the SM Higgs and can traverse up to large distances before decaying. 
Additionally, the Br($h\to \phi\phi$) is independent of the mixing angle. Therefore, even for larger decay lengths, the production rate of $\phi$ doesn't decrease.
This makes the dedicated detectors sensitive to LLPs with large $c\tau$'s in this scenario, which was not obtained in \textbf{Case A}. 
For instance, the C, D, and F-type detectors, which were not very efficient for LLPs produced in \textbf{Case A}, gain sensitivity for \textbf{Case B}.

We have observed that out of all the dedicated detectors, A1, B4, and G2 are the most sensitive to LLPs with high decay lengths. A1 is advantageous because of its excellent coverage, while G2 benefits from a high decay volume of $100\times 100\times 100 ~\text{m}^3$. If the A1 detector cannot be constructed due to space constraints, the B4 configuration should be considered for a cylindrical geometry but with half the coverage of A1. For the benchmarks with high values of $c\tau$, we calculate the number of LLPs detected in A1, B4, and G2. The results are shown in Table\,\ref{tab:det_eff}. For BPA1, 38 events can be observed at A1, 21 events can be observed at B4, and 87 events can be observed at G2. We recall from Table\,\ref{tab:decay_events_a} that with a 100\% detector acceptance, only 22 events were detected in VTX and DCH combined, while 13 events were detected in the MS. The G2 detector provides significant improvement in this situation. Similarly, in BPA5, the number of events detected within the IDEA detector was low (5 in MS, 8 in VTX+DCH), while the dedicated detector G2 can detect 15 events. BPB6, having a decay length of 1\,m, is observed best at A1. Of the 403 events decaying outside the detector (see Table\,\ref{tab:decay_events_b}), 153 events can be observed at A1.
Detectors B4 and G2 can also improve the sensitivity to this benchmark as compared to the analyses performed for decays within the IDEA detector.

\begin{table}[htb!]
\centering
\resizebox{0.6\textwidth}{!}{
\begin{tabular}{|c|c|c|c|c|c|}
\hline
\multirow{2}{*}{Benchmarks} & \multicolumn{5}{c|}{Dedicated detector analysis} \\ \cline{2-6} 
 & $m_{\phi}$ (GeV) & $c\tau$ (mm) & A1 & B4 & G2 \\ \hline\hline
BPA1 & 0.4 & 39666.6 & 38 & 21 & 87 \\
BPA5 & 3.5 & 10285.4 & 11 & 6 & 15 \\
BPB6 & 6.0 & 1000 & 153 & 83 & 14 \\
\hline\hline
\end{tabular}
}
\caption{Number of events detected in A1, B4, and G2 detectors assuming 100\% efficiency for our chosen benchmark points from \textbf{Case A} and \textbf{Case B}.}
\label{tab:det_eff}
\end{table}

\begin{figure}[htb!]
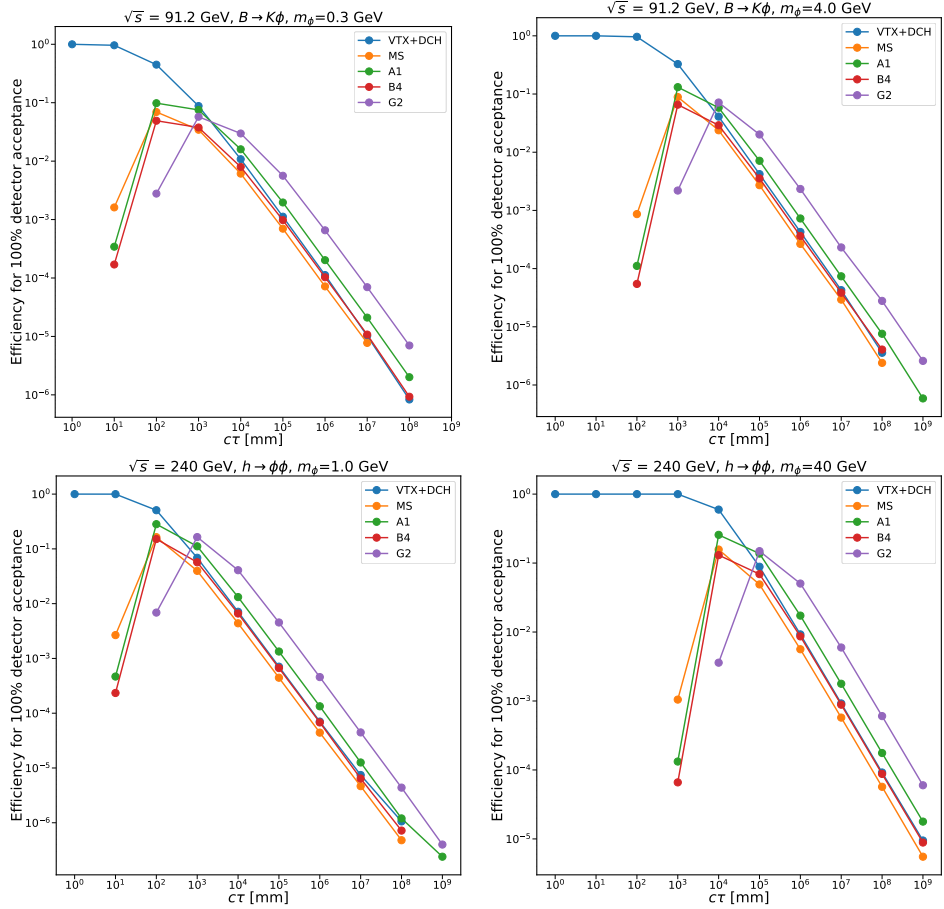

    \centering
    \includegraphics[width=0.4\textwidth]{Eff_plot_b_0.pdf}~
    \includegraphics[width=0.4\textwidth]{Eff_plot_b_5.pdf}\\
    \includegraphics[width=0.4\textwidth]{Eff_plot_h_0.pdf}~
    \includegraphics[width=0.4\textwidth]{Eff_plot_h_5.pdf}
    \caption{Efficiencies of the VTX+DCH systems, the Muon System (MS) and dedicated detectors A1, B4, and G2 plotted as a function of $c\tau$ for different masses of LLP in \textbf{Case A} scenario \textit{(top panel)} and in \textbf{Case B} scenario \textit{(bottom panel)}.}
    \label{fig:dedi_best}
\end{figure}

To study the sensitivity of the dedicated detectors across low to high lifetimes, we vary the decay length of two LLPs of masses 0.3\,GeV and 4.0\,GeV generated via $B\to K\phi$, and two LLPs of masses 1.0\,GeV and 40\,GeV generated via $Zh, h\to \phi\phi$ process. Fig.\,\ref{fig:dedi_best} shows the efficiencies of these detectors for two different masses of LLPs as a function of $c\tau$ for $B\to K\phi$ and $h\to \phi\phi$ events, respectively. Also shown are the efficiencies obtained in VTX+DCH and Muon Systems of the IDEA detector. In \textbf{Case A}, the detectors A1 and G2 show higher efficiency than the VTX+DCH and MS components when the decay length is $\sim10^4$ mm. The efficiency of B4 is comparable to that of VTX and DCH combined. For a 0.4 (4.4)\,GeV LLP, the efficiency of the A1 detector is 3 (2.5) times greater than that of the MS at a $c\tau$ of $10^6$ mm, while the G2 detector reaches up to an efficiency 10 times that of MS for $c\tau$ of $10^8$ mm. Improvement in the efficiency by a factor of 10 is also obtained for the G2 detector in the $h\to\phi\phi$ scenario.

As discussed, the DELIGHT B configuration (G2 detector) benefits from its large decay volume. Compared to all the other box-shaped detectors used in this study, it is the only one that achieves a tenfold improvement in efficiency over the IDEA detector. The substantial enhancement points us towards the need for a large detector volume to capture LLPs. Therefore, if a dedicated detector is to be constructed for LLP detection at FCC-ee, its advantage would be maximized only if it is designed with a sufficiently large size, similar to DELIGHT B. However, constructing such a large detector will be ultimately subject to space constraints and shall require a detailed feasibility study.
From Fig.\,\ref{fig:dedi_best}, we observe that for long-lived particles with relatively short decay lengths, the IDEA detector performs extremely well. For intermediate to large decay lengths, the performance of IDEA is comparable to that of several dedicated detector proposals. We emphasize that further optimization of the IDEA detector should be guided by LLP-motivated physics and design considerations.

\medskip

\noindent
\textbf{A discussion on backgrounds}
\medskip

One possible source of background in the dedicated detectors is the flux of high-energy muons and neutrinos originating from the IP at FCC-ee. Another source might be low-energy neutrons, photons, and electrons originating from the secondary interactions of high-energy muons and neutrinos outside the main detector, as also discussed in Ref.\,\cite{Aielli:2019ivi} in the context of CODEX-b. One way to mitigate the background from high-energy muons, and any secondary neutral hadrons originating from its interactions outside the main detector, is the use of lead and concrete shielding before the placement of the dedicated LLP detector. The muons that overcome the shielding barrier can be vetoed using either the information from the MS or by placing scintillator layers in front of the detector decay volume. The neutrinos, on the other hand, pose a significant challenge considering the huge flux $\sim\mathcal{O}(10^{12})$ from $Z\to \nu\bar{\nu}$ decays and the little to no interaction with the shielding material. However, if the LLP detector volume is filled with low-density gas, the probability of neutrino interactions within the decay volume becomes vanishingly small, thus effectively rendering the neutrino background negligible. A detailed study of the possible backgrounds using the \textsc{Geant-4}\,\cite{GEANT4:2002zbu} simulation and exploring their mitigation is left for a future study.

\section{Conclusions}
\label{sec:concl}

In the diverse landscape of current experiments and future proposals aimed at searching for light new physics, understanding the role of next-generation colliders is crucial. The most likely candidate is an electron-positron machine. Among the many options available, we focus here on the FCC-ee. However, the characteristics of these colliders are similar, so our discussion applies to all of them. To assess the potential of the FCC-ee for detecting light, long-lived BSM particles, the first question to address is: \textit{Will ongoing experiments, along with existing proposals for beam dump or dedicated LLP detectors, leave any regions that can still be probed by the FCC-ee?} The answer to this question will also depend on how many of the proposed experiments receive approval.
Even if one or more of the beam dump, collider, or dedicated LLP detectors observes a signal in the near 
future, \textit{can the FCC-ee contribute to model identification and the determination of BSM particle properties?}

To illustrate these issues, we consider the dark Higgs model as an example. 
The dark Higgs boson, $\phi$, is long-lived and can be produced either from $B$-meson decays or Higgs boson 
decays at the FCC-ee. Since the LLP in this model decays to 
SM particles via its mixing with the SM Higgs boson, there 
is a wide variety of decay modes depending on the mass of 
the mediator. The excellent particle identification 
capabilities of the FCC-ee detectors can provide 
sensitivity to decay modes involving mesons, and this can in turn help in the determination of various branching fractions of $\phi$, in case of discovery.
In the present work, we select benchmarks that span various possibilities and explore the questions discussed above.
We divide the benchmarks into two cases $-$ without and with a large trilinear scalar coupling.
While $B\rightarrow K\phi$ is the dominant production mode in the first case, in the case of large trilinear coupling, $h\to\phi\phi$ is the dominant production mode. We study the benchmarks from the first case at the $\sqrt{s} = 91.2$\,GeV run of FCC-ee, and the second case at the $HZ$ production peak of $\sqrt{s} =240$\,GeV.
For each benchmark, we perform analyses based on the detector element where the LLP decays and the final state it decays into. 
We also discuss briefly the mass and lifetime range that might be explored in the production of $\phi$ from the process $Z\to Z^*\phi$ at the FCC-ee.

We note that the primary background for light long-lived particles consists of SM long-lived hadrons. 
A comprehensive list of these hadrons, detailing their decay 
lengths and modes, shows that many of them significantly affect the analysis of some of our benchmark points due to the high production rates of the SM backgrounds.
For example, the presence of $K_S$, $K_L$, and $\Lambda^0$ is a major challenge in the search for LLPs below a mass of 1\,GeV decaying to pions or kaons within the VTX or the DCH detectors.
The muon spectrometer provides a much cleaner environment and, hence, a better prospect for these LLP benchmarks.

We quote the expected number of signal events at FCC-ee after the analysis cuts for each benchmark for their dominant decay modes.
For the case where the LLP comes from the Higgs boson decay, we do not use any specific cut on the decay products of the associated $Z$ boson. 
Instead, our analysis strategy demands the observation of both the LLP decay vertices to reduce the SM backgrounds.
A similar analysis can be performed for other production modes of the Higgs boson, like the vector boson fusion process at higher energies of $e^+ e^-$ machines.

Finally, we extensively study possible dedicated LLP detector options for the FCC-ee.
We consider nine different types of detector configurations with varying dimensions and positions around the FCC-ee collider complex.
For LLPs coming from $B$-meson decays, the sensitivity is limited due to the dependence of the branching fraction on the mixing angle, unlike LLPs coming from the Higgs boson decay.
We observe that the detector configurations A1, B4 and G2 perform the best among the dedicated detectors we considered.
The detector configuration G2 is the DELIGHT B configuration that we proposed for FCC-hh in Ref.\,\cite{Bhattacherjee:2021rml}, just rotated in the azimuthal direction.
We have observed in Refs.\,\cite{Bhattacherjee:2021rml,Bhattacherjee:2023plj} that the DELIGHT B detector configuration performs well for LLPs produced at the FCC-hh. Therefore, this can serve as a shared detector concept for both FCC-ee and FCC-hh.

In scenarios with negligible trilinear coupling, benchmarks BPA1 and BPA5 have large decay lengths and, therefore, are best observed in our proposed dedicated detectors around the FCC-ee interaction point, followed by the muon spectrometer of the IDEA detector. In the large trilinear coupling regime, benchmarks BPB1 and BPB2 produce about 100 and 10 signal events, respectively, in the pion/kaon decay modes within the inner detectors. Our proposed dedicated detectors A1 and B4 for FCC-ee perform the best for BPB6. 

We also provide the efficiencies for each of our analyses with varying mass and decay length of the LLP for any non-minimal extension of this model, where the relation between mass and sin$^2\,\theta\times c\tau$ gets modified as compared to the minimal model.
These efficiencies are also useful for translating the results for arbitrary branching fractions for the various decay modes of $\phi$, given there is a theory uncertainty on this for $\phi$ having masses between $\sim 1-4$\,GeV.

It should be noted that our benchmark analyses are background-free. While we do not consider the systematic uncertainties related to the background, the statistical uncertainty must still be taken into account. The final results, particularly for benchmarks yielding very few events (such as BPA1), should be verified with higher MC statistics and by including background contributions from material interactions.

The IDEA detector at FCC-ee stands as a powerful probe for LLP discovery, particularly for LLPs with short decay lengths, which are extremely difficult to probe at the LHC amidst the overwhelming hadronic activity. In conclusion, our study highlights the importance of future lepton colliders in enhancing the search for light long-lived particles.

\section*{Acknowledgement}

BB acknowledges the MATRICS Grant (MTR/2022/000264) of the Science and Engineering Research Board (SERB), Government of India. The work of BB is also supported by the Core Research Grant CRG/2022/001922 of the Science and Engineering Research Board (SERB), Government of India. BB and CB are grateful to the Center for High Energy Physics, Indian Institute of Science, for the cluster facility.
HKD would like to thank the Nikhef Theory Group for its 
kind hospitality while part of this work was being 
completed. The work of NG was supported by the Japan Society for the Promotion of Science(JSPS) as a part of the JSPS Postdoctoral Program (Standard), grant number: JP24KF0189, and by the World Premier International Research Center Initiative (WPI), MEXT, Japan (Kavli IPMU). NG would also like to thank CHEP, IISC for financial support where part of the work was done. SM is supported by a Grant-in-Aid for Scientific Research from the Ministry of Education, Culture, Sports, Science, and Technology (MEXT), Japan: 23K20232 (20H01895), 24H00244 (20H00153), and 24H02244 by the JSPS Core-to-Core Program: JPJSCCA20200002 and by the World Premier International Research Center Initiative (WPI), MEXT, Japan (Kavli IPMU).

\clearpage

\appendix

\section{Configurations of dedicated detectors}
\label{app:dedi_det}

In this section, we present the illustrations of the different dedicated detectors proposed in this study, in Section \ref{sec:dedicated}. The detectors are categorized into various types. The figures show the model of each type of detector, and the tables list the various parameters of the dedicated detectors.

\begin{figure}[hbt!]
  \begin{minipage}[b]{.55\linewidth}
    \centering
    \includegraphics[width=\linewidth]{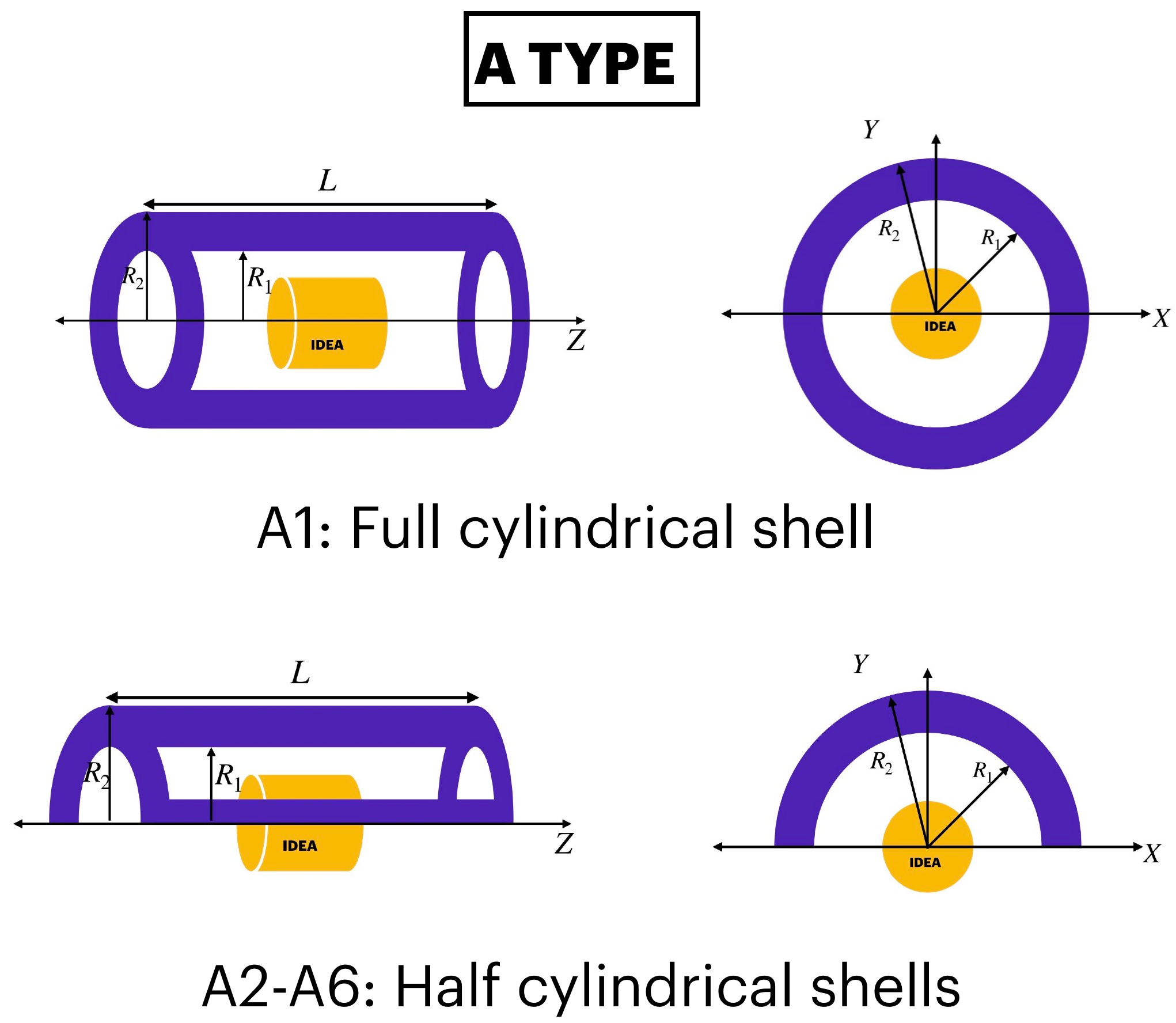}
    \captionof{figure}{``A'' series detectors $-$ illustration}
    \label{fig:det_A}
  \end{minipage}\hfill
  \begin{minipage}[b]{.4\linewidth}
    \centering
    \resizebox{\linewidth}{!}{
    \begin{tabular}{|c|c|c|c|}
    \hline
    Detector & $R_1$ (m) & $R_2$ (m) & $L$ (m) \\
    \hline\hline
    A1 & 6 & 11 & 20 \\
    A2 & 6 & 11 & 10 \\
    A3 & 7 & 10 & 10 \\
    A4 & 9 & 10 & 10 \\
    A5 & 9 & 10 & 6 \\
    A6 & 7 & 10 & 6 \\
    \hline\hline
    \end{tabular}}\\
    \vspace{1cm}
    \captionof{table}{``A'' series detectors $-$ parameters}
    \label{tab:det_A}
  \end{minipage}
\end{figure}

\begin{figure}[hbt!]
  \begin{minipage}[b]{.55\linewidth}
    \centering
    \includegraphics[width=\linewidth]{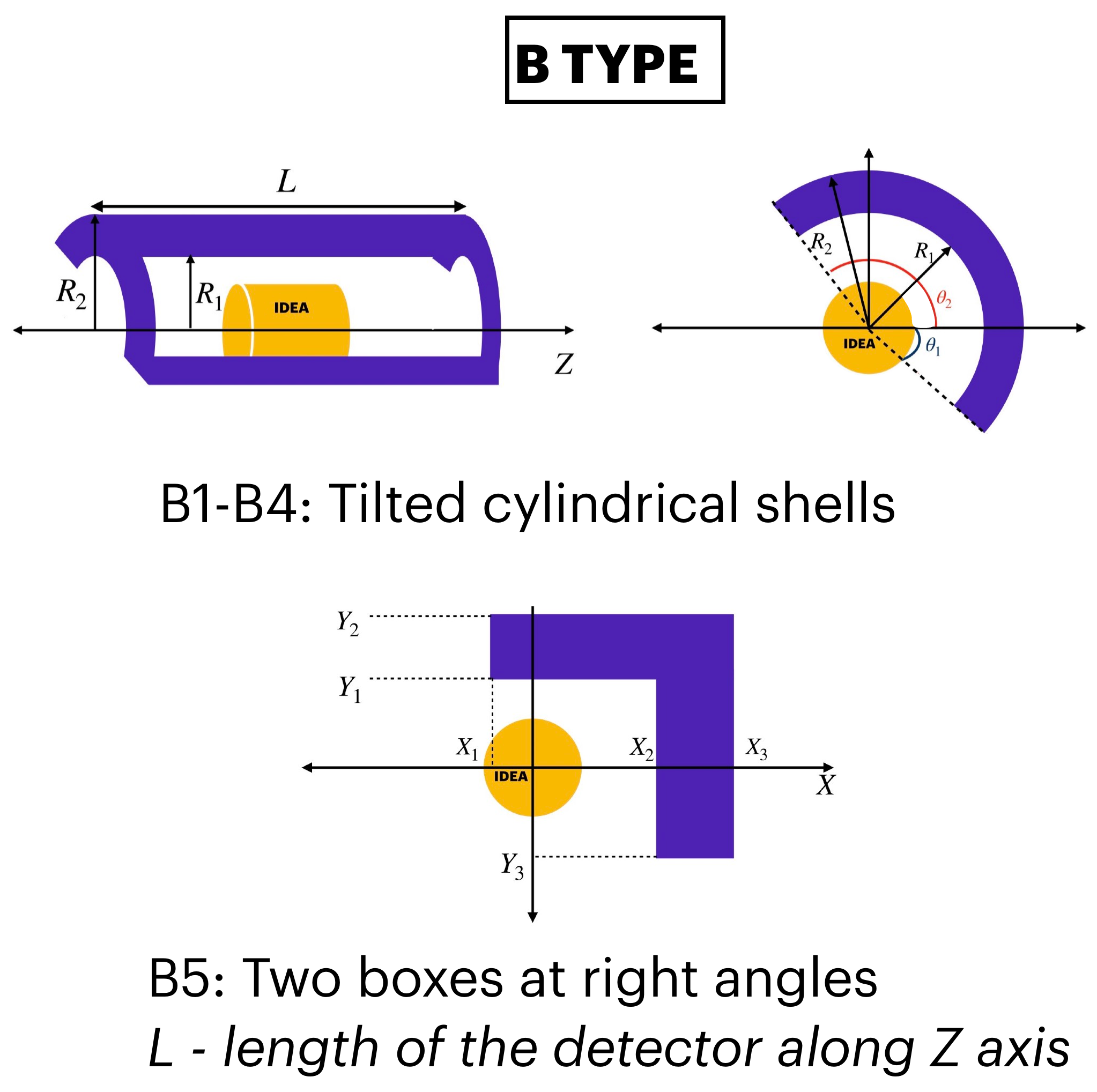}
    \captionof{figure}{``B'' series detectors $-$ \\
    illustration}
    \label{fig:det_B}
  \end{minipage}\hfill
  \begin{minipage}[b]{.45\linewidth}
    \centering
    \resizebox{\linewidth}{!}{
    \begin{tabular}{|c|c|c|c|c|c|}
    \hline
    \multirow{2}{*}{Detector} & $R_1$ & $R_2$ & $L$ & $\theta_1$ & $\theta_2$ \\
    & (m) & (m) & (m) & ($\degree$) & ($\degree$) \\
    \hline\hline
    B1 & 9 & 10 & 6 & \multirow{4}{*}{-45} & \multirow{4}{*}{135} \\
    B2 & 7 & 10 & 12 & & \\
    B3 & 6 & 11 & 12 & & \\
    B4 & 6 & 11 & 20 & & \\
    \hline\hline
    \end{tabular}}\\
    \vspace{0.5cm}
    \begin{eqnarray}
    \text{B5:} & X_1 = -3\,{\rm m}, X_2 = 7\,{\rm m}, X_3 = 12\,{\rm m}, \nonumber\\
       & Y_1 = 6\,{\rm m}, Y_2 = 11\,{\rm m}, Y_3 = -6\,{\rm m}, \nonumber \\
       & L = 20\,{\rm m} \nonumber
    \end{eqnarray}
    \captionof{table}{``B'' series detectors $-$ \\
    parameters}
    \label{tab:det_B}
  \end{minipage}
\end{figure}

\begin{figure}[hbt!]
  \begin{minipage}[b]{.55\linewidth}
    \centering
    \includegraphics[width=\linewidth]{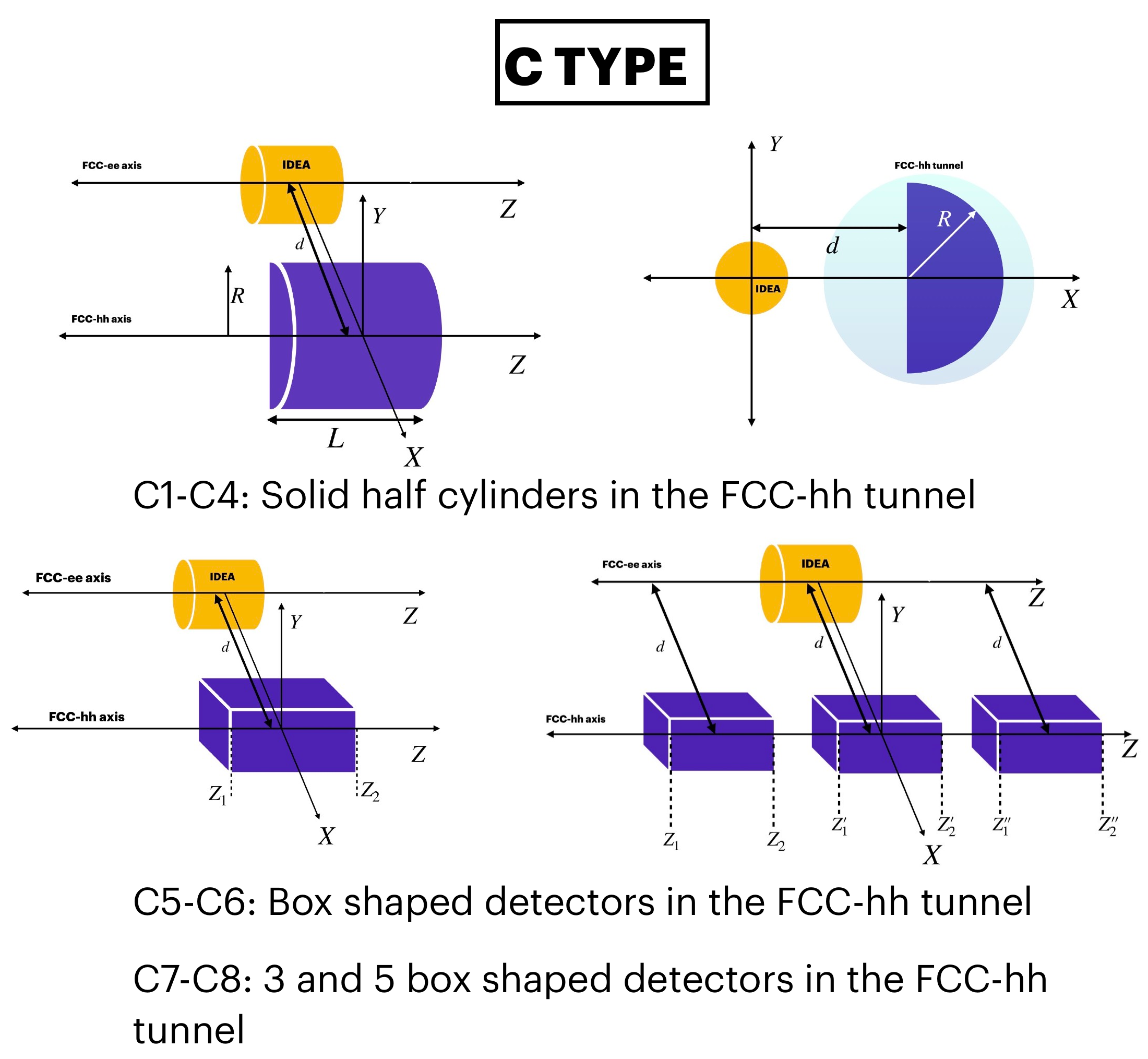}
    \captionof{figure}{``C'' series detectors $-$ \\
    illustration}
    \label{fig:det_C}
  \end{minipage}\hfill
  \begin{minipage}[b]{.4\linewidth}
  \resizebox{\linewidth}{!}{
    \begin{tabular}{|c|c|c|c|}
    \hline
    Detector & $d$ (m) & $R$ (m) & $L$ (m) \\
    \hline\hline
    C1 & 10 & 10 & 10 \\
    C2 & 10 & 10 & 20 \\
    C3 & 12 & 5 & 20 \\
    C4 & 12 & 5 & 10 \\
    \hline\hline
    \end{tabular}}\\
    \vspace{0.5cm}
    \begin{eqnarray}
    \text{C5 (C6):} & X_1 = 10\,{\rm m}, X_2 = 20 (15)\,{\rm m}, \nonumber\\
    & Y_1 = -5\,{\rm m}, Y_2 = 10 (5)\,{\rm m}, \nonumber\\
    & Z_1 = -10 (-5)\,{\rm m}, \nonumber\\
    & Z_2 = 10 (5)\,{\rm m}, d = 10\,{\rm m}  \nonumber   
    \end{eqnarray}
    \captionof{table}{``C'' series detectors $-$ \\
    parameters}
    \label{tab:det_C_1}
   \end{minipage}  
\end{figure}    

\begin{table}[hbt!]
    \centering
    \resizebox{\linewidth}{!}{
    \begin{tabular}{|c|c|c|c|c|c|c|c|c|c|c|c|c|c|c|c|}
    \hline
    \multirow{2}{*}{Detector} & $d$ & $X_1$ & $X_2$ & $Y_1$ & $Y_2$ & $Z_1$ & $Z_2$ & $Z'_1$ & $Z'_2$ & $Z''_1$ & $Z''_2$ & $Z'''_1$ & $Z'''_2$ & $Z''''_1$ & $Z''''_2$ \\
    & (m) & (m) & (m) & (m) & (m) & (m) & (m) & (m) & (m) & (m) & (m) & (m) & (m) & (m) & (m) \\
    \hline\hline
    C7 & 10 & 10 & 15 & $-5$ & 5 & $-15$ & $-10$ & $-5$ & 5 & 10 & 15 & $-$ & $-$ & $-$ & $-$ \\
    C8 & 10 & 10 & 15 & $-5$ & 5 & $-25$ & $-20$ & $-15$ & $-10$ & $-5$ & 5 & 10 & 15 & 20 & 25 \\
    \hline\hline
    \end{tabular}}
    \caption{``C'' series detectors $-$
    parameters (contd.)}
    \label{tab:det_C_2}
\end{table}

\begin{figure}[hbt!]
    \centering
    \includegraphics[width=\linewidth]{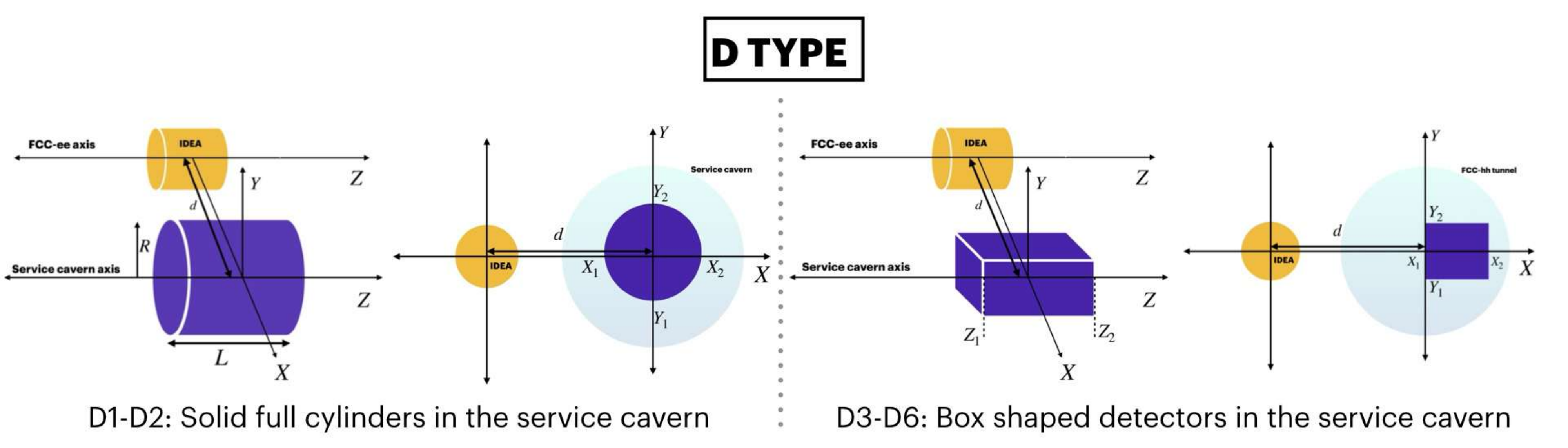}
    \captionof{figure}{``D'' series detectors $-$ illustration}
    \label{fig:det_D}
\end{figure}

\begin{table}
    \centering
    \resizebox{0.4\linewidth}{!}{
    \begin{tabular}{|c|c|c|c|}
    \hline
    Detector & $d$ (m) & $R$ (m) & $L$ (m) \\
    \hline\hline
    D1 & 80 & 10 & 20 \\
    D2 & 80 & 5 & 10 \\
    \hline\hline
    \end{tabular}}\\
    \vspace{0.5cm}
    \resizebox{0.8\linewidth}{!}{
    \begin{tabular}{|c|c|c|c|c|c|c|c|}
    \hline
    Detector & $d$ (m) & $X_1$ (m) & $X_2$ (m) & $Y_1$ (m) & $Y_2$ (m) & $Z_1$ (m) & $Z_2$ (m) \\
    \hline\hline
    D3 & \multirow{4}{*}{80} & 80 & 85 & $-5$ & 5 & $-5$ & 5 \\
    D4 & & 80 & 85 & $-5$ & 5 & $-10$ & 10 \\
    D5 & & 80 & 85 & $-5$ & 5 & $-15$ & 15 \\
    D6 & & 80 & 90 & $-5$ & 10 & $-15$ & 15 \\
    \hline\hline
    \end{tabular}}
    \captionof{table}{``D'' series detectors $-$
    parameters}
    \label{tab:det_D}
\end{table}

\begin{figure}[hbt!]
  \begin{minipage}[b]{.5\linewidth}
    \centering
    \includegraphics[width=\linewidth]{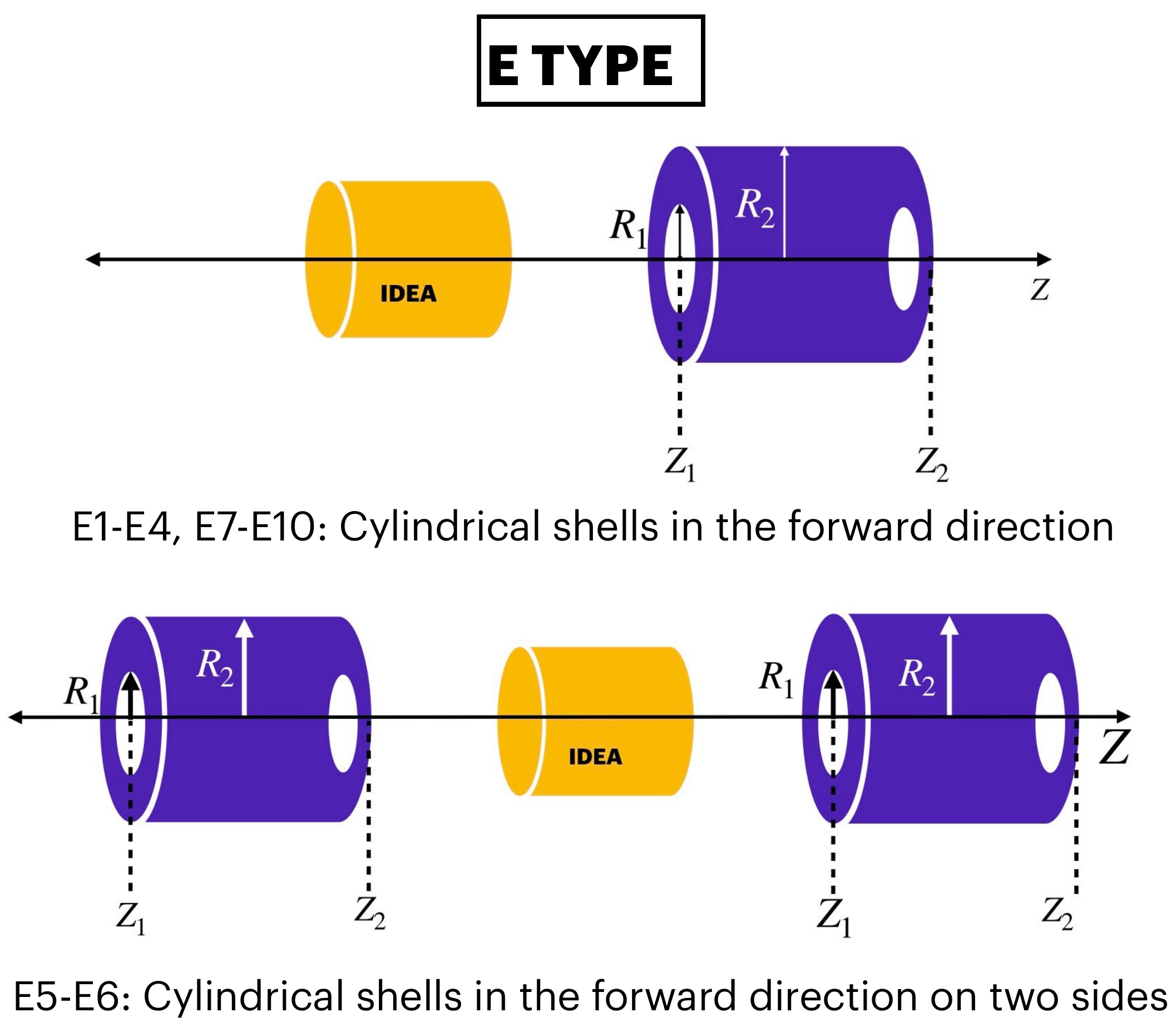}
    \captionof{figure}{``E'' series detectors $-$ \\
    illustration}
    \label{fig:det_E}
  \end{minipage}\hfill
  \begin{minipage}[b]{.45\linewidth}
    \centering
    \resizebox{\linewidth}{!}{
    \begin{tabular}{|c|c|c|c|c|}
    \hline
    Detector & $R_1$ (m) & $R_2$ (m) & $Z_1$ (m) & $Z_2$ (m)\\
    \hline\hline
    E1 & 2 & 5 & 10 & 15 \\
    E2 & 5 & 10 & 10 & 25 \\
    E3 & 5 & 10 & 15 & 20 \\
    E4 & 5 & 10 & 20 & 30 \\
    E5 & 5 & 10 & $\pm$10 & $\pm$25 \\
    E6 & 5 & 10 & $\pm$15 & $\pm$25 \\
    E7 & 0.1 & 5 & 10 & 15 \\
    E8 & 1 & 6 & 15 & 25 \\
    E9 & 2 & 10 & 10 & 20 \\
    E10 & 5 & 10 & 10 & 20 \\
    \hline\hline
    \end{tabular}}
    \captionof{table}{``E'' series detectors $-$ \\
    parameters}
    \label{tab:det_E}
  \end{minipage}
\end{figure}

\begin{figure}[hbt!]
  \begin{minipage}[b]{.5\linewidth}
    \centering
    \includegraphics[width=\linewidth]{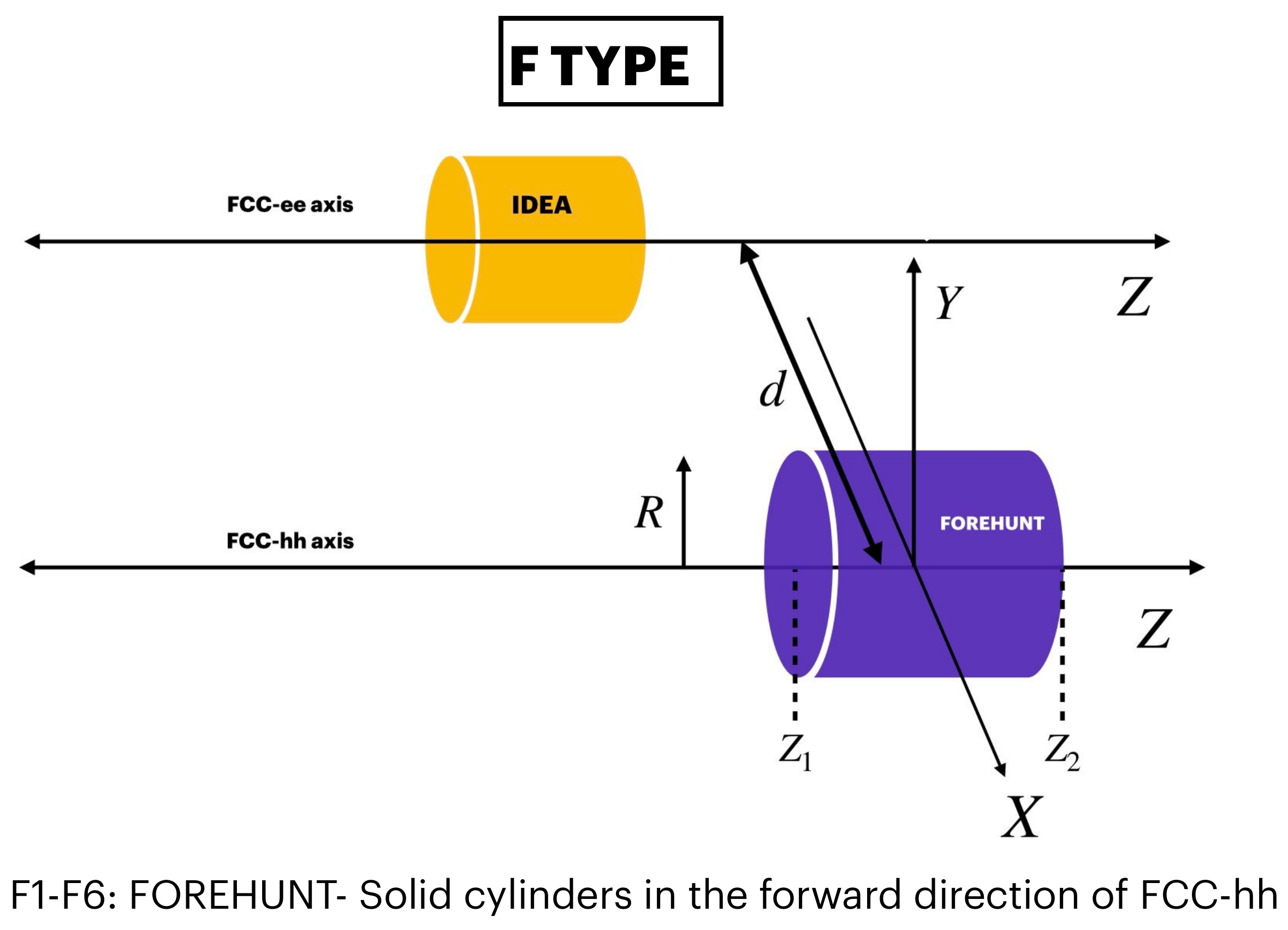}
    \captionof{figure}{``F'' series detectors $-$ \\
    illustration}
    \label{fig:det_F}
  \end{minipage}\hfill
  \begin{minipage}[b]{.45\linewidth}
    \centering
    \resizebox{\linewidth}{!}{
    \begin{tabular}{|c|c|c|c|c|}
    \hline
    Detector & $d$ (m) & $R$ (m) & $Z_1$ (m) & $Z_2$ (m)\\
    \hline\hline
    F1 & \multirow{6}{*}{10} & 1 & 50 & 60 \\
    F2 & & 2 & 50 & 70 \\
    F3 & & 5 & 50 & 75 \\
    F4 & & 2 & 75 & 100 \\
    F5 & & 5 & 75 & 100 \\
    F6 & & 5 & 100 & 120 \\
    \hline\hline
    \end{tabular}}\\
    \vspace{1cm}
    \captionof{table}{``F'' series detectors $-$ \\
    parameters}
    \label{tab:det_F}
  \end{minipage}
\end{figure}

\begin{figure}[hbt!]
    \centering
    \includegraphics[width=0.8\linewidth]{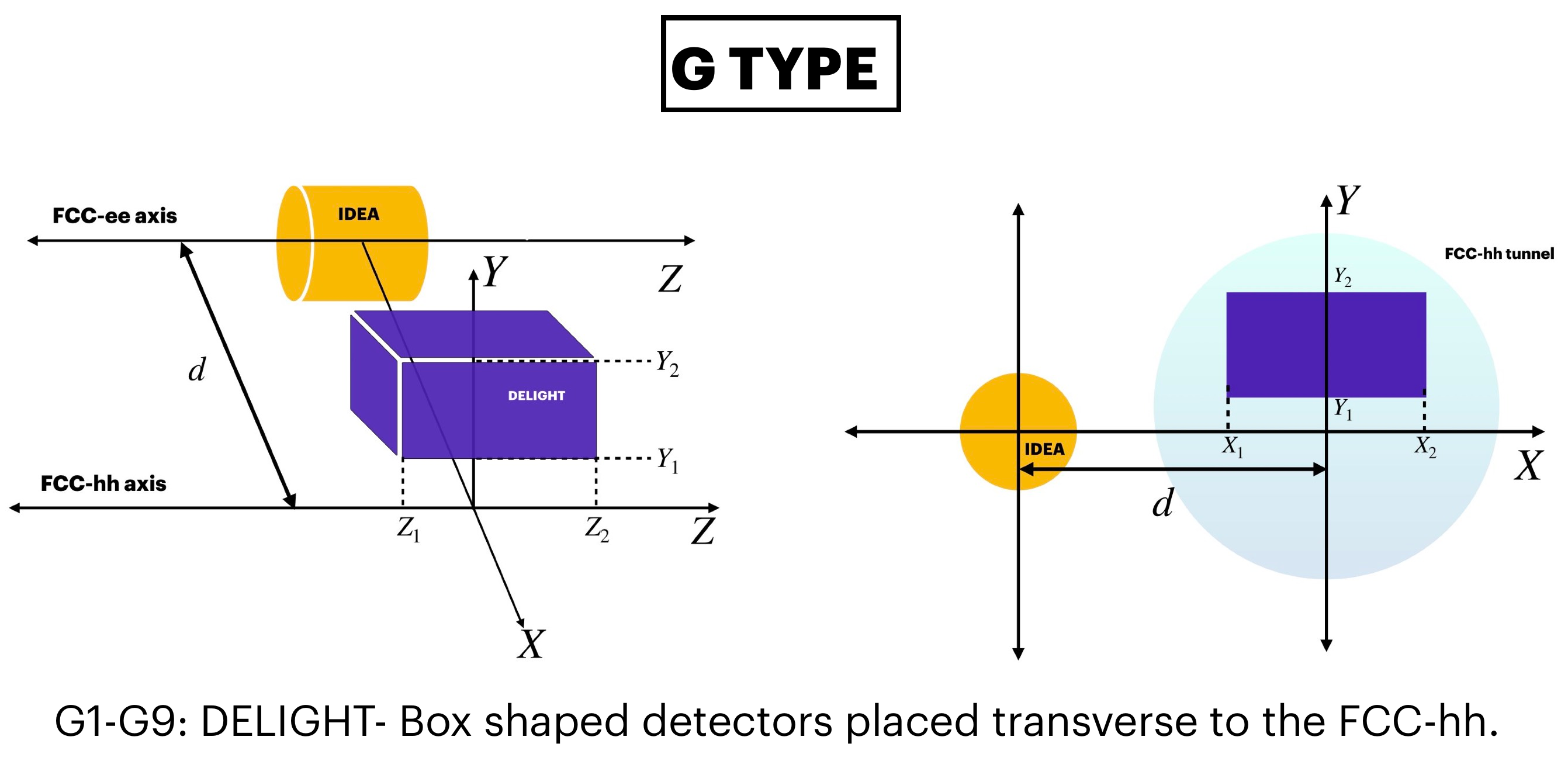}
    \captionof{figure}{``G'' series detectors $-$
    illustration}
    \label{fig:det_G}
\end{figure}

\begin{table}[hbt!]  
    \centering
    \resizebox{0.7\linewidth}{!}{
    \begin{tabular}{|c|c|c|c|c|c|c|c|}
    \hline
    Detector & $d$ (m) & $X_1$ (m) & $X_2$ (m) & $Y_1$ (m) & $Y_2$ (m) & $Z_1$ (m) & $Z_2$ (m) \\
    \hline\hline
    G1 & \multirow{9}{*}{10} & $-50$ & 50 & 25 & 50 & $-50$ & 50 \\
    G2 & & $-50$ & 50 & 25 & 125 & $-50$ & 50 \\
    G3 & & $-25$ & 25 & 25 & 225 & $-25$ & 25 \\
    G4 & & $-25$ & 25 & 25 & 50 & $-25$ & 25 \\
    G5 & & $-12.5$ & 12.5 & 25 & 50 & $-12.5$ & 12.5 \\
    G6 & & $-25$ & 25 & 25 & 75 & $-25$ & 25 \\
    G7 & & 0 & 25 & 0 & 30 & $-50$ & 50 \\
    G8 & & $-25$ & 25 & 25 & 125 & $-25$ & 25 \\
    G9 & & 25 & 75 & $-25$ & 25 & $-25$ & 25 \\
    \hline\hline
    \end{tabular}}
    \captionof{table}{``G'' series detectors $-$
    parameters}
    \label{tab:det_G}
\end{table}

\begin{figure}[hbt!]
  \begin{minipage}[b]{.47\linewidth}
    \centering
    \includegraphics[width=\linewidth]{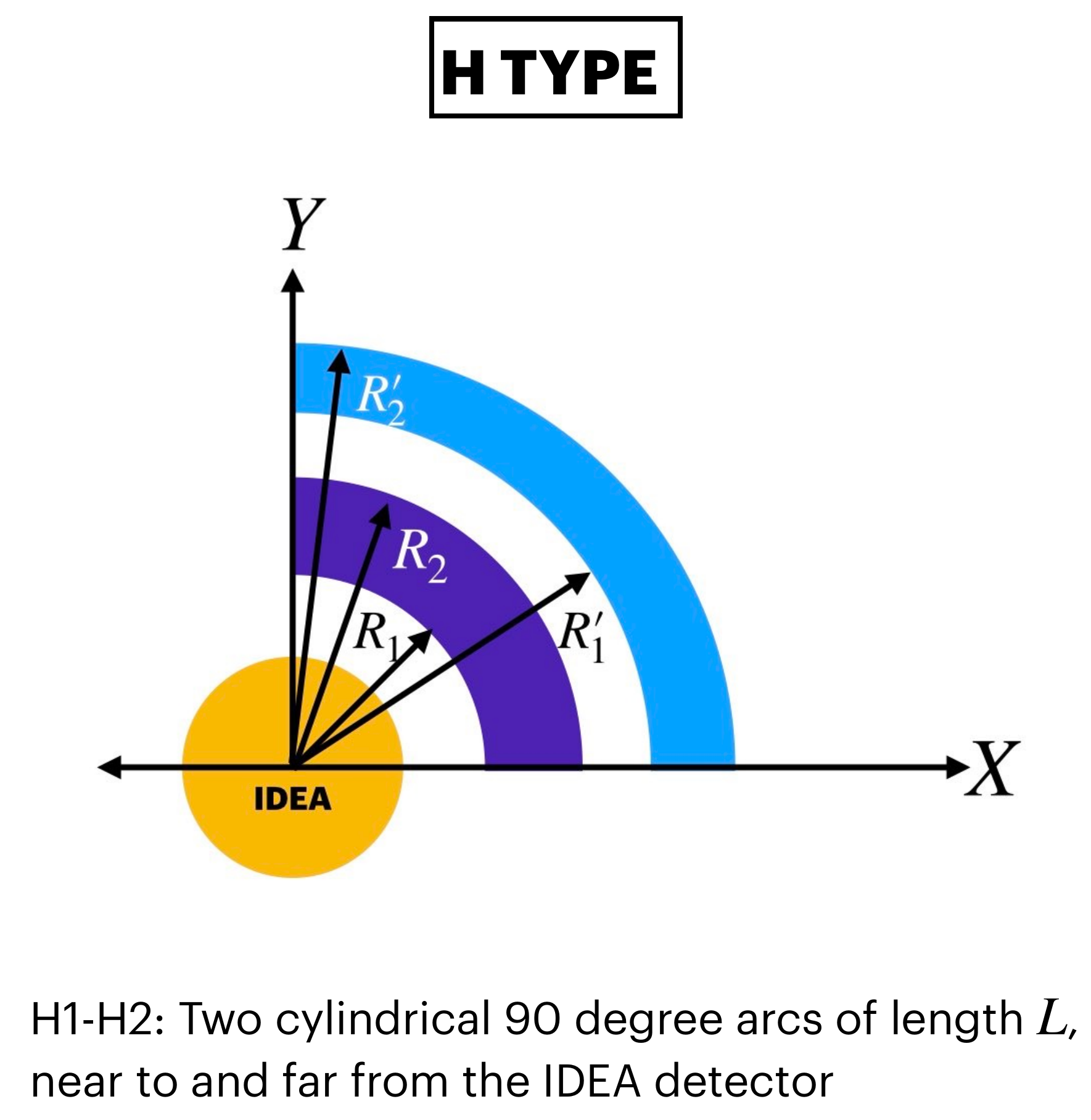}
    \captionof{figure}{``H'' series detectors $-$ \\
    illustration}
    \label{fig:det_H}
  \end{minipage}\hfill
  \begin{minipage}[b]{.5\linewidth}
    \centering
    \resizebox{\linewidth}{!}{
    \begin{tabular}{|c|c|c|c|c|c|}
    \hline
    Detector & $R_1$ (m) & $R_2$ (m) & $R'_1$ (m) & $R'_2$ (m) & $L$ (m) \\
    \hline\hline
    H1 & 9 & 10 & 12 & 13 & 6 \\
    H2 & 9 & 14 & 18 & 23 & 6 \\
    \hline\hline
    \end{tabular}}\\
    \vspace{3cm}
    \captionof{table}{``H'' series detectors $-$ \\
    parameters}
    \label{tab:det_H}
  \end{minipage}
\end{figure}

\begin{figure}[hbt!]
    \centering
    \includegraphics[width=0.7\linewidth]{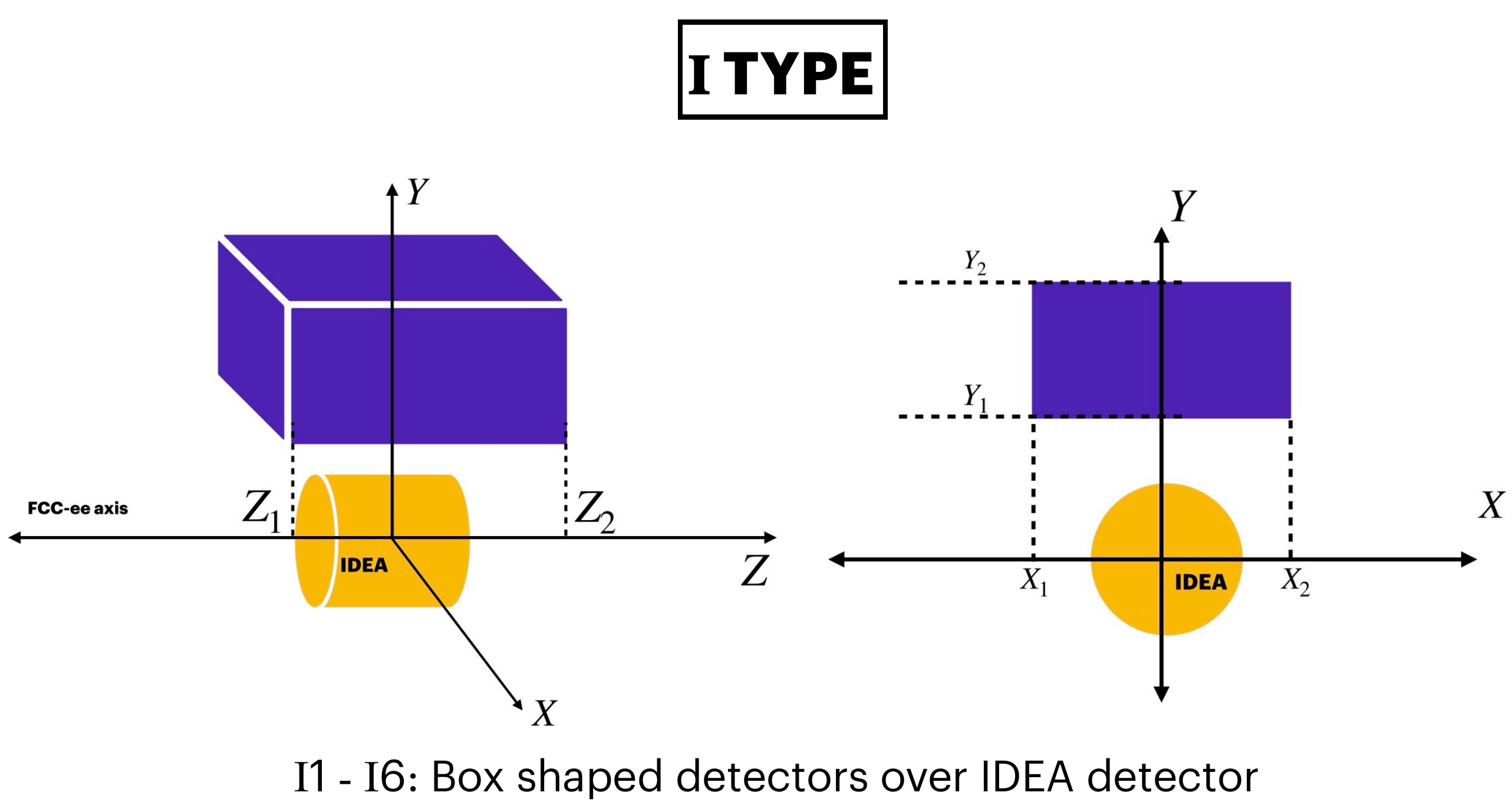}
    \captionof{figure}{``I'' series detectors $-$ illustration}
    \label{fig:det_I}
\end{figure}
\begin{table}
    \centering
    \resizebox{0.7\linewidth}{!}{
    \begin{tabular}{|c|c|c|c|c|c|c|}
    \hline
    Detector & $X_1$ (m) & $X_2$ (m) & $Y_1$ (m) & $Y_2$ (m) & $Z_1$ (m) & $Z_2$ (m) \\
    \hline\hline
    I1 & $-50$ & 50 & 100 & 110 & $-50$ & 50 \\
    I2 & $-50$ & 50 & 100 & 150 & $-50$ & 50 \\
    I3 & $-25$ & 25 & 50 & 70 & $-25$ & 25 \\
    I4 & $-25$ & 25 & 30 & 70 & $-25$ & 25 \\
    I5 & $-3.5$ & 3.5 & 30 & 70 & $-7$ & 7 \\
    I6 & $-7$ & 7 & 30 & 70 & $-3.5$ & 3.5 \\
    \hline\hline
    \end{tabular}}
    \captionof{table}{``I'' series detectors $-$ parameters}
    \label{tab:det_I}
\end{table}

\clearpage

\section{Efficiency maps of dedicated detectors in the case of $B\to K\phi$}
\label{app:dedi_BKphi}

\begin{figure}[htb!]
    \centering
    \includegraphics[width=0.45\linewidth]{Eff_A1_b.pdf}~
    \includegraphics[width=0.45\linewidth]{Eff_A2_b.pdf}\\
    \includegraphics[width=0.45\linewidth]{Eff_A3_b.pdf}~
    \includegraphics[width=0.45\linewidth]{Eff_A4_b.pdf}\\
    \includegraphics[width=0.45\linewidth]{Eff_A5_b.pdf}~
    \includegraphics[width=0.45\linewidth]{Eff_A6_b.pdf}
    \caption{Efficiency map of the A-type dedicated detectors in the $m_{\phi}-c\tau$ plane, for the LLPs produced via the process $B\to K\phi$.}
    \label{fig:eff_a_A}
\end{figure}

\begin{figure}[htb!]
    \centering
    \includegraphics[width=0.45\linewidth]{Eff_B1_b.pdf}~
    \includegraphics[width=0.45\linewidth]{Eff_B2_b.pdf}\\
    \includegraphics[width=0.45\linewidth]{Eff_B3_b.pdf}~
    \includegraphics[width=0.45\linewidth]{Eff_B4_b.pdf}\\
    \includegraphics[width=0.45\linewidth]{Eff_B5_b.pdf}
    \caption{Efficiency map of the B-type dedicated detectors in the $m_{\phi}-c\tau$ plane, for the LLPs produced via the process $B\to K\phi$.}
    \label{fig:eff_b_A}
\end{figure}

\begin{figure}[htb!]
    \centering
    \includegraphics[width=0.45\linewidth]{Eff_C1_b.pdf}~
    \includegraphics[width=0.45\linewidth]{Eff_C2_b.pdf}\\
    \includegraphics[width=0.45\linewidth]{Eff_C3_b.pdf}~
    \includegraphics[width=0.45\linewidth]{Eff_C4_b.pdf}\\
    \includegraphics[width=0.45\linewidth]{Eff_C5_b.pdf}~
    \includegraphics[width=0.45\linewidth]{Eff_C6_b.pdf}\\
    \includegraphics[width=0.45\linewidth]{Eff_C7_b.pdf}~
    \includegraphics[width=0.45\linewidth]{Eff_C8_b.pdf}
    \caption{Efficiency map of the C-type dedicated detectors in the $m_{\phi}-c\tau$ plane, for the LLPs produced via the process $B\to K\phi$.}
    \label{fig:eff_c_A}
\end{figure}

\begin{figure}[htb!]
    \centering
    \includegraphics[width=0.45\linewidth]{Eff_D1_b.pdf}~
    \includegraphics[width=0.45\linewidth]{Eff_D2_b.pdf}\\
    \includegraphics[width=0.45\linewidth]{Eff_D3_b.pdf}~
    \includegraphics[width=0.45\linewidth]{Eff_D4_b.pdf}\\
    \includegraphics[width=0.45\linewidth]{Eff_D5_b.pdf}~
    \includegraphics[width=0.45\linewidth]{Eff_D6_b.pdf}
    \caption{Efficiency map of the D-type dedicated detectors in the $m_{\phi}-c\tau$ plane, for the LLPs produced via the process $B\to K\phi$.}
    \label{fig:eff_d_A}
\end{figure}

\begin{figure}[htb!]
    \centering
    \includegraphics[width=0.45\linewidth]{Eff_E1_b.pdf}~
    \includegraphics[width=0.45\linewidth]{Eff_E2_b.pdf}\\
    \includegraphics[width=0.45\linewidth]{Eff_E3_b.pdf}~
    \includegraphics[width=0.45\linewidth]{Eff_E4_b.pdf}\\
    \includegraphics[width=0.45\linewidth]{Eff_E5_b.pdf}~
    \includegraphics[width=0.45\linewidth]{Eff_E6_b.pdf}\\
    \includegraphics[width=0.45\linewidth]{Eff_E7_b.pdf}~
    \includegraphics[width=0.45\linewidth]{Eff_E8_b.pdf}\\
    \includegraphics[width=0.45\linewidth]{Eff_E9_b.pdf}~
    \includegraphics[width=0.45\linewidth]{Eff_E10_b.pdf}
    \caption{Efficiency map of the E-type dedicated detectors in the $m_{\phi}-c\tau$ plane, for the LLPs produced via the process $B\to K\phi$.}
    \label{fig:eff_e_A}
\end{figure}

\begin{figure}[htb!]
    \centering
    \includegraphics[width=0.45\linewidth]{Eff_F1_b.pdf}~
    \includegraphics[width=0.45\linewidth]{Eff_F2_b.pdf}\\
    \includegraphics[width=0.45\linewidth]{Eff_F3_b.pdf}~
    \includegraphics[width=0.45\linewidth]{Eff_F4_b.pdf}\\
    \includegraphics[width=0.45\linewidth]{Eff_F5_b.pdf}~
    \includegraphics[width=0.45\linewidth]{Eff_F6_b.pdf}
    \caption{Efficiency map of the F-type dedicated detectors in the $m_{\phi}-c\tau$ plane, for the LLPs produced via the process $B\to K\phi$.}
    \label{fig:eff_f_A}
\end{figure}

\begin{figure}[htb!]
    \centering
    \includegraphics[width=0.45\linewidth]{Eff_G1_b.pdf}~
    \includegraphics[width=0.45\linewidth]{Eff_G2_b.pdf}\\
    \includegraphics[width=0.45\linewidth]{Eff_G3_b.pdf}~
    \includegraphics[width=0.45\linewidth]{Eff_G4_b.pdf}\\
    \includegraphics[width=0.45\linewidth]{Eff_G5_b.pdf}~
    \includegraphics[width=0.45\linewidth]{Eff_G6_b.pdf}\\
    \includegraphics[width=0.45\linewidth]{Eff_G7_b.pdf}~
    \includegraphics[width=0.45\linewidth]{Eff_G8_b.pdf}\\
    \includegraphics[width=0.45\linewidth]{Eff_G9_b.pdf}
    \caption{Efficiency map of the G-type dedicated detectors in the $m_{\phi}-c\tau$ plane, for the LLPs produced via the process $B\to K\phi$.}
    \label{fig:eff_g_A}
\end{figure}

\begin{figure}[htb!]
    \centering
    \includegraphics[width=0.45\linewidth]{Eff_H1_b.pdf}~
    \includegraphics[width=0.45\linewidth]{Eff_H2_b.pdf}
    \caption{Efficiency map of the H-type dedicated detectors in the $m_{\phi}-c\tau$ plane, for the LLPs produced via the process $B\to K\phi$.}
    \label{fig:eff_h_A}
\end{figure}

\begin{figure}[htb!]
    \centering
    \includegraphics[width=0.45\linewidth]{Eff_I1_b.pdf}~
    \includegraphics[width=0.45\linewidth]{Eff_I2_b.pdf}\\
    \includegraphics[width=0.45\linewidth]{Eff_I3_b.pdf}~
    \includegraphics[width=0.45\linewidth]{Eff_I4_b.pdf}\\
    \includegraphics[width=0.45\linewidth]{Eff_I5_b.pdf}~
    \includegraphics[width=0.45\linewidth]{Eff_I6_b.pdf}
    \caption{Efficiency map of the I-type dedicated detectors in the $m_{\phi}-c\tau$ plane, for the LLPs produced via the process $B\to K\phi$.}
    \label{fig:eff_i_A}
\end{figure}

\clearpage

\section{Efficiency maps of dedicated detectors in the case of $h\to \phi\phi$}
\label{app:dedi_hphiphi}

\begin{figure}[htb!]
    \centering
    \includegraphics[width=0.45\linewidth]{Eff_A1_h.pdf}~
    \includegraphics[width=0.45\linewidth]{Eff_A2_h.pdf}\\
    \includegraphics[width=0.45\linewidth]{Eff_A3_h.pdf}~
    \includegraphics[width=0.45\linewidth]{Eff_A4_h.pdf}\\
    \includegraphics[width=0.45\linewidth]{Eff_A5_h.pdf}~
    \includegraphics[width=0.45\linewidth]{Eff_A6_h.pdf}
    \caption{Efficiency map of the A-type dedicated detectors in the $m_{\phi}-c\tau$ plane, for the LLPs produced via the process $h\to \phi\phi$.}
    \label{fig:eff_a_B}
\end{figure}

\begin{figure}[htb!]
    \centering
    \includegraphics[width=0.45\linewidth]{Eff_B1_h.pdf}~
    \includegraphics[width=0.45\linewidth]{Eff_B2_h.pdf}\\
    \includegraphics[width=0.45\linewidth]{Eff_B3_h.pdf}~
    \includegraphics[width=0.45\linewidth]{Eff_B4_h.pdf}\\
    \includegraphics[width=0.45\linewidth]{Eff_B5_h.pdf}
    \caption{Efficiency map of the B-type dedicated detectors in the $m_{\phi}-c\tau$ plane, for the LLPs produced via the process $h\to \phi\phi$.}
    \label{fig:eff_b_B}
\end{figure}

\begin{figure}[htb!]
    \centering
    \includegraphics[width=0.45\linewidth]{Eff_C1_h.pdf}~
    \includegraphics[width=0.45\linewidth]{Eff_C2_h.pdf}\\
    \includegraphics[width=0.45\linewidth]{Eff_C3_h.pdf}~
    \includegraphics[width=0.45\linewidth]{Eff_C4_h.pdf}\\
    \includegraphics[width=0.45\linewidth]{Eff_C5_h.pdf}~
    \includegraphics[width=0.45\linewidth]{Eff_C6_h.pdf}\\
    \includegraphics[width=0.45\linewidth]{Eff_C7_h.pdf}~
    \includegraphics[width=0.45\linewidth]{Eff_C8_h.pdf}
    \caption{Efficiency map of the C-type dedicated detectors in the $m_{\phi}-c\tau$ plane, for the LLPs produced via the process $h\to \phi\phi$.}
    \label{fig:eff_c_B}
\end{figure}

\begin{figure}[htb!]
    \centering
    \includegraphics[width=0.45\linewidth]{Eff_D1_h.pdf}~
    \includegraphics[width=0.45\linewidth]{Eff_D2_h.pdf}\\
    \includegraphics[width=0.45\linewidth]{Eff_D3_h.pdf}~
    \includegraphics[width=0.45\linewidth]{Eff_D4_h.pdf}\\
    \includegraphics[width=0.45\linewidth]{Eff_D5_h.pdf}~
    \includegraphics[width=0.45\linewidth]{Eff_D6_h.pdf}
    \caption{Efficiency map of the D-type dedicated detectors in the $m_{\phi}-c\tau$ plane, for the LLPs produced via the process $h\to \phi\phi$.}
    \label{fig:eff_d_B}
\end{figure}

\begin{figure}[htb!]
    \centering
    \includegraphics[width=0.45\linewidth]{Eff_E1_h.pdf}~
    \includegraphics[width=0.45\linewidth]{Eff_E2_h.pdf}\\
    \includegraphics[width=0.45\linewidth]{Eff_E3_h.pdf}~
    \includegraphics[width=0.45\linewidth]{Eff_E4_h.pdf}\\
    \includegraphics[width=0.45\linewidth]{Eff_E5_h.pdf}~
    \includegraphics[width=0.45\linewidth]{Eff_E6_h.pdf}\\
    \includegraphics[width=0.45\linewidth]{Eff_E7_h.pdf}~
    \includegraphics[width=0.45\linewidth]{Eff_E8_h.pdf}\\
    \includegraphics[width=0.45\linewidth]{Eff_E9_h.pdf}~
    \includegraphics[width=0.45\linewidth]{Eff_E10_h.pdf}
    \caption{Efficiency map of the E-type dedicated detectors in the $m_{\phi}-c\tau$ plane, for the LLPs produced via the process $h\to \phi\phi$.}
    \label{fig:eff_e_B}
\end{figure}

\begin{figure}[htb!]
    \centering
    \includegraphics[width=0.45\linewidth]{Eff_F1_h.pdf}~
    \includegraphics[width=0.45\linewidth]{Eff_F2_h.pdf}\\
    \includegraphics[width=0.45\linewidth]{Eff_F3_h.pdf}~
    \includegraphics[width=0.45\linewidth]{Eff_F4_h.pdf}\\
    \includegraphics[width=0.45\linewidth]{Eff_F5_h.pdf}~
    \includegraphics[width=0.45\linewidth]{Eff_F6_h.pdf}
    \caption{Efficiency map of the F-type dedicated detectors in the $m_{\phi}-c\tau$ plane, for the LLPs produced via the process $h\to \phi\phi$.}
    \label{fig:eff_f_B}
\end{figure}

\begin{figure}[htb!]
    \centering
    \includegraphics[width=0.45\linewidth]{Eff_G1_h.pdf}~
    \includegraphics[width=0.45\linewidth]{Eff_G2_h.pdf}\\
    \includegraphics[width=0.45\linewidth]{Eff_G3_h.pdf}~
    \includegraphics[width=0.45\linewidth]{Eff_G4_h.pdf}\\
    \includegraphics[width=0.45\linewidth]{Eff_G5_h.pdf}~
    \includegraphics[width=0.45\linewidth]{Eff_G6_h.pdf}\\
    \includegraphics[width=0.45\linewidth]{Eff_G7_h.pdf}~
    \includegraphics[width=0.45\linewidth]{Eff_G8_h.pdf}\\
    \includegraphics[width=0.45\linewidth]{Eff_G9_h.pdf}
    \caption{Efficiency map of the G-type dedicated detectors in the $m_{\phi}-c\tau$ plane, for the LLPs produced via the process $h\to \phi\phi$.}
    \label{fig:eff_g_B}
\end{figure}

\begin{figure}[htb!]
    \centering
    \includegraphics[width=0.45\linewidth]{Eff_H1_h.pdf}~
    \includegraphics[width=0.45\linewidth]{Eff_H2_h.pdf}
    \caption{Efficiency map of the H-type dedicated detectors in the $m_{\phi}-c\tau$ plane, for the LLPs produced via the process $h\to \phi\phi$.}
    \label{fig:eff_h_B}
\end{figure}

\begin{figure}[htb!]
    \centering
    \includegraphics[width=0.45\linewidth]{Eff_I1_h.pdf}~
    \includegraphics[width=0.45\linewidth]{Eff_I2_h.pdf}\\
    \includegraphics[width=0.45\linewidth]{Eff_I3_h.pdf}~
    \includegraphics[width=0.45\linewidth]{Eff_I4_h.pdf}\\
    \includegraphics[width=0.45\linewidth]{Eff_I5_h.pdf}~
    \includegraphics[width=0.45\linewidth]{Eff_I6_h.pdf}
    \caption{Efficiency map of the I-type dedicated detectors in the $m_{\phi}-c\tau$ plane, for the LLPs produced via the process $h\to \phi\phi$.}
    \label{fig:eff_i_B}
\end{figure}

\clearpage

\providecommand{\href}[2]{#2}\begingroup\raggedright\endgroup

\end{document}